\documentclass[useAMS,usenatbib,twocolumn]{mn2e}
\usepackage{amsmath}
\usepackage{amssymb}
\usepackage{graphicx}
\usepackage{color}
\usepackage{amsmath,amsbsy}

\title[]{Star formation efficiencies of molecular clouds in a galactic center environment}
\author[Bertram et al.]{Erik~Bertram$^1$, Simon~C.~O.~Glover$^1$, Paul~C.~Clark$^{1,2}$, Ralf~S.~Klessen$^{1,3,4}$\\
$^1$Zentrum f\"ur Astronomie der Universit\"at Heidelberg, Institut f\"ur Theoretische  Astrophysik, Albert-Ueberle-Str.~2, 69120 Heidelberg, Germany\\
$^2$School of Physics and Astronomy, Cardiff University, CF24 3AA, UK \\
$^3$Department of Astronomy and Astrophysics, University of California, 1156 High Street, Santa Cruz, CA 95064, USA \\
$^4$Kavli Institute for Particle Astrophysics and Cosmology, Stanford University, SLAC, Menlo Park, CA 94025, USA \\
}

\begin{document}

\maketitle

\abstract
We use the {\sc Arepo} moving mesh code to simulate the evolution of molecular clouds exposed to a harsh environment similar to that found in the galactic center (GC), in an effort to understand why the star formation efficiency (SFE) of clouds in this environment is so small. Our simulations include a simplified treatment of time-dependent chemistry and account for the highly non-isothermal nature of the gas and the dust. We model clouds with a total mass of $1.3 \times 10^{5} \: {\rm M_{\odot}}$ and explore the effects of varying the mean cloud density and the virial parameter, $\alpha = E_{\rm kin} / | E_{\rm pot}|$. We vary the latter from $\alpha = 0.5$ to $\alpha = 8.0$, and so many of the clouds that we simulate are gravitationally unbound. We expose our model clouds to an interstellar radiation field (ISRF) and cosmic ray flux (CRF) that are both a factor of 1000 higher than the values found in the solar neighbourhood. As a reference, we also run simulations with local solar neighbourhood values of the ISRF and the CRF in order to better constrain the effects of the extreme conditions in the GC on the SFE. Despite the harsh environment and the large turbulent velocity dispersions adopted, we find that all of the simulated clouds form stars within less than a gravitational free-fall time. Increasing the virial parameter from $\alpha = 0.5$ to $\alpha = 8.0$ decreases the SFE by a factor $\sim 4-10$, while increasing the ISRF/CRF by a factor of 1000 decreases the SFE again by a factor $\sim2-6$. However, even in our most unbound clouds, the SFE remains higher than that inferred for real GC clouds. We therefore conclude that high levels of turbulence and strong external heating are not enough by themselves to lead to a persistently low SFE at the center of the Galaxy.
\endabstract


\section{Introduction}
\label{sec:introduction}

Understanding star formation is an important task in theoretical astrophysics \citep[see e.g.,][]{MacLowAndKlessen2004,ScaloAndElmegreen2004,ElmegreenAndScalo2004,McKeeAndOstriker2007,Ballesteros-ParedesEtAl2007}. Stars form in molecular clouds (MC) in the interstellar medium (ISM) due to gravitational contraction of overdense gas regions \citep[see also the lecture notes by][]{KlessenAndGlover2014}. Several studies in the past have revealed that the number of stars formed per unit time (referred to as the star formation rate or SFR) is proportional to the amount of gas in the star-forming region \citep[see e.g.,][]{Schmidt1959,Kennicutt1998} and that the SFR surface density shows a power-law dependence on the gas surface density. This is known as the Schmidt-Kennicutt relation and is an important empirical result in star formation theory. It appears to hold in the disk of our Milky Way and also in distant galaxies \citep{KennicuttAndEvans2012}. However, there is still some debate about whether a universal relationship holds for all galaxies or not \citep[see, e.g.][]{ShettyEtAl2013,ShettyEtAl2014a}.

As shown by \citeauthor{LongmoreEtAl2013a}~(2013a), the inner 500~pc region of our Galaxy, known as the Central Molecular Zone (CMZ), contains the largest reservoir of dense gas in the Milky Way, with densities of order of several $10^3\,$cm$^{-3}$ and a total mass of about $10^7\,$M$_{\odot}$. This would be enough to form several Orion-like clusters, but the measured SFR is significantly smaller (\citeauthor{LongmoreEtAl2013a}~2013a). This is a direct contradiction to the Schmidt-Kennicutt relation, which would predict a high SFR in the center of our Galaxy. The key questions are, why on one hand the observed SFR is suppressed by a factor of $\gtrsim10$ (\citeauthor{LongmoreEtAl2013a}~2013a) and on the other hand which physical processes regulate star formation in the central few hundred parsecs of the Milky Way (see, e.g. \citeauthor{LongmoreEtAl2013a}~2013a, \citeauthor{KruijssenEtAl2014}~2014, \citeauthor{KrumholzEtAl2015}~2015).

Several studies have tried to solve this problem of star formation in such an extreme environment, which challenges current star formation theories. For example, \citet{KruijssenEtAl2014} studied the impact of several mechanisms on the SFR on different physical scales, e.g. the very strong radiation field, magnetic fields, turbulent pressure, galactic tides or feedback. They argue that star formation could be episodic due to a gradual build-up of dense gas by spiral instabilities or that variations in the rates of gas flows into the CMZ might significantly alter the star formation process. \citeauthor{LongmoreEtAl2013a}~(2013a) state, for example, that the order of magnitude higher internal velocity dispersion could disrupt compact regions before they are able to go into gravitational collapse. However, the question of what physical processes are most important for regulating the SFR in the galactic center (GC) remains unresolved.

One important way in which we can distinguish between different models for the regulation of star formation in the GC is by examining whether they produce low star formation efficiencies by reducing the formation of stars in the individual dense clouds present in the CMZ, or whether they instead merely regulate the rate at which molecular gas is allowed to form these dense clouds. The presence of a number of massive, dense clouds in the CMZ that do not appear to be currently forming stars provides a hint that the star formation efficiencies of individual clouds in the CMZ might be low. An example of such a cloud, which is almost devoid of star formation, is also known as ``The Brick'' (see e.g. \citealt{GuestenEtAl1981,LisEtAl1994,LisAndMenten1998,LisEtAl2001,MolinariEtAl2011,ImmerEtAl2012,LongmoreEtAl2012,KauffmannEtAl2013,ClarkEtAl2013,JohnstonEtAl2014,RathborneEtAl2014,KruijssenEtAl2015,PillaiEtAl2015}, as well as \citeauthor{LongmoreEtAl2013b}~2013b for a discussion of several other dense starless clouds close to the Brick that have similar properties). On the other hand, it is also possible that we are just observing the Brick and its neighbouring starless clouds at a very early time in their evolution, before they have started to form stars. This question is difficult to resolve observationally, but numerical simulations can help us to understand which of the possibilities is more likely.

In this paper we investigate the impact of a strong interstellar radiation field (ISRF) and a high cosmic ray flux (CRF) on the SFR in clouds with different initial number densities and varying levels of turbulence. We change the amount of turbulent kinetic energy $E_{\text{kin}}$ with respect to the potential energy $E_{\text{pot}}$, as parameterized by $\alpha = E_{\text{kin}} / |E_{\text{pot}}|$ (see Section \ref{subsec:quantities}). We focus on the question of whether such a high ISRF/CRF combined with a high level of turbulence could be the main physical drivers to suppress star formation in the GC. Therefore, we adopt environmental conditions that are similar to those experienced by a typical GC cloud. In particular, we adopted values for the ISRF strength and the CRF comparable to those inferred for the Brick. We model the behaviour of different MCs exposed to this environment using the A{\sc repo} moving mesh code \citep{Springel2010} and explore the effect of changing their density and turbulent velocity dispersion. For technical reasons, the clouds we model have lower densities than typical GC clouds, and so should be less likely to form stars than real GC clouds. Despite this and despite the harshness of their environment, it proves to be very difficult to suppress star formation within the clouds, as we will see in the following sections.

The structure of our paper is as follows. In Section \ref{sec:methodsandsims} we present the simulations and the methods used in this paper. In Section \ref{sec:results} we present the results of our studies with different virial $\alpha$ parameters and various initial number densities. We discuss our results in Section \ref{sec:discussion} and present our conclusions in Section \ref{sec:summary}.

\section{Methods and simulations}
\label{sec:methodsandsims}

\subsection{Computational method}
\label{subsec:commethod}

Our simulations are performed using the moving mesh code A{\sc repo} \citep{Springel2010}, which uses an unstructured mesh defined by the Voronoi tessellation of a set of discrete points. We make use of a detailed atomic and molecular cooling function, described in detail in \citet{GloverEtAl2010} and \citet{GloverAndClark2012b}, and a simplified treatment of the molecular chemistry of the gas. Our chemical treatment is based on the work of \citet{NelsonAndLanger1997} and \citet{GloverAndMacLow2007}, and allows us to follow the formation and destruction of H$_{2}$ and CO self-consistently within our simulations. Full details of the chemical model with a description of how the chemistry interacts with the ISRF via the T{\sc ree}C{\sc ol} algorithm can be found in \citet{ClarkEtAl2012}. 
Examples of the use of our chemical model with the A{\sc repo} code can be found in \citeauthor{SmithEtAl2014a}~(2014a) and \citeauthor{SmithEtAl2014b}~(2014b).

We assume that the gas has a uniform solar metallicity and adopt the standard ratio of helium to hydrogen, and abundances of carbon and oxygen taken from \citet{SembachEtAl2000}, i.e. $x_{\text{C}} = 1.4 \times 10^{-4}$ and $x_{\text{O}} = 3.2 \times 10^{-4}$, where $x_{\text{C}}$ and $x_{\text{O}}$ are the fractional abundances by number of carbon and oxygen relative to hydrogen. However, we have to keep in mind that the CMZ has actually a super-solar metallicity. Nevertheless, we use a uniform solar value in order to be conservative regarding the cooling and star formation rates in our runs. At the start of the simulations, hydrogen, helium and oxygen are in atomic form, while carbon is assumed to be in singly ionized form, as C$^{+}$. We also adopt the standard local value for the dust-to-gas ratio of 1:100 \citep{GloverEtAl2010}, and assume that the dust properties do not vary with the gas density. The cosmic ray ionization rate of atomic hydrogen is set to $\zeta = 3 \times 10^{-14} \: {\rm s^{-1}}$ \citep{ClarkEtAl2013}, which is a factor of $\sim1000$ higher than the value in the solar neighbourhood \citep{Yusef-ZadehEtAl2007}. For the incident ultraviolet radiation field, we adopt the same spectral slope as given in \citet{Draine1978}. We denote the strength of the Draine ISRF as $G_0 = 1$ and perform simulations with a field strength $G_0 = 1000$ \citep{ClarkEtAl2013}. The Draine field has a strength $G_0 = 1.7$ in \citet{Habing1968} units, corresponding to an integrated flux of $2.7 \times 10^{-3}\,$erg\,cm$^{-2}$s$^{-1}$. Furthermore, as a reference, we also run simulations with local solar neighbourhood values of the ISRF and the CRF in order to explore the effect of a different radiation field on the SFE. In this case, we set the field strength of the ISRF to $G_0 = 1$ and the cosmic ray ionization rate of atomic hydrogen to $\zeta = 3 \times 10^{-17} \: {\rm s^{-1}}$. Our simulations use a Jeans refinement criterion, which is active over the whole simulation period in order to accurately refine dense and collapsed gas regions in the box. We use a constant number of 8 cells per Jeans length, which is sufficient to avoid artificial fragmentation \citep{TrueloveEtAl1998,GreifEtAl2011}.

\subsection{Sink particles}
\label{subsec:sinkmethod}

Furthermore, we make use of a sink particle implementation \citep{GreifEtAl2011} based on the prescription in \citet{BateEtAl1995} and \citet{JappsenEtAl2005}, to track the star formation process during the simulations. Before a cell is turned into a sink particle, it undergoes a series of tests. First, the particle must reach a critical density threshold of $n_{\text{thresh}} \approx 10^7\,$cm$^{-3}$. This value was chosen to be higher than the typical post-shock densities found in the clouds to ensure that we do not attempt to form sinks in regions that are not gravitationally collapsing. For our more extreme run with $n_0 = 10^4\,$cm$^{-3}$ we adopt a higher threshold of $n_{\text{thresh}} \approx 10^9\,$cm$^{-3}$ in order to account for the two orders of magnitude higher density on average. The second test is to check whether the new sink particle is sufficiently far away from any other sink particle, measured in terms of one accretion radius $r_{\text{acc}}$. The third is to check whether the size of the cell is less than the accretion radius of the sink particle that it will become.

We set the accretion radius to a constant value of $r_{\text{acc}}\sim0.01\,$pc, roughly corresponding to the scale of the thermal Jeans length for a sink particle formation threshold of $n_{\text{thresh}} \approx 10^7\,$cm$^{-3}$ at a mean gas temperature of $\sim20\,$K. For our model with a threshold of $n_{\text{thresh}} \approx 10^9\,$cm$^{-3}$, we adopt the same value for the accretion radius. This is also a reasonable estimate for this higher density model, since more checks will guarantee that the gas is bound and collapsing onto the sink particle. In addition, we explore whether a different value of $r_{\text{acc}}$ might have a significant impact on the SFEs by running a simulation in which we increased $r_{\text{acc}}$ by a factor of 10. We find that in this case, the SFE increases by around a factor of $1.1-1.3$. In view of the various error sources for estimating a star formation efficiency per free-fall time (see also the discussion in Section \ref{subsec:analysis}), we find that a different accretion radius does not significantly affect the results of this paper. This is because we are mainly interested in estimating the total mass that goes into gravitational collapse instead of measuring precise SFEs.

Once these preliminary criteria are fulfilled, more checks guarantee that this potential sink particle is in a correct dynamical state. In a first test, we require that the particle is sub-virial, i.e. it needs to fulfill the condition $\alpha \le 0.5$, where $\alpha$ is the ratio of kinetic and gravitational energy. Second, we ensure that $\alpha + \beta \le 1$, where $\beta$ is the ratio of rotational and gravitational energy. Third, we require $\text{div} (\mathbf{a}) < 0$, where $\mathbf{a}$ is the acceleration, which ensures that the particle is not tidally disrupted or bouncing. If all these conditions are achieved, the local gas condensation can become a sink.

\subsection{Important quantities}
\label{subsec:quantities}

In this study, we use the virial $\alpha$ parameter to regulate the amount of turbulent kinetic energy in the simulation domain. We define this as
\begin{equation}
\alpha = \frac{E_{\text{kin}}}{|E_{\text{pot}}|},
\end{equation}
where $E_{\text{kin}}$ and $E_{\text{pot}}$ denote the total kinetic and potential energy in the box at the start of the simulation, given via
\begin{equation}
E_{\text{kin}} = \frac{1}{2}M_{\text{tot}}\sigma_v^2
\end{equation}
and
\begin{equation}
E_{\text{pot}} = -\frac{3GM_{\text{tot}}^2}{5R}.
\end{equation}
In this context, $\sigma_v$ denotes the turbulent 3D velocity dispersion, $M_{\text{tot}}$ and $R$ the total mass and radius of the (initially) uniform sphere and $G$ the gravitational constant. Furthermore, we use
\begin{equation}
M_{\text{tot}} = \frac{4}{3}\pi R^3 \rho,
\end{equation}
where $\rho$ is the initial mass density. However, we note that the virial parameter is also often defined via $\alpha = 2E_{\text{kin}} / |E_{\rm pot}|$ in the literature, which is different from the notation used above. In our definition, a value of $\alpha = 0.5$ defines virialized clouds, $\alpha = 1.0$ denotes clouds with energy equipartition and $\alpha > 1.0$ describes clouds that are (highly) unbound. In addition, we can estimate the crossing time from these quantities via $t_{\text{cross}} \approx R / \sigma_v$.

Furthermore, we quantify the amount of gas mass being converted to stars (i.e. sink particles) due to gravitational collapse within one free-fall time $t_{\text{ff}}$ as
\begin{equation}
\epsilon_{\text{ff}} = t_{\text{ff}} \cdot \frac{\dot M_{*}}{M_{\text{tot}}},
\end{equation}
where $\dot M_{*}$ is the star formation rate in the computational domain averaged over $t_{\text{ff}}$ \citep{KrumholzAndMcKee2005,KrumholzAndTan2007,Murray2011}. The parameter $\epsilon_{\text{ff}}$ is thus a measure of the star formation efficiency (SFE) of each model. The free-fall timescale itself is defined via
\begin{equation}
t_{\text{ff}} = \sqrt{\frac{3\pi}{32G\rho}}.
\end{equation}
In this paper, we compute $\epsilon_{\text{ff}}$ for all clouds individually and compare our results to the average SFE inferred for the galactic center region. It is therefore important to emphasize that our study is aimed at testing the idea that the SFE at the galactic center is low because the SFE of the individual dense clouds is low. With our isolated cloud models, we cannot test the competing idea that the SFE of the region is low because it is difficult to form dense clouds there.

\subsection{Initial conditions and model parameters}
\label{subsec:ics}

\begin{table*}
\begin{tabular}{l|c|c|c|c|c|c|c}
\hline\hline
Model name & Virial $\alpha$ & Initial density $n_0$ & Radius $R$ & Velocity $\sigma_v$ & Free-fall time $t_{\text{ff}}$ & Crossing time $t_{\text{cross}}$ & ISRF \& CRF \\
 & & [cm$^{-3}$] & [pc] & [km/s] & [Myr] & [Myr] \\
\hline
GC-0.5-100 & 0.5 & 100 & 19.1 & 3.7 & 4.4 & 5.1 & 1000 \\
GC-1.0-100 & 1.0 & 100 & 19.1 & 5.2 & 4.4 & 3.4 & 1000 \\
GC-2.0-100 & 2.0 & 100 & 19.1 & 7.4 & 4.4 & 2.5  & 1000 \\
GC-4.0-100 & 4.0 & 100 & 19.1 & 10.5 & 4.4 & 1.8 & 1000 \\
GC-8.0-100 & 8.0 & 100 & 19.1 & 14.7 & 4.4 & 1.3 & 1000 \\
\hline
GC-0.5-1000 & 0.5 & 1000 & 8.9 & 5.4 & 1.4 & 1.6 & 1000 \\
GC-1.0-1000 & 1.0 & 1000 & 8.9 & 7.6 & 1.4 & 1.1 & 1000 \\
GC-2.0-1000 & 2.0 & 1000 & 8.9 & 10.8 & 1.4 & 0.8 & 1000 \\
GC-4.0-1000 & 4.0 & 1000 & 8.9 & 15.3 & 1.4 & 0.6 & 1000 \\
GC-8.0-1000 & 8.0 & 1000 & 8.9 & 21.6 & 1.4 & 0.4 & 1000 \\
\hline
GC-16.0-10000 & 16.0 & 10000 & 4.1 & 44.9 & 0.4 & 0.09 & 1000\\
\hline
SOL-0.5-1000 & 0.5 & 1000 & 8.9 & 5.4 & 1.4 & 1.6 & 1 \\
SOL-8.0-1000 & 8.0 & 1000 & 8.9 & 21.6 & 1.4 & 0.4 & 1 \\
\hline
\hline
The Brick & 1.0 & $7.3 \times 10^4$ & 2.8 & 16.0 & 0.34 & 0.17 & $\sim100-1000$ \\
\hline
\end{tabular}
\caption{Initial conditions for our different cloud models. For each run, we list the virial $\alpha$ parameter, the initial number density, the spherical cloud radius, the velocity dispersion, the free-fall time, the crossing time and the scaling factor of the interstellar radiation field (ISRF) and the cosmic ray flux (CRF) relative to the solar neighourhood value. For comparison, we also list the parameters for the GC cloud G0.253+0.016, known as ``the Brick'', which are given in Table 2 in \citet{LongmoreEtAl2012}.}
\label{tab:setup}
\end{table*}

For simplicity, we assume that the cloud is initially spherical and embedded in a low-density environment. Furthermore, we apply periodic boundary conditions. We perform two sets of runs in which we vary the virial $\alpha$ parameter, using values $\alpha = 0.5, 1.0, 2.0, 4.0\,$ and $8.0$. In the first set of runs, we take an initial hydrogen nuclei number density of $n_0 = 100\,$cm$^{-3}$ for the cloud. In the second set of runs, we take a number density of $1000\,$cm$^{-3}$. In all of our simulations, the density of the gas surrounding the cloud is $\approx1$\,cm$^{-3}$. We use a total mass of $M_{\text{tot}} = 1.3 \times 10^5\,$M$_{\odot}$ and an initial total number of $2 \times 10^6$ cells for the whole box, including the low-density regions. The initial cell mass within the uniform cloud corresponds to $\approx2\,$M$_{\odot}$.

Given these parameters, we can compute the radii and the initial velocity dispersions of the clouds. The cubic side lengths of the total simulation domain are set to $5 \times$ the individual cloud radii. All clouds are initially at rest and placed in the center of the box. We use an initial random velocity field with a power spectrum of $P(k) \propto k^{-4}$ that consists of a natural mixture of solenoidal and compressive modes, which decays throughout the simulation. The timesteps between the individual snapshots are $\Delta t_{\text{snap}} \approx 22\,$kyr and $\Delta t_{\text{snap}} \approx 7\,$kyr for models with initial number densities of $n_0 = 100\,$cm$^{-3}$ and $1000\,$cm$^{-3}$, respectively, corresponding to $\sim200$ snapshots per simulation in total.

In addition, we have also modelled an even more extreme cloud with an initial number density of $n_0 = 10^4\,$cm$^{-3}$ and a virial $\alpha$ parameter of $\alpha = 16.0$. In this case, we set $\Delta t_{\text{snap}} \approx 2\,$kyr for the same total number of time snapshots. However, running further simulations with $n_0 \ge 10^4\,$cm$^{-3}$ and $\alpha < 16.0$ is computationally prohibitive, owing to the rapid rate at which stars form in these models. For this reason we only focus on this one more extreme numerical model. Nevertheless, as we discuss in more detail later, such a high $\alpha$ run alone is still enough for our study, since even in this extreme case, the cloud still forms a substantial number of stars. Furthermore, regarding the densities in the CMZ, we have to keep in mind that a significant number of GC clouds have densities higher than those modelled in our simulations, which would lead to even larger turbulent velocities in our simulations for the same constant total mass. Table \ref{tab:setup} gives an overview about the different numerical models and their initial conditions.

\section{Results}
\label{sec:results}

\subsection{Analysis of the model clouds}
\label{subsec:analysis}

\begin{figure*}
\centerline{
\includegraphics[height=0.25\linewidth]{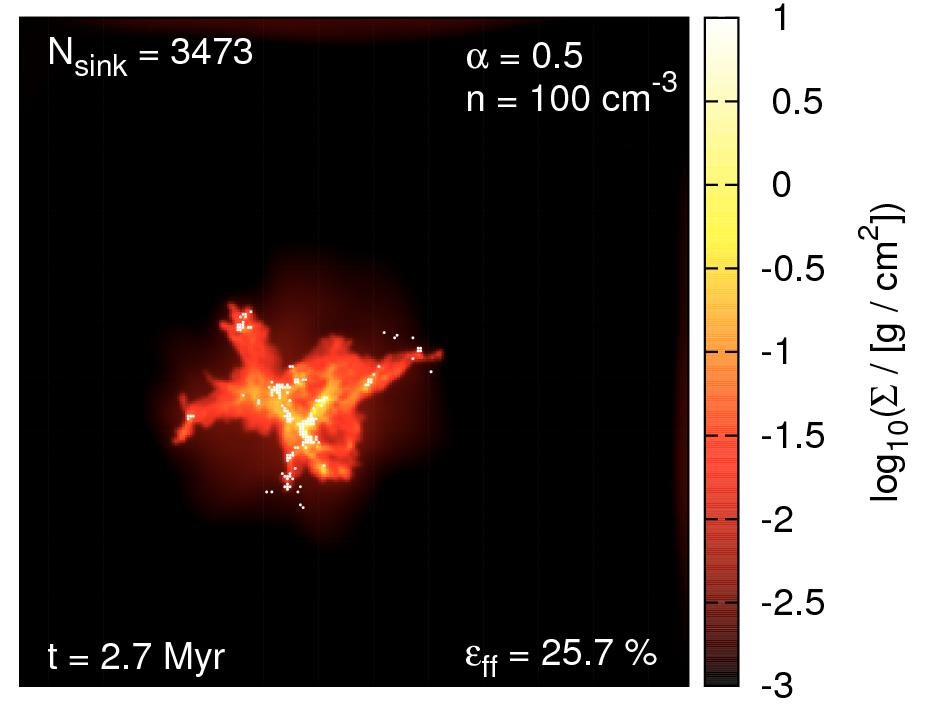}
\includegraphics[height=0.25\linewidth]{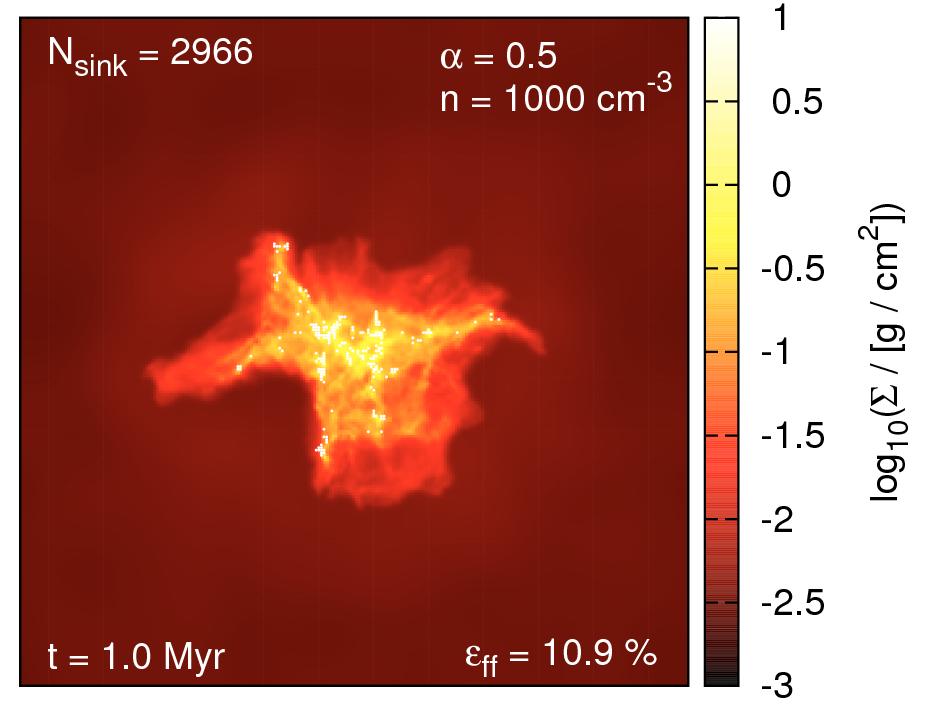}
}
\centerline{
\includegraphics[height=0.25\linewidth]{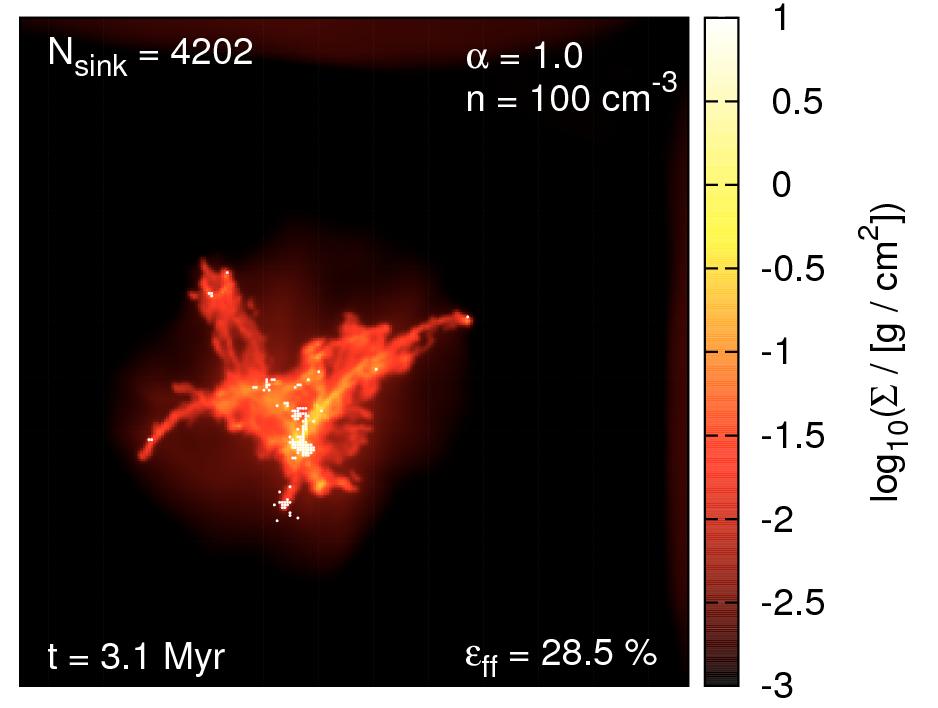}
\includegraphics[height=0.25\linewidth]{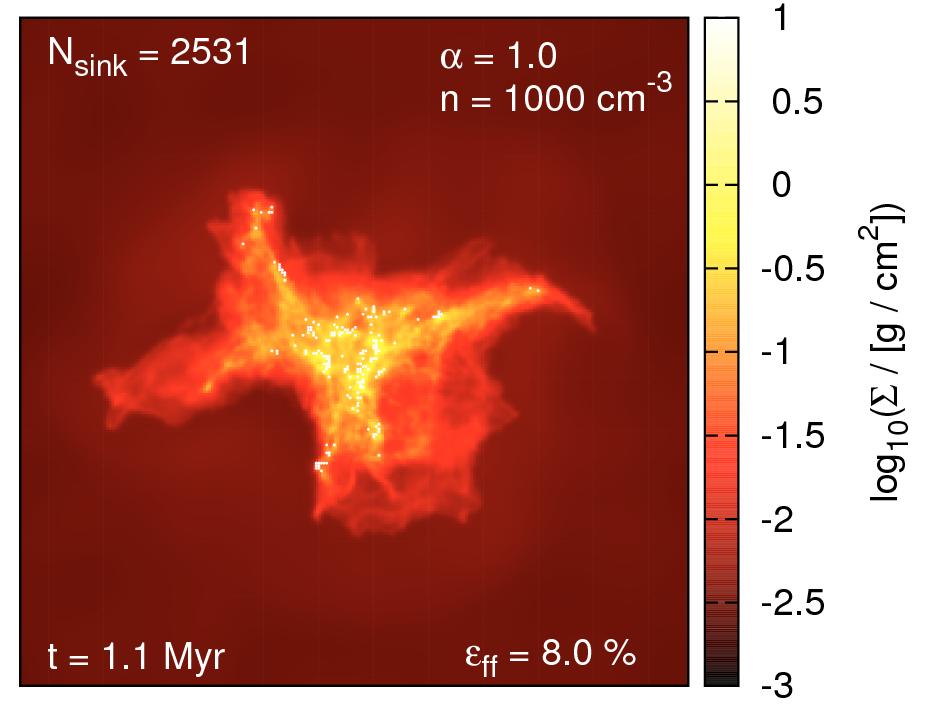}
}
\centerline{
\includegraphics[height=0.25\linewidth]{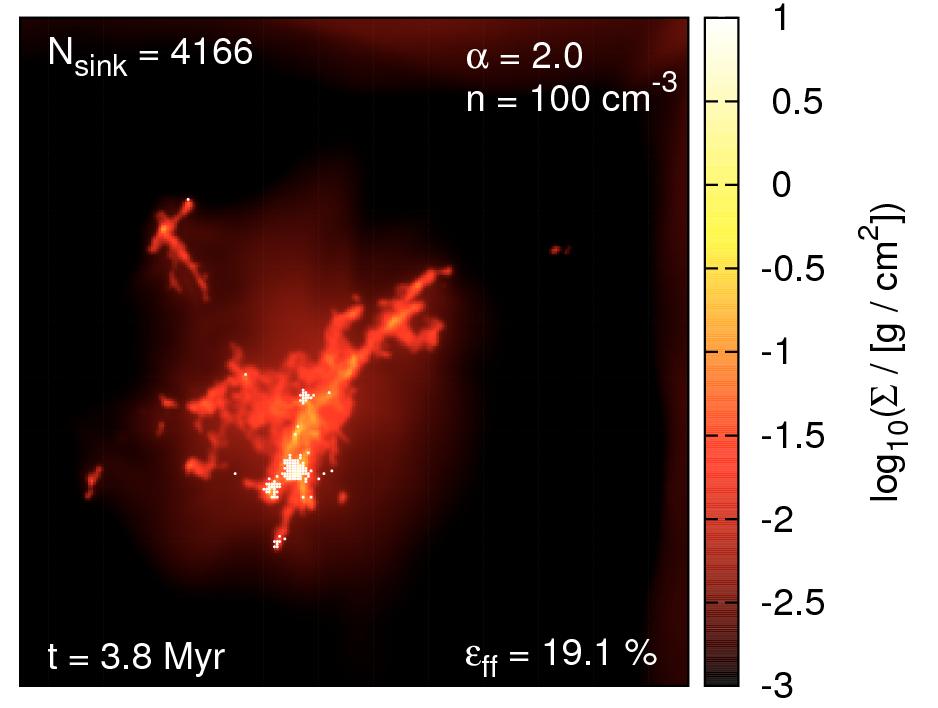}
\includegraphics[height=0.25\linewidth]{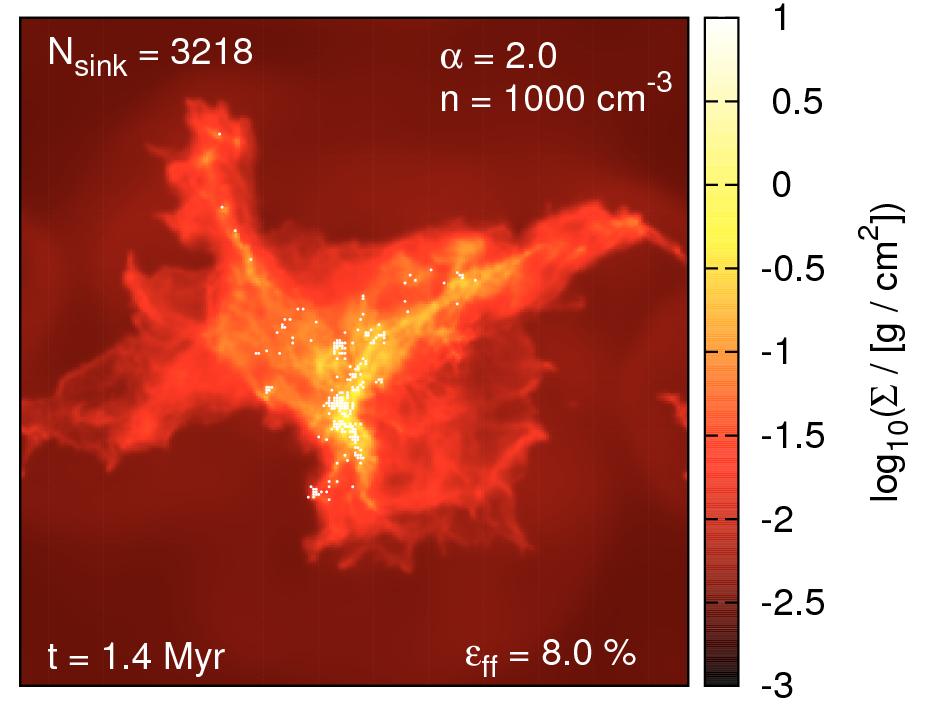}
}
\centerline{
\includegraphics[height=0.25\linewidth]{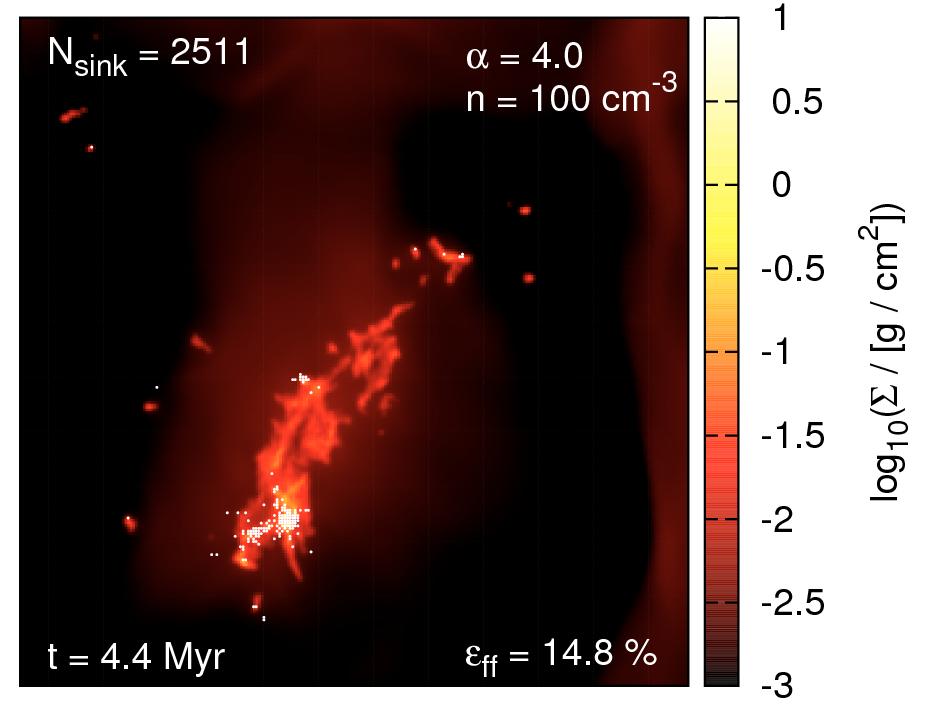}
\includegraphics[height=0.25\linewidth]{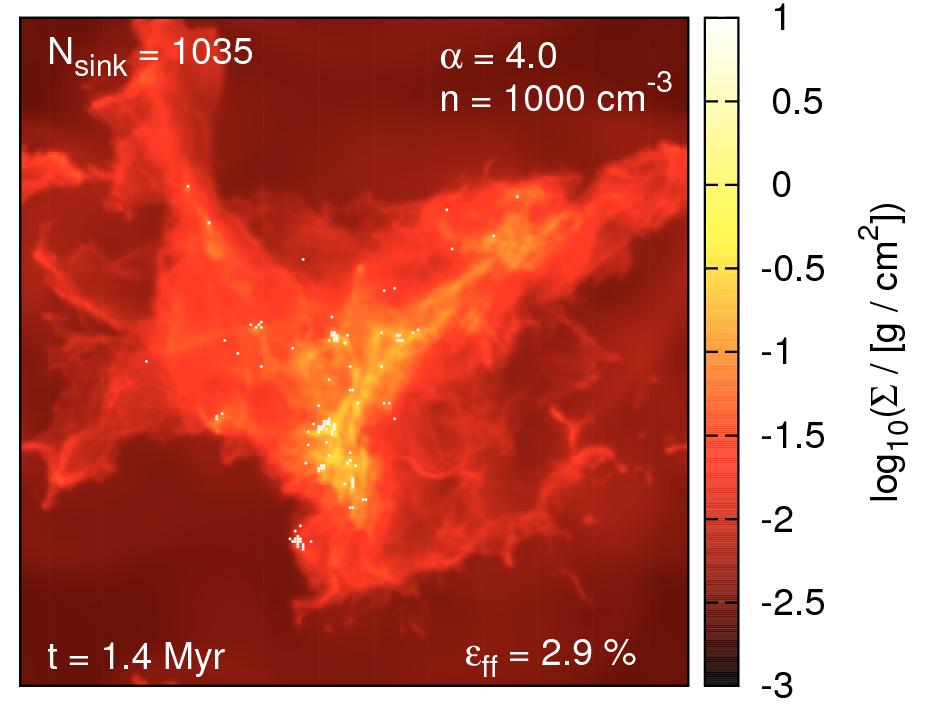}
}
\centerline{
\includegraphics[height=0.25\linewidth]{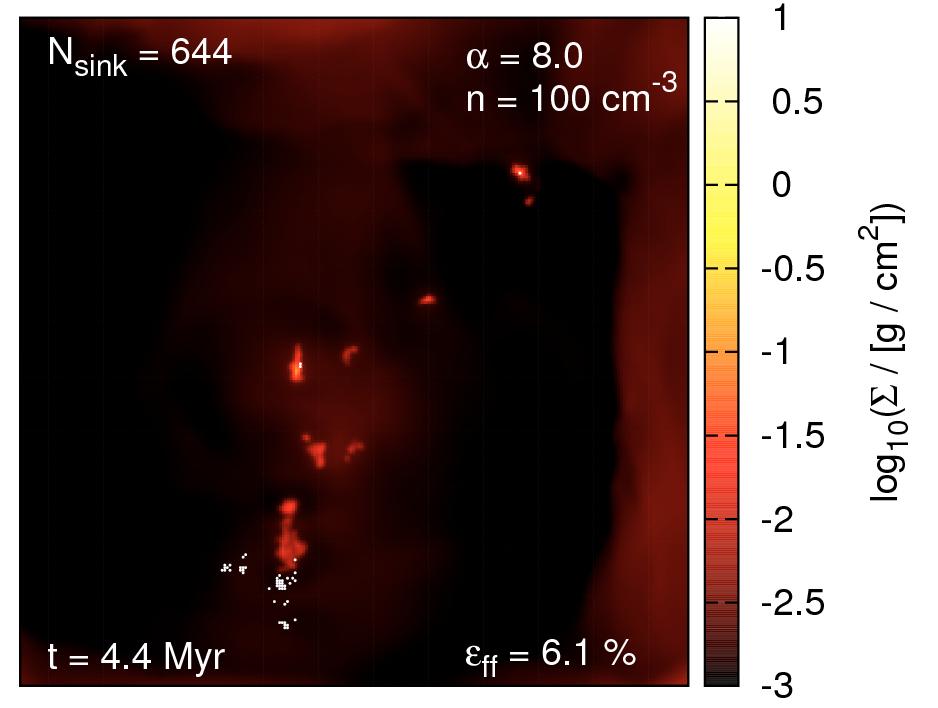}
\includegraphics[height=0.25\linewidth]{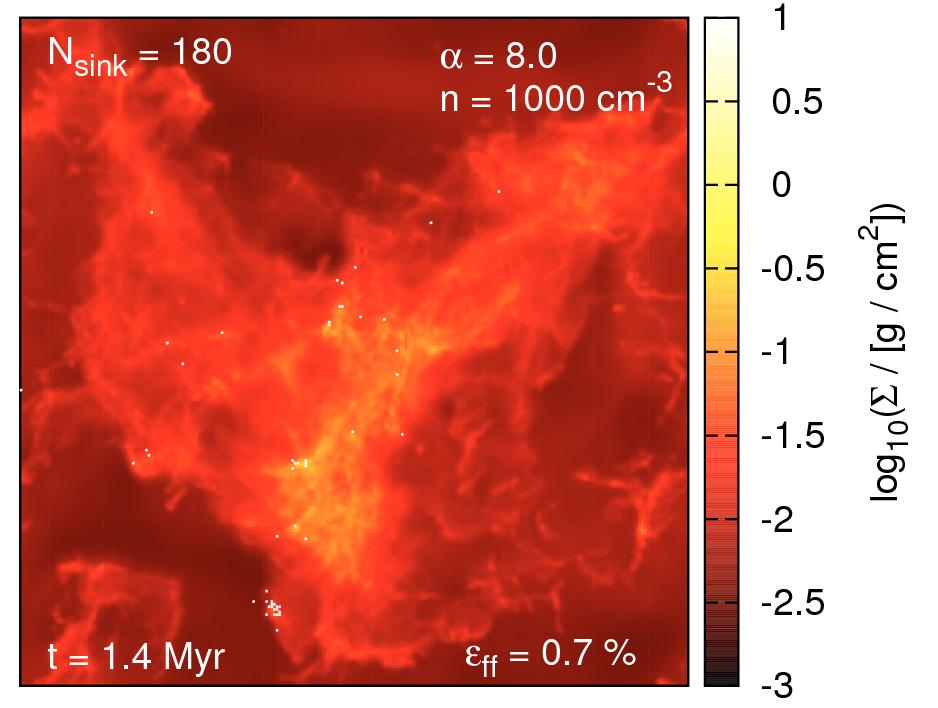}
}
\caption{Logarithmic column density maps in z-direction for all cloud models with initial number densities of $n_0 = 100\,$cm$^{-3}$ (left column) and $1000\,$cm$^{-3}$ (right column) for different virial parameters $\alpha = 0.5, 1.0, 2.0, 4.0, 8.0$ (from top to bottom), showing an extract of the central cloud regions. In each plot we give the number of sink particles, the estimated star formation efficiency per free-fall time and the simulation time. Sink particles are formed during the simulations and marked with white dots in each map. The side length of each box shown above corresponds to 74.5\,pc and 44.5\,pc for the $100\,$cm$^{-3}$ and $1000\,$cm$^{-3}$ models, respectively.}
\label{fig:images}
\end{figure*}

\begin{figure}
\centerline{
\includegraphics[height=0.74\linewidth]{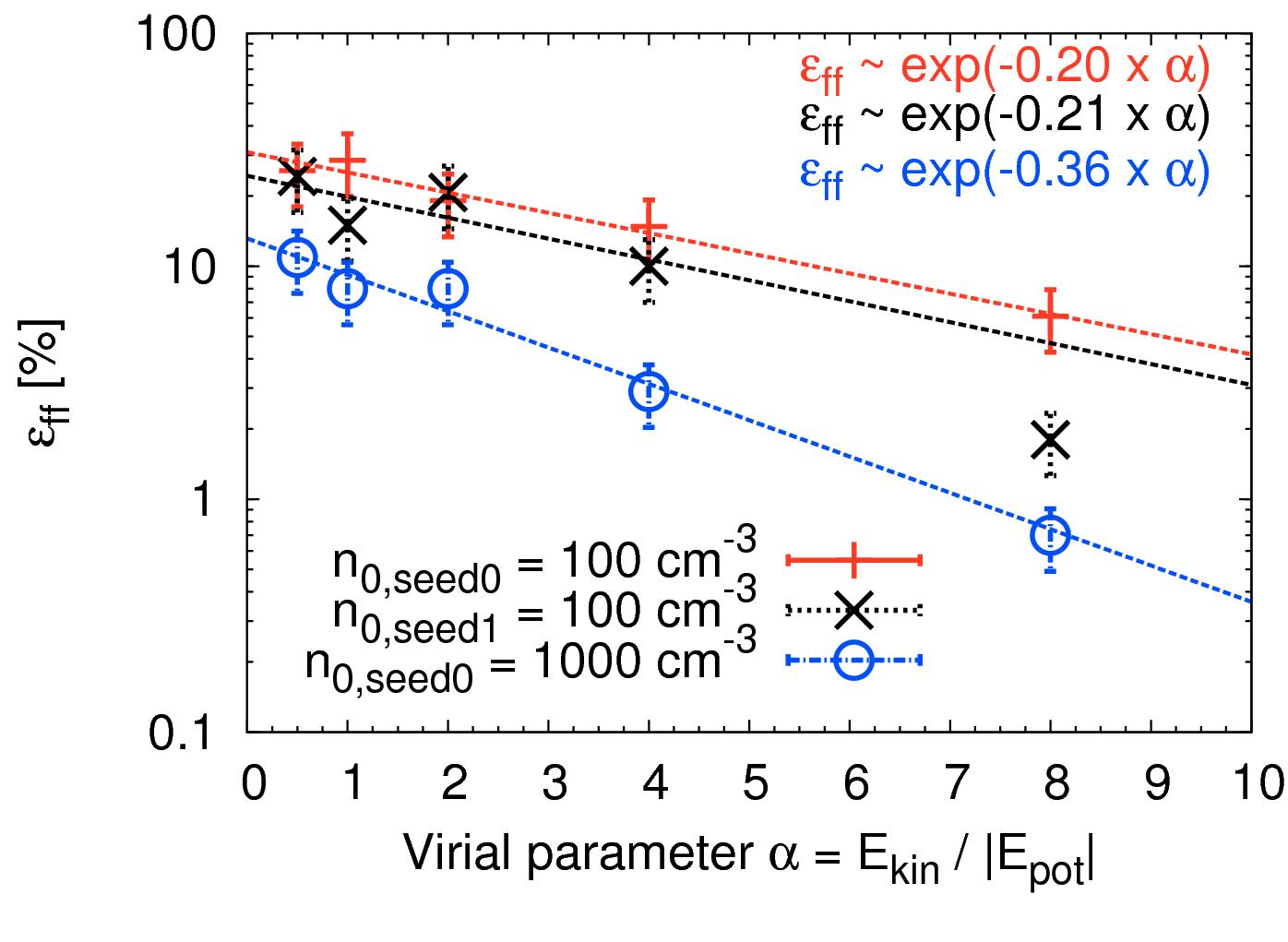} \\
}
\caption{Star formation efficiencies per free-fall time against the corresponding virial $\alpha$ parameter for our fiducial models with $n_{0,\text{seed0}} = 100\,$cm$^{-3}$ and $n_{0,\text{seed0}} = 1000\,$cm$^{-3}$ (Table \ref{tab:SFE}) and for one model using a different seed of the turbulent velocity field with $n_{0,\text{seed1}} = 100\,$cm$^{-3}$ (Table \ref{tab:SFE_seed}). We generally find a decreasing trend of the SFEs with higher $\alpha$ for all density models. Although there are slight differences in the individual SFEs between the two models with $n_{0,\text{seed0}} = 100\,$cm$^{-3}$ and $n_{0,\text{seed1}} = 100\,$cm$^{-3}$, we nevertheless find the same general trends of decreasing $\epsilon_{\text{ff}}$ with increasing $\alpha$. To illustrate the different trends, we also fit exponential functions $\propto \exp(-c\alpha)$ to the models and list the corresponding slopes in the plot. Error bars indicate variations of the SFE of $\sim30\%$, which we conservatively estimate in Section \ref{subsec:analysis}.}
\label{fig:SFEvsAlpha}
\end{figure}

\begin{figure*}
\centerline{
\includegraphics[height=0.03\linewidth]{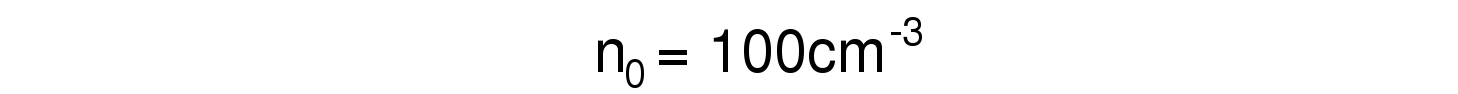}
\includegraphics[height=0.03\linewidth]{images/n100.jpg}
}
\centerline{
\includegraphics[height=0.237\linewidth]{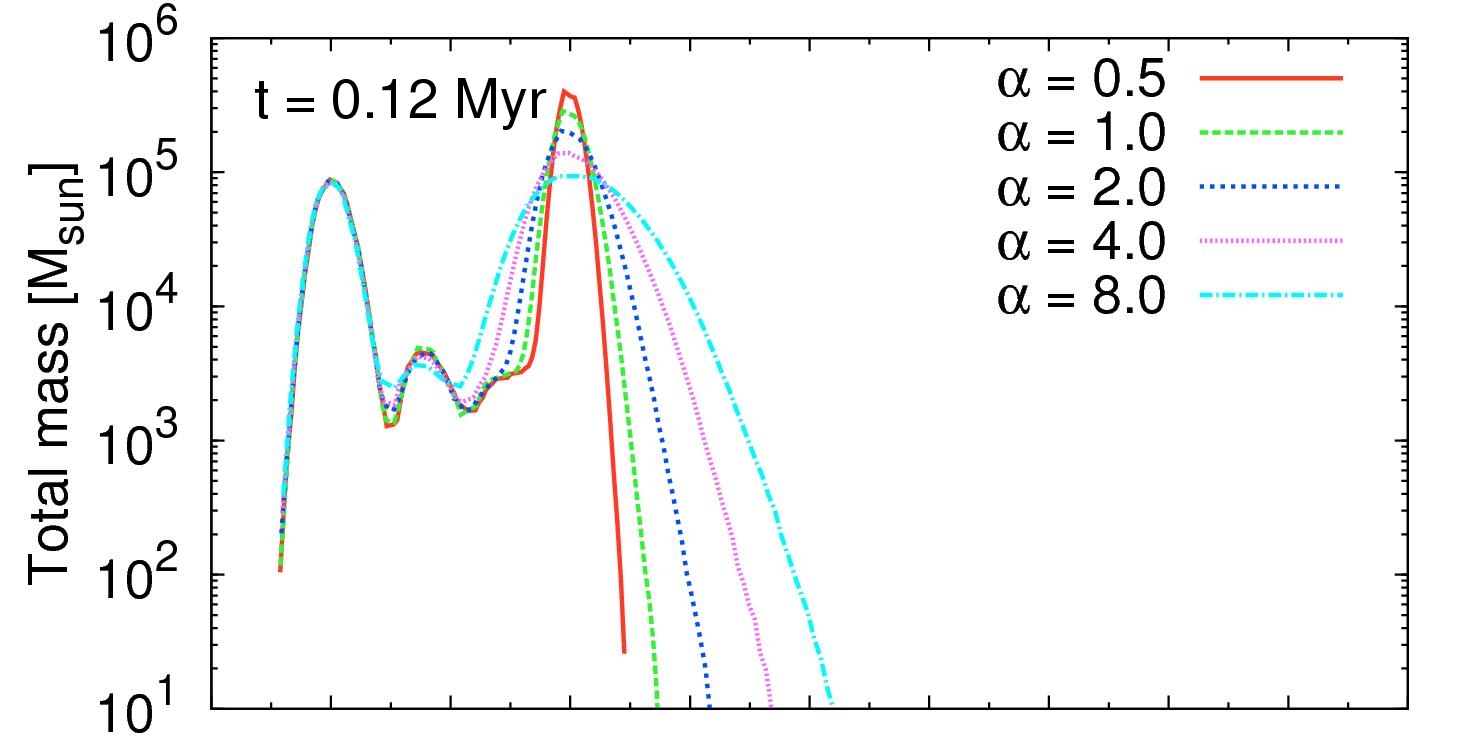}
\includegraphics[height=0.237\linewidth]{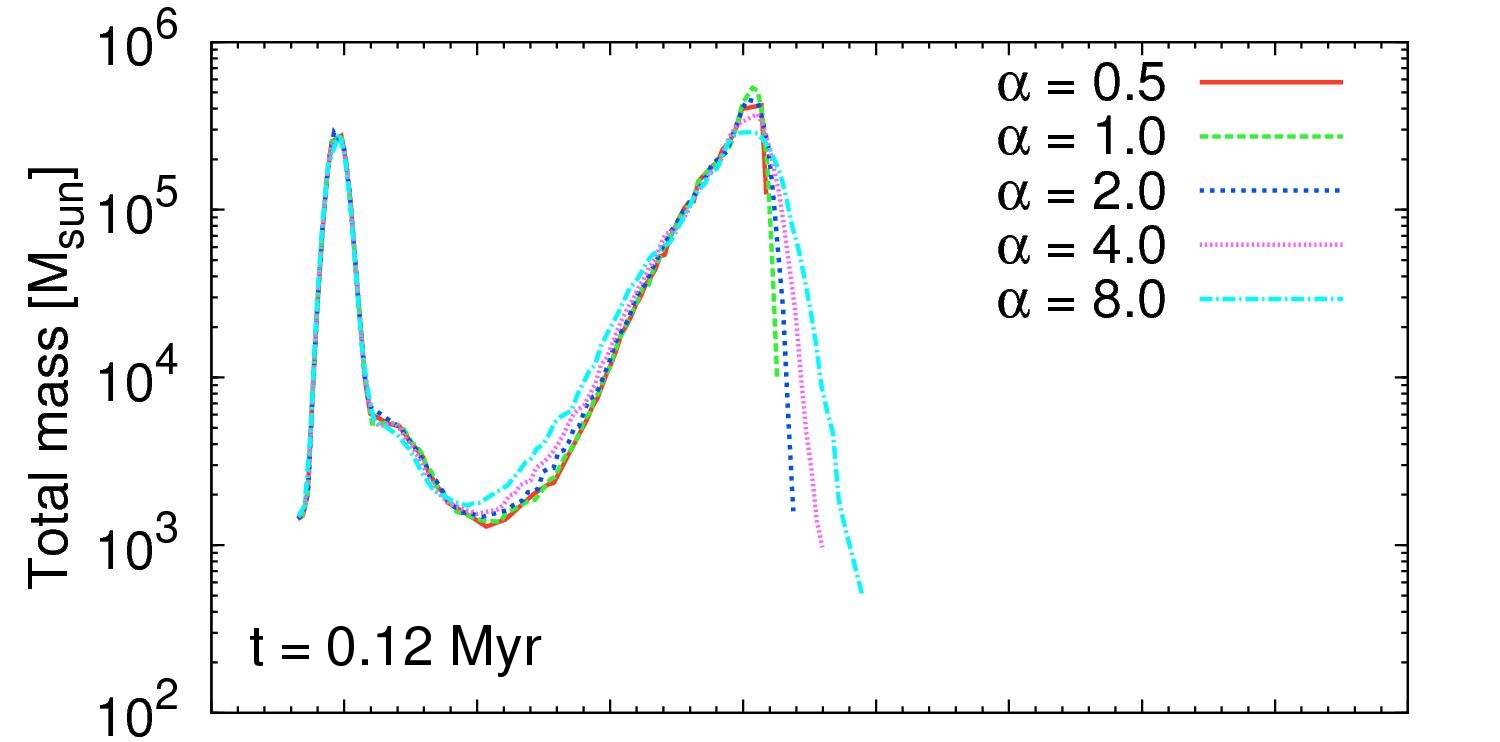}
}
\centerline{
\includegraphics[height=0.237\linewidth]{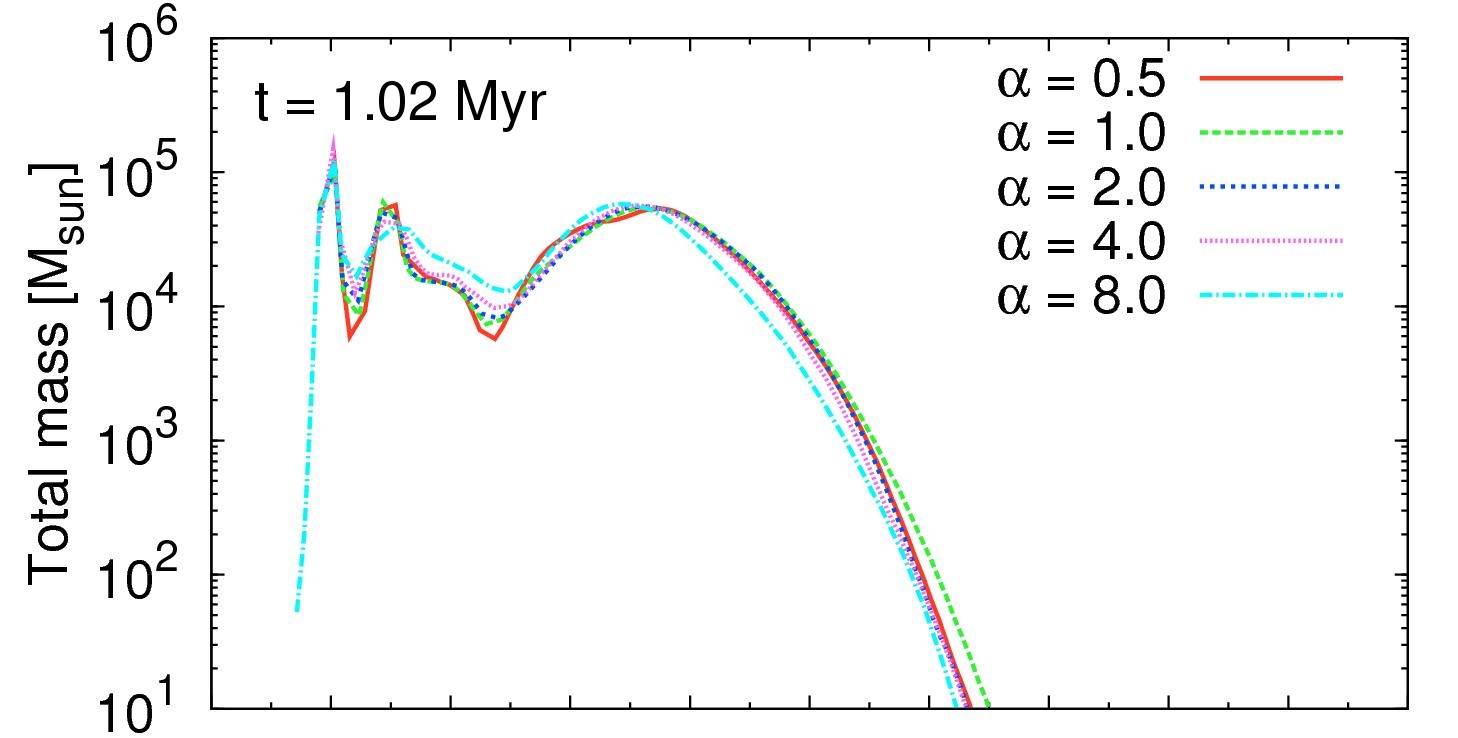}
\includegraphics[height=0.237\linewidth]{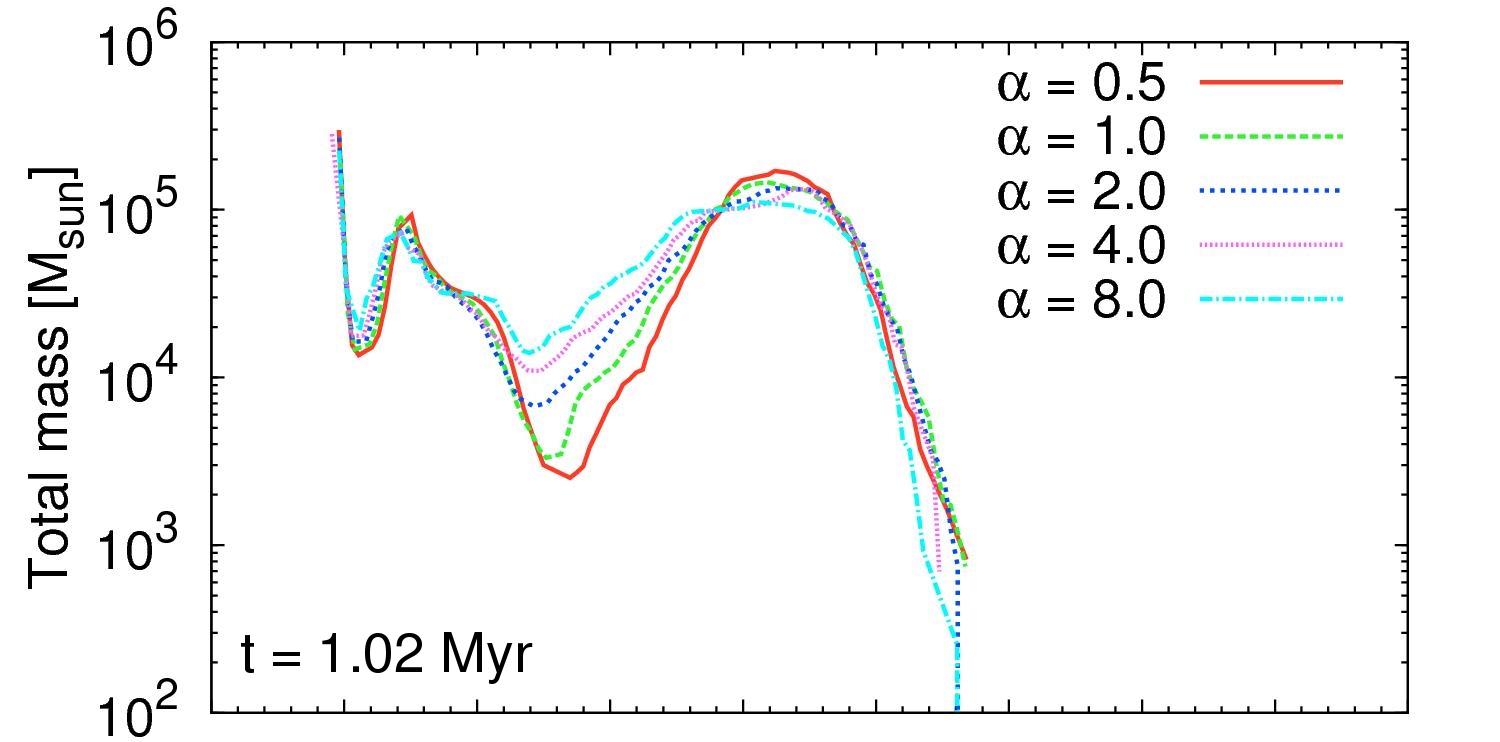}
}
\centerline{
\includegraphics[height=0.237\linewidth]{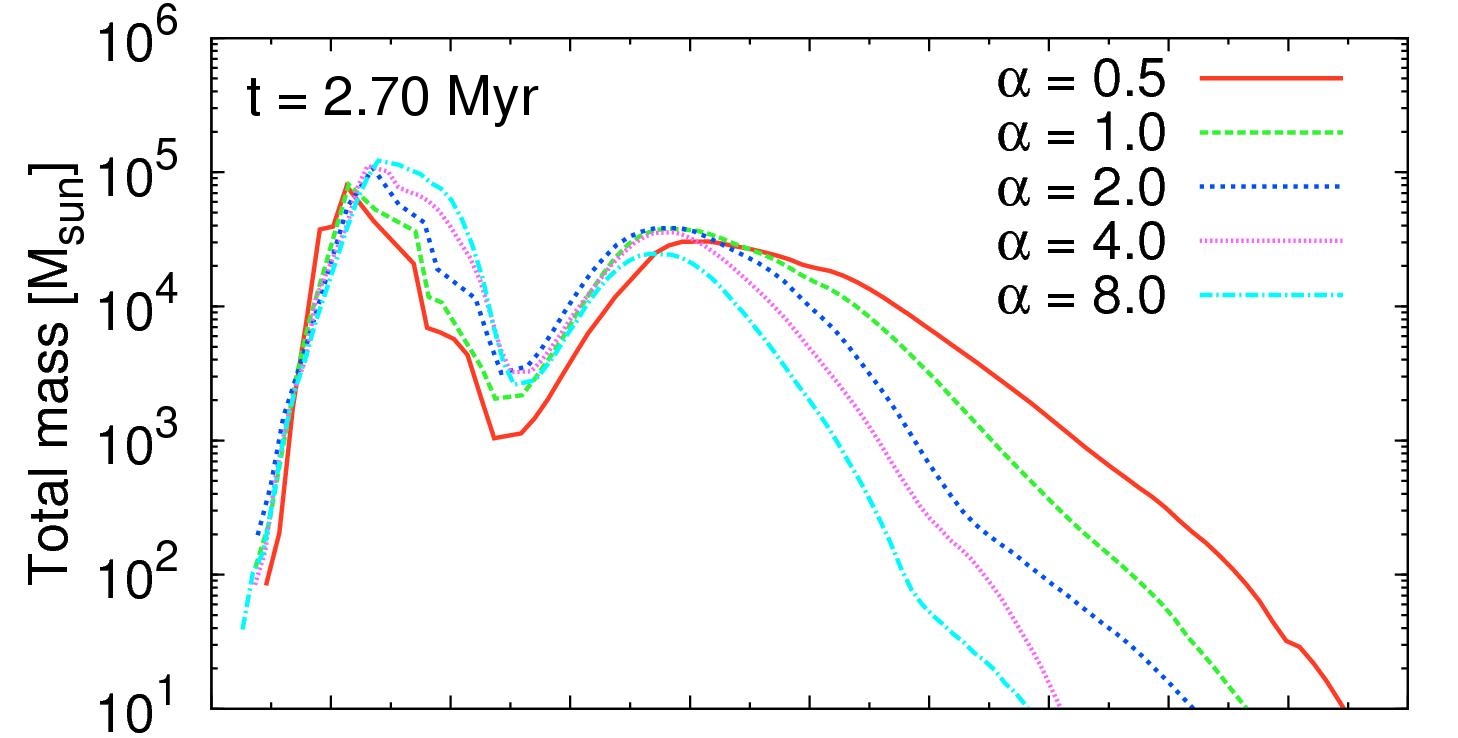}
\includegraphics[height=0.237\linewidth]{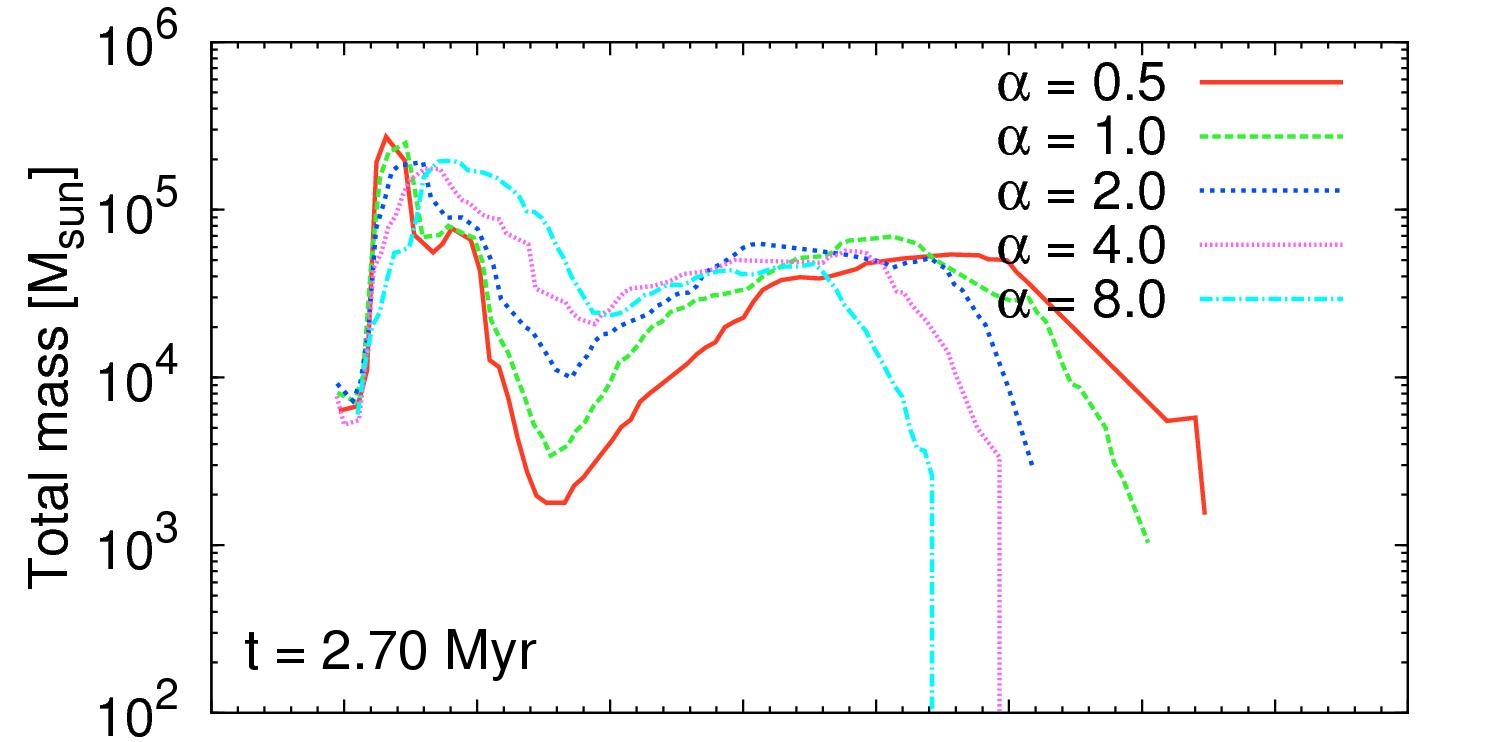}
}
\centerline{
\includegraphics[height=0.237\linewidth]{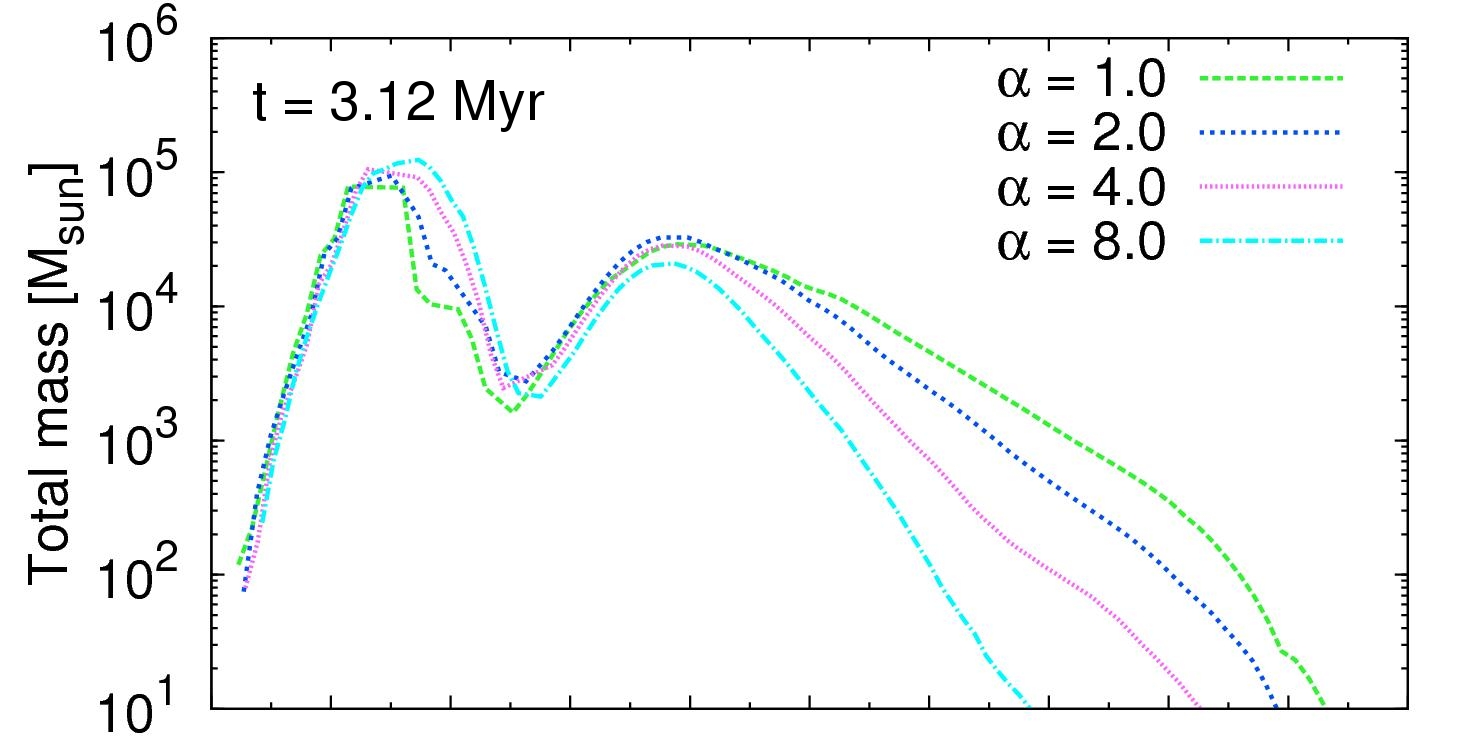}
\includegraphics[height=0.237\linewidth]{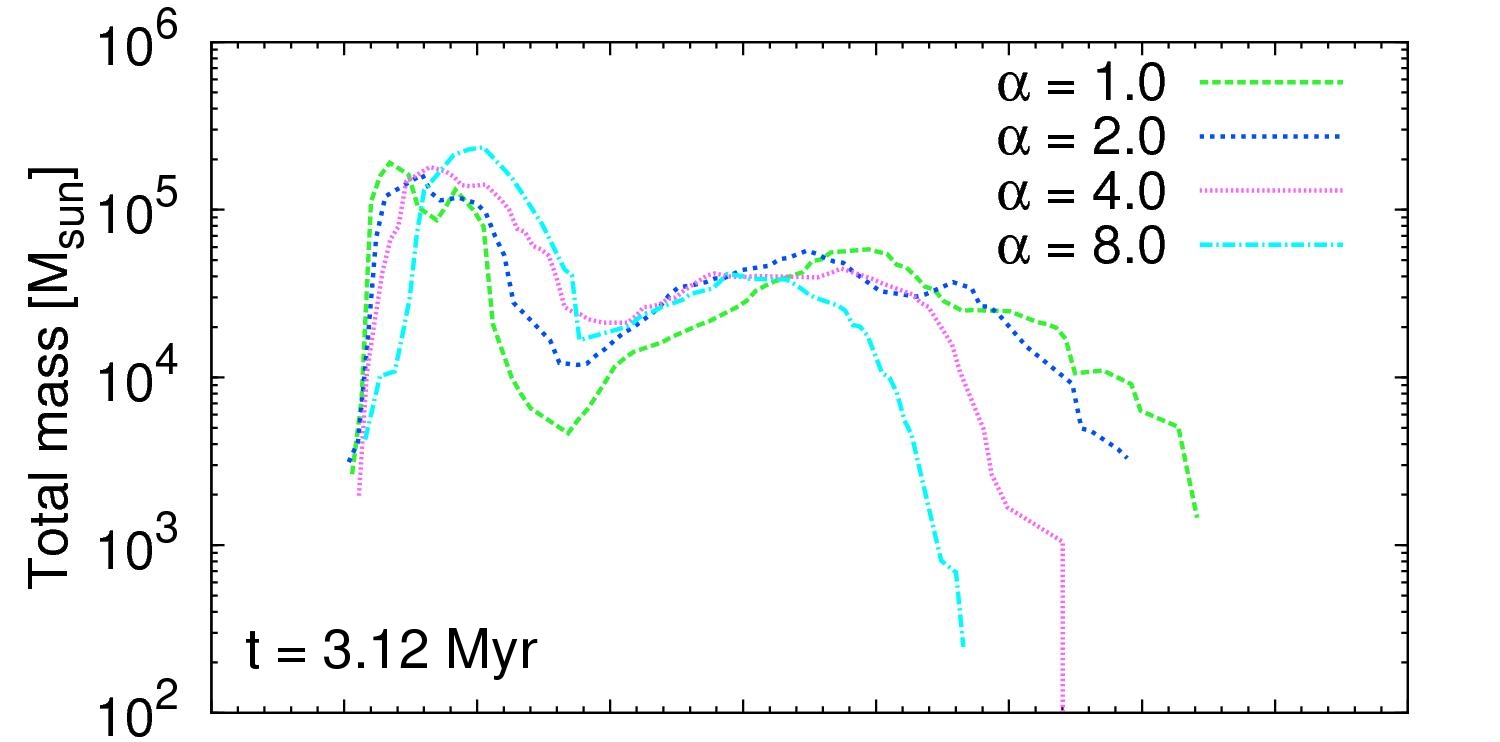}
}
\centerline{
\includegraphics[width=0.47\linewidth]{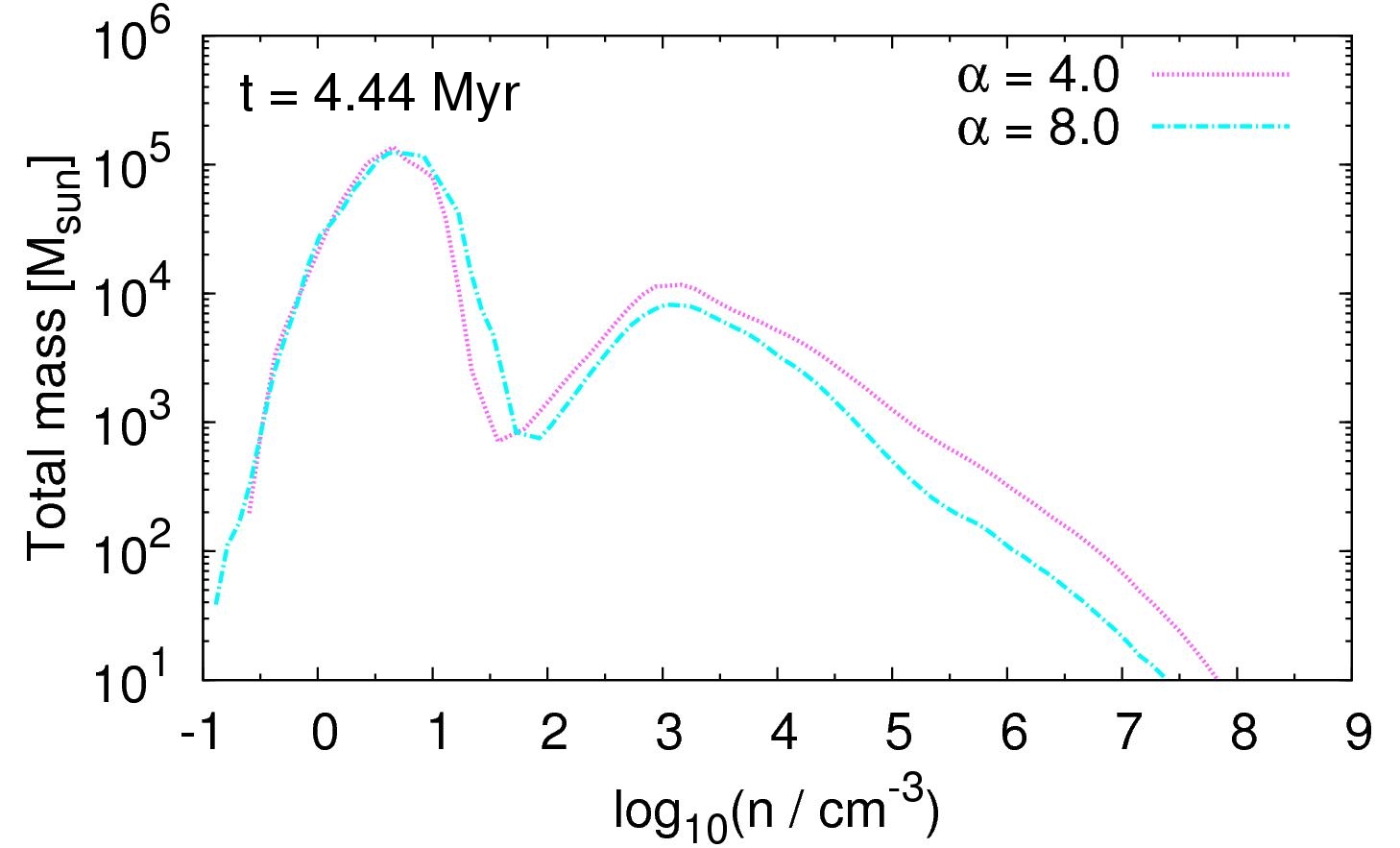}
\includegraphics[width=0.48\linewidth]{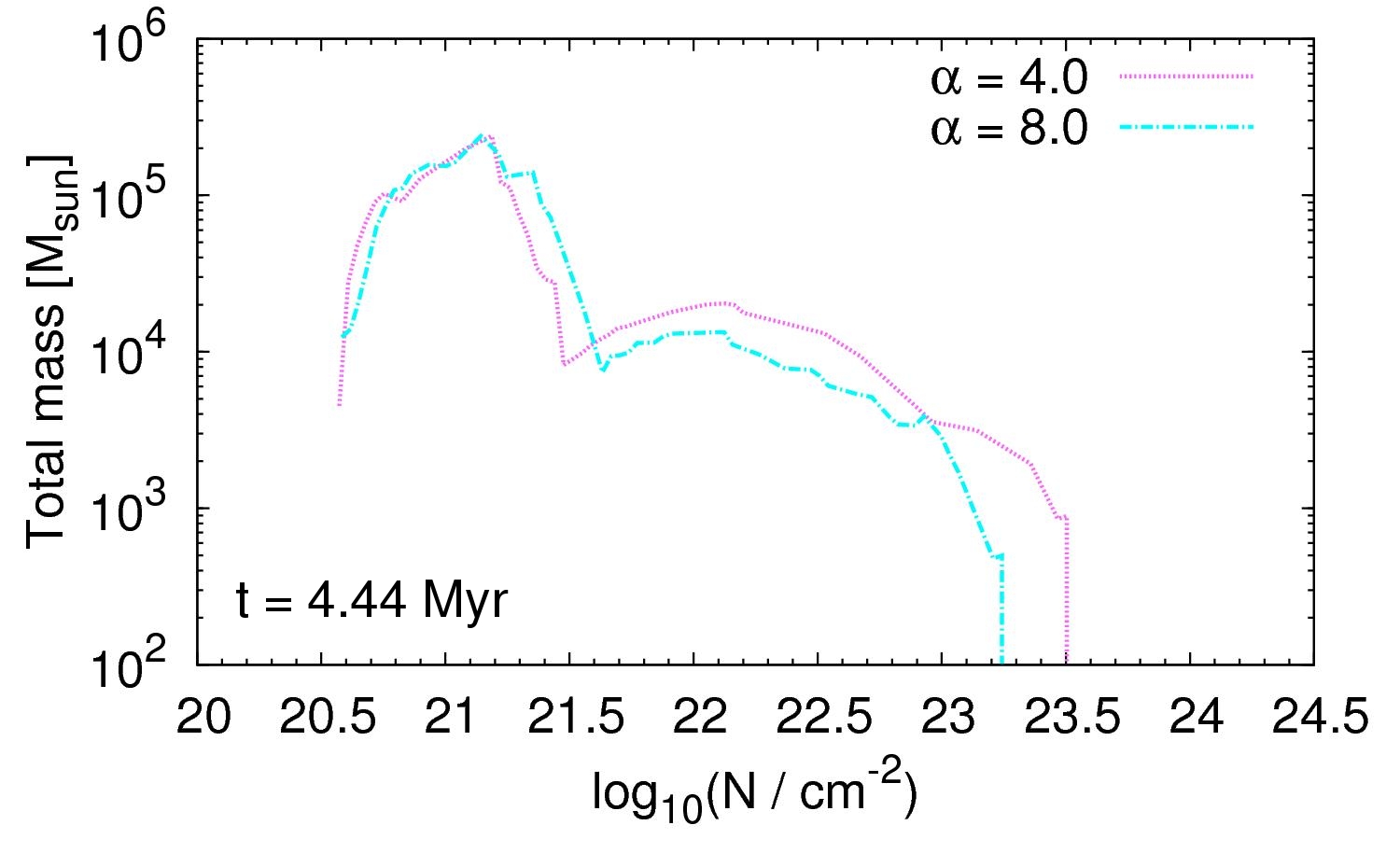}
}
\caption{Left column: Mass-weighted volume density PDFs at different times for our five virial $\alpha$ parameters, $\alpha = 0.5, 1.0, 2.0, 4.0$ and $8.0$. Right column: Same as left column, but with column density PDFs. In order to compute the column density PDFs, our simulations were projected onto a regular $1024^2$ map. All plots are computed for our models using an initial number density of $n_0 = 100\,$cm$^{-3}$.}
\label{fig:PDF100}
\end{figure*}

\begin{figure*}
\centerline{
\includegraphics[height=0.03\linewidth]{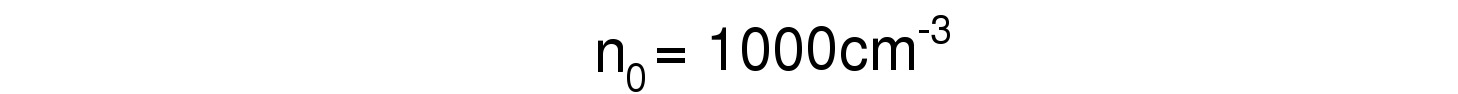}
\includegraphics[height=0.03\linewidth]{images/n1000.jpg}
}
\centerline{
\includegraphics[height=0.237\linewidth]{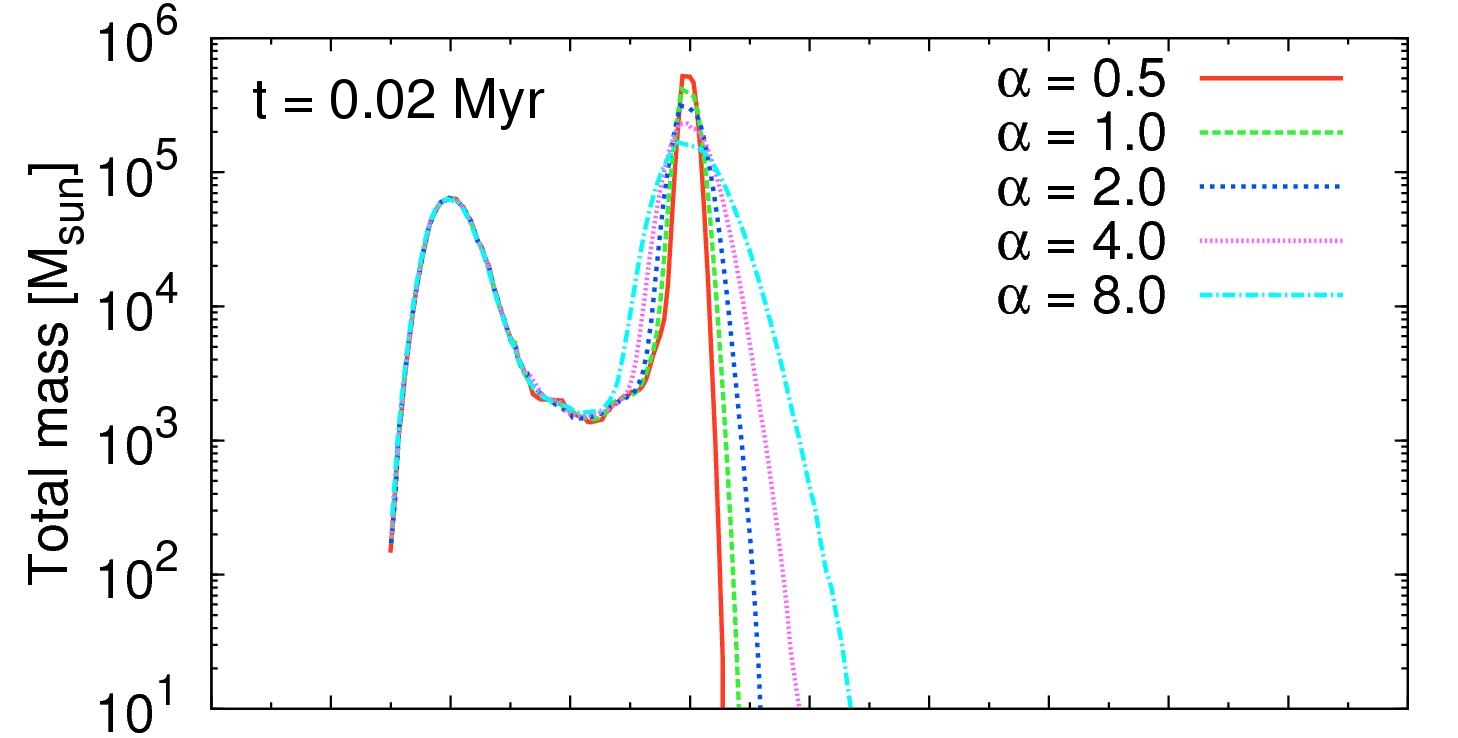}
\includegraphics[height=0.237\linewidth]{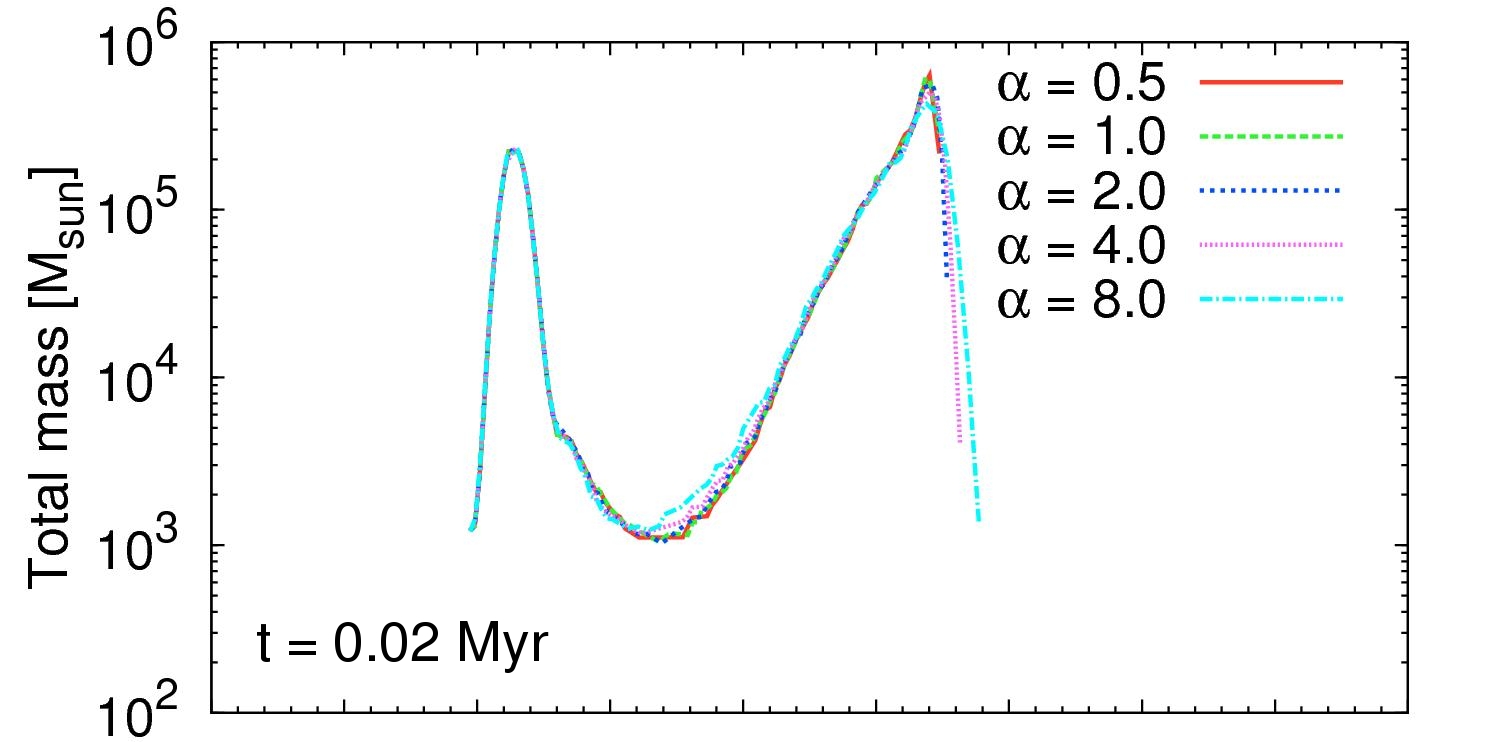}
}
\centerline{
\includegraphics[height=0.237\linewidth]{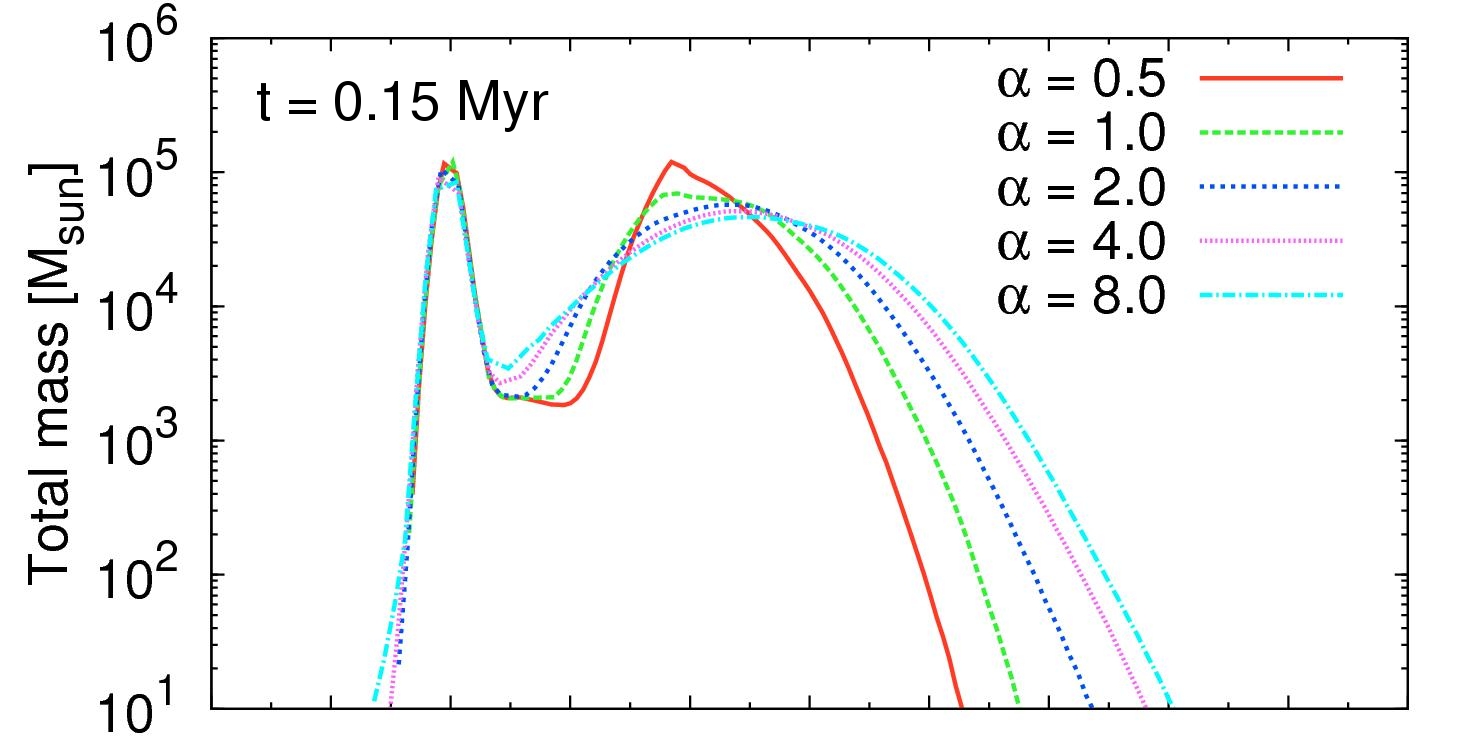}
\includegraphics[height=0.237\linewidth]{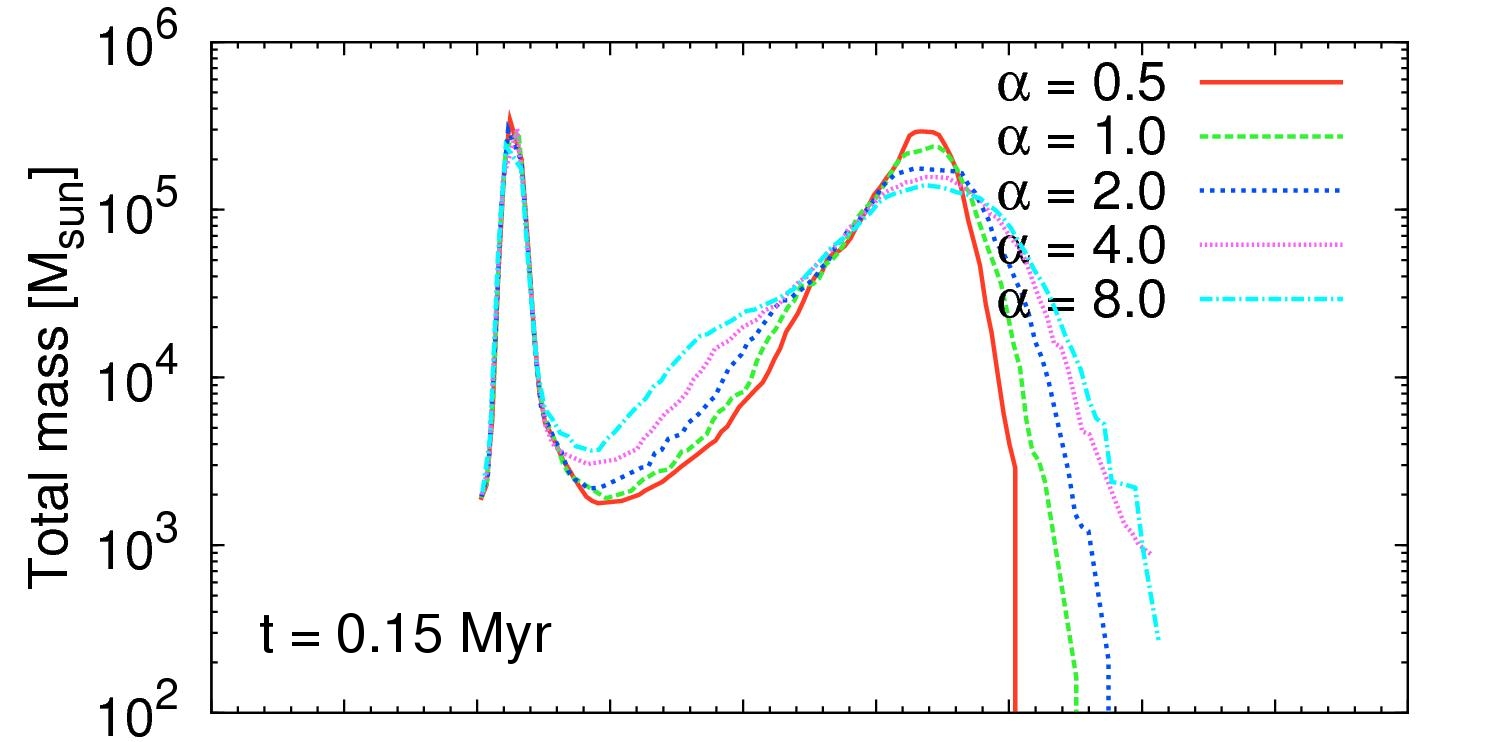}
}
\centerline{
\includegraphics[height=0.237\linewidth]{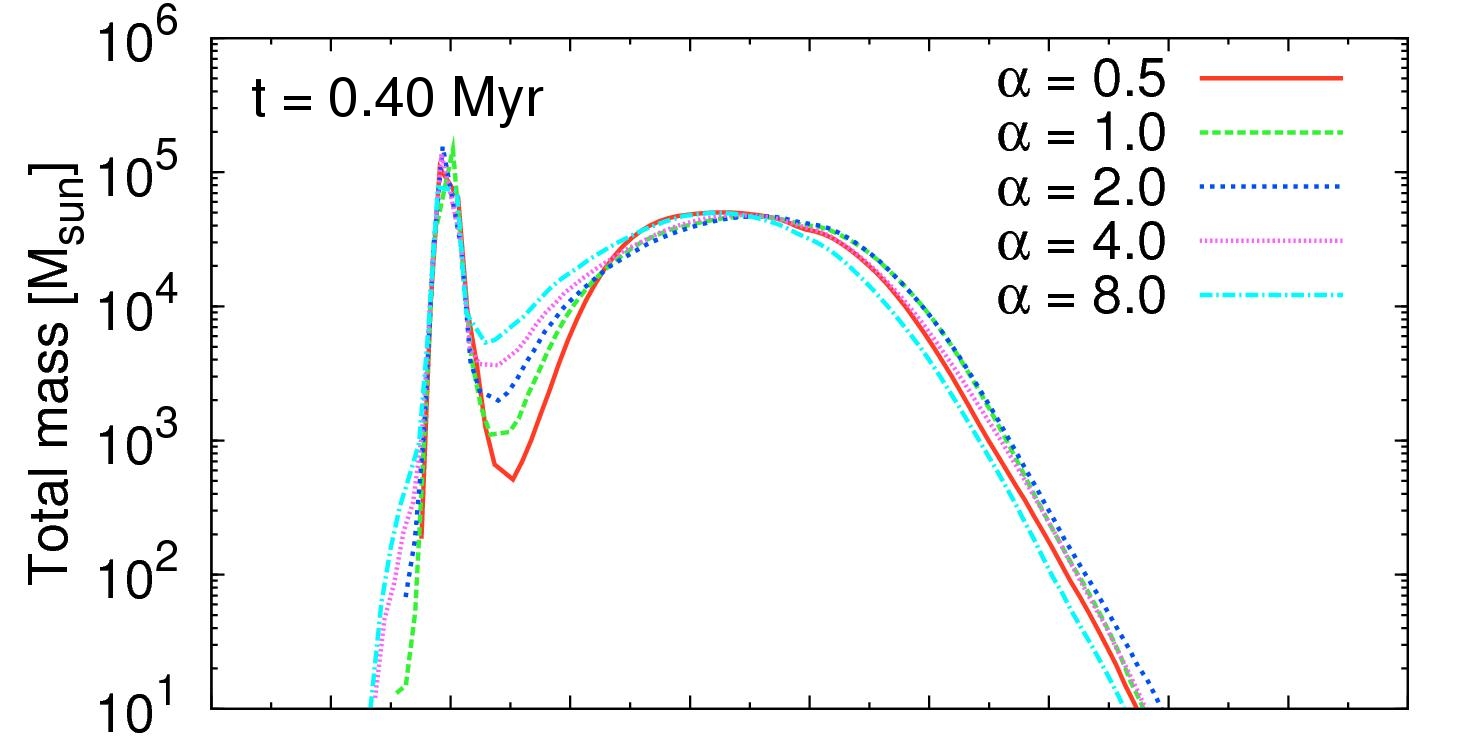}
\includegraphics[height=0.237\linewidth]{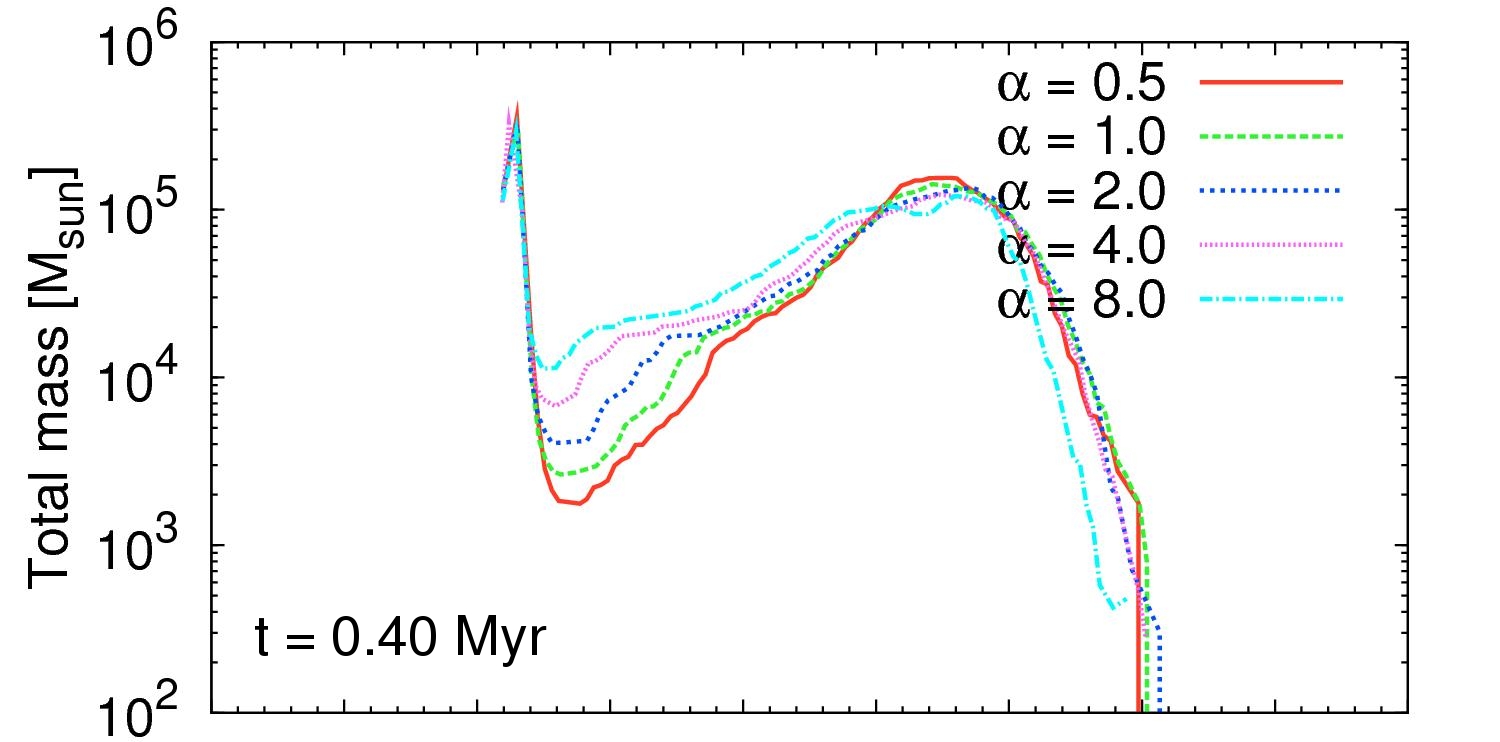}
}
\centerline{
\includegraphics[height=0.237\linewidth]{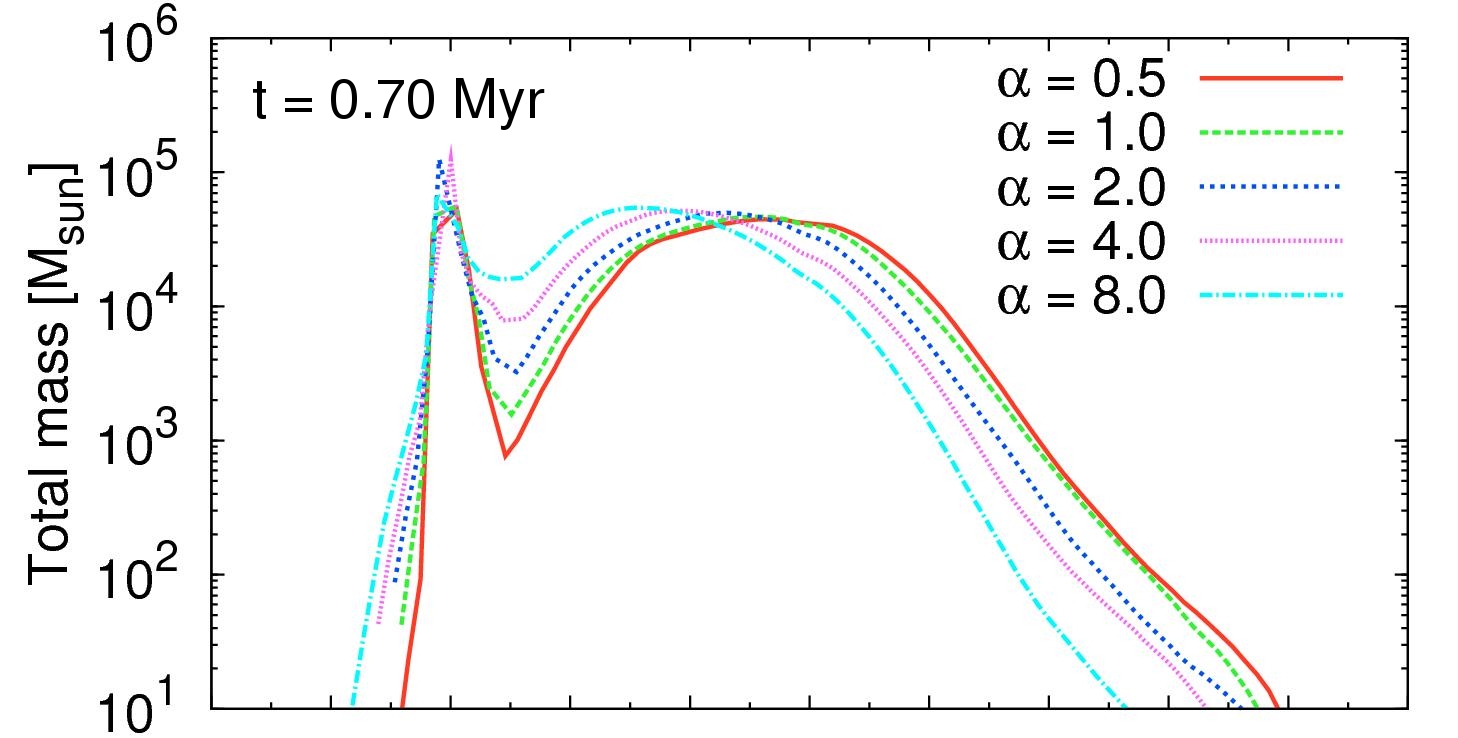}
\includegraphics[height=0.237\linewidth]{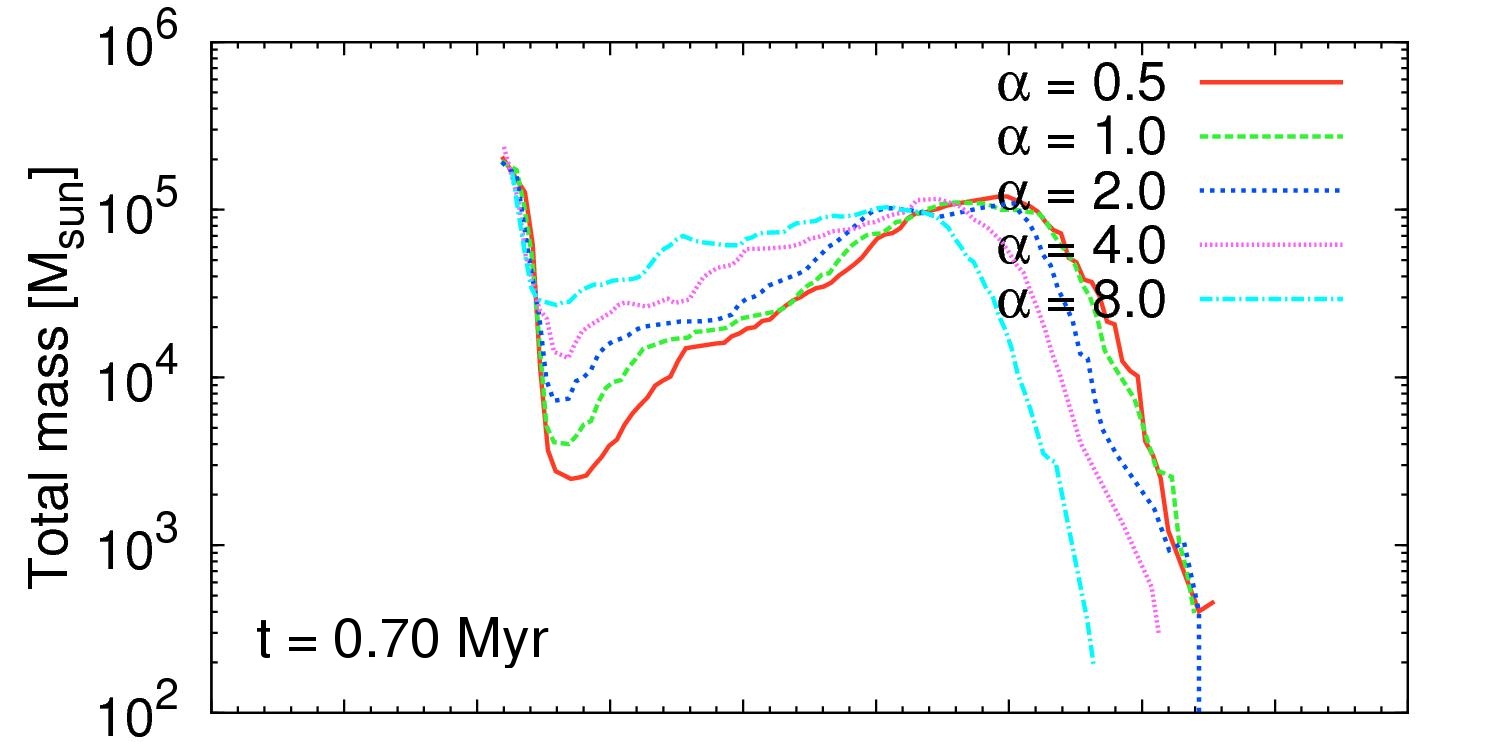}
}
\centerline{
\includegraphics[width=0.47\linewidth]{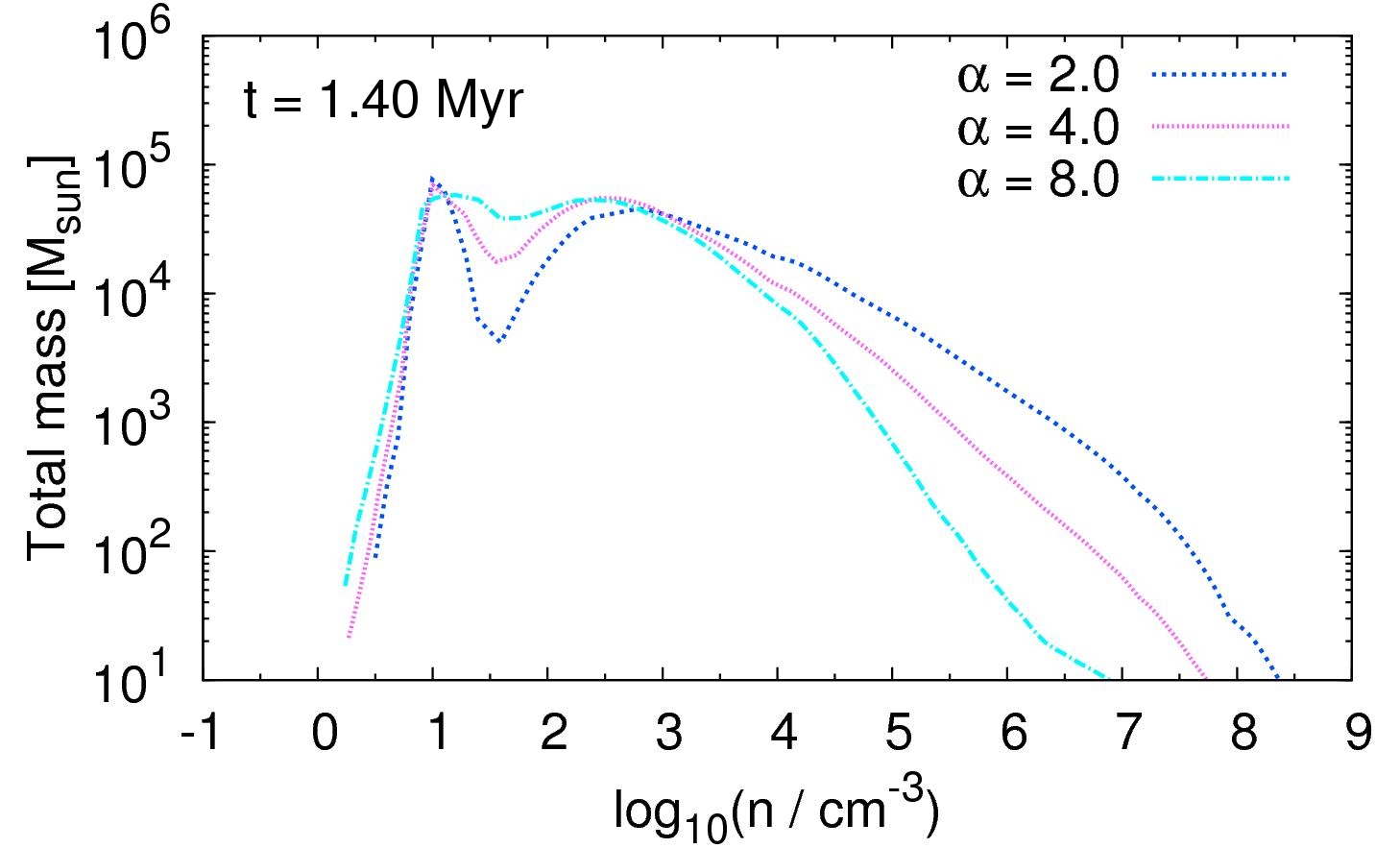}
\includegraphics[width=0.48\linewidth]{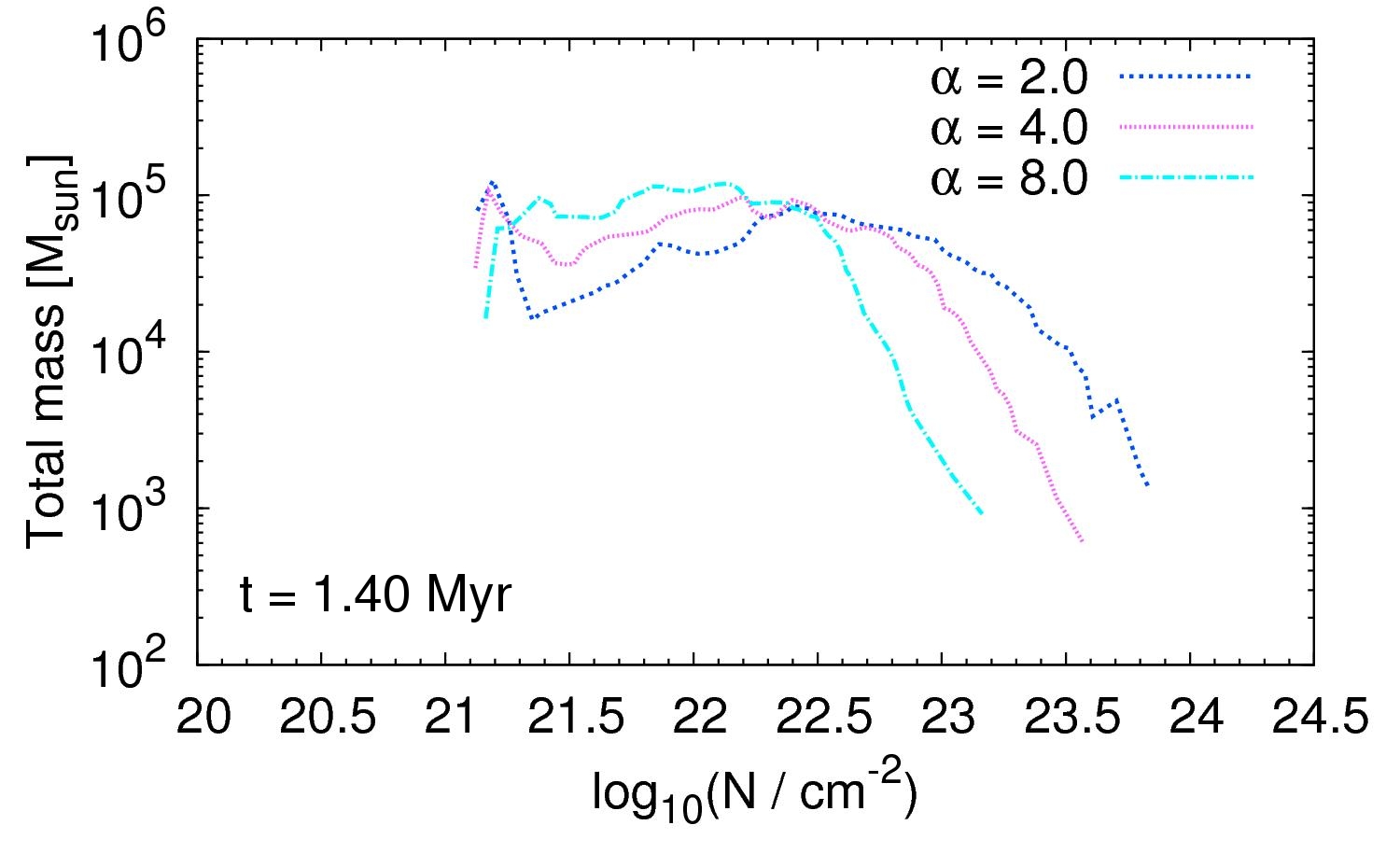}
}
\caption{Same as Fig. \ref{fig:PDF100}, but for our models using an initial number density of $n_0 = 1000\,$cm$^{-3}$.}
\label{fig:PDF1000}
\end{figure*}

\begin{figure*}
\centerline{
\includegraphics[height=0.03\linewidth]{images/n100.jpg}
\includegraphics[height=0.03\linewidth]{images/n1000.jpg}
}
\centerline{
\includegraphics[height=0.237\linewidth]{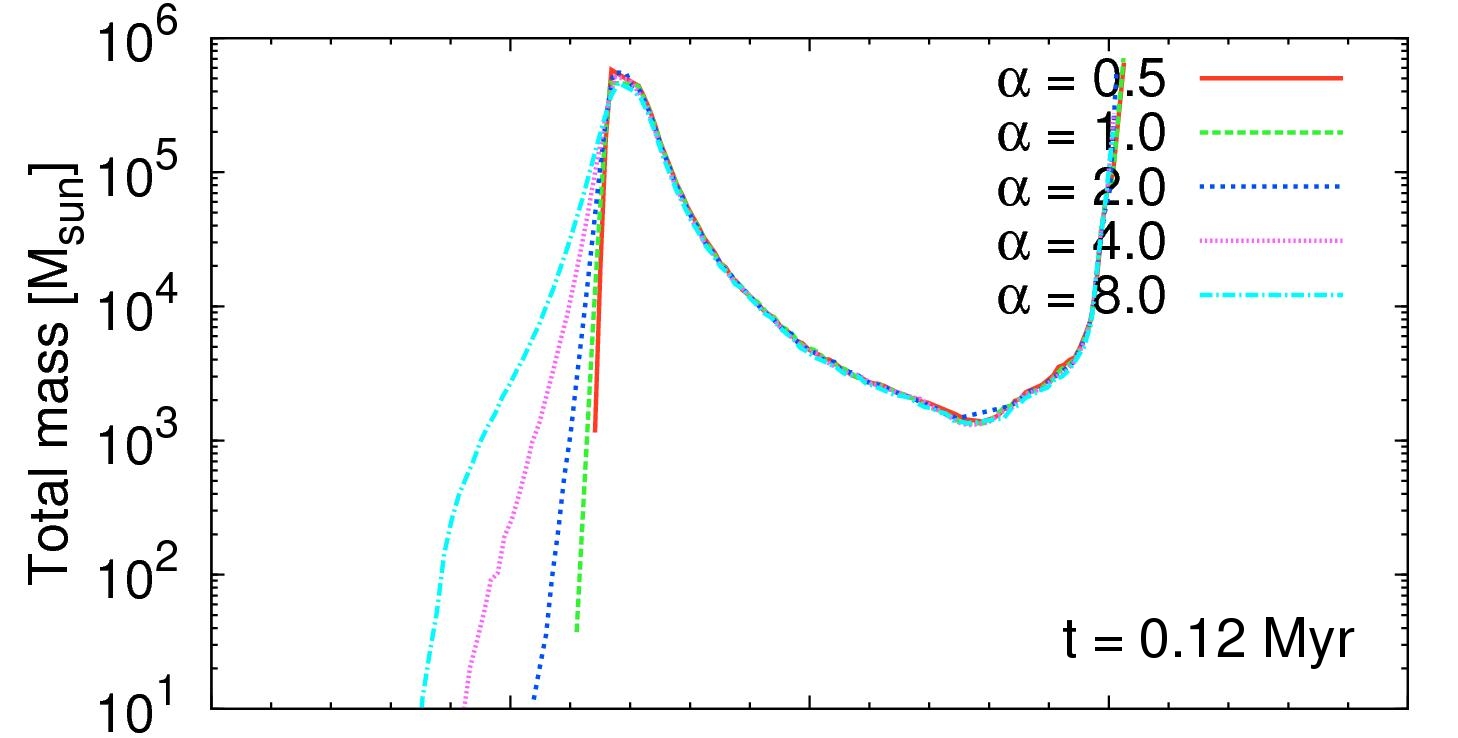}
\includegraphics[height=0.237\linewidth]{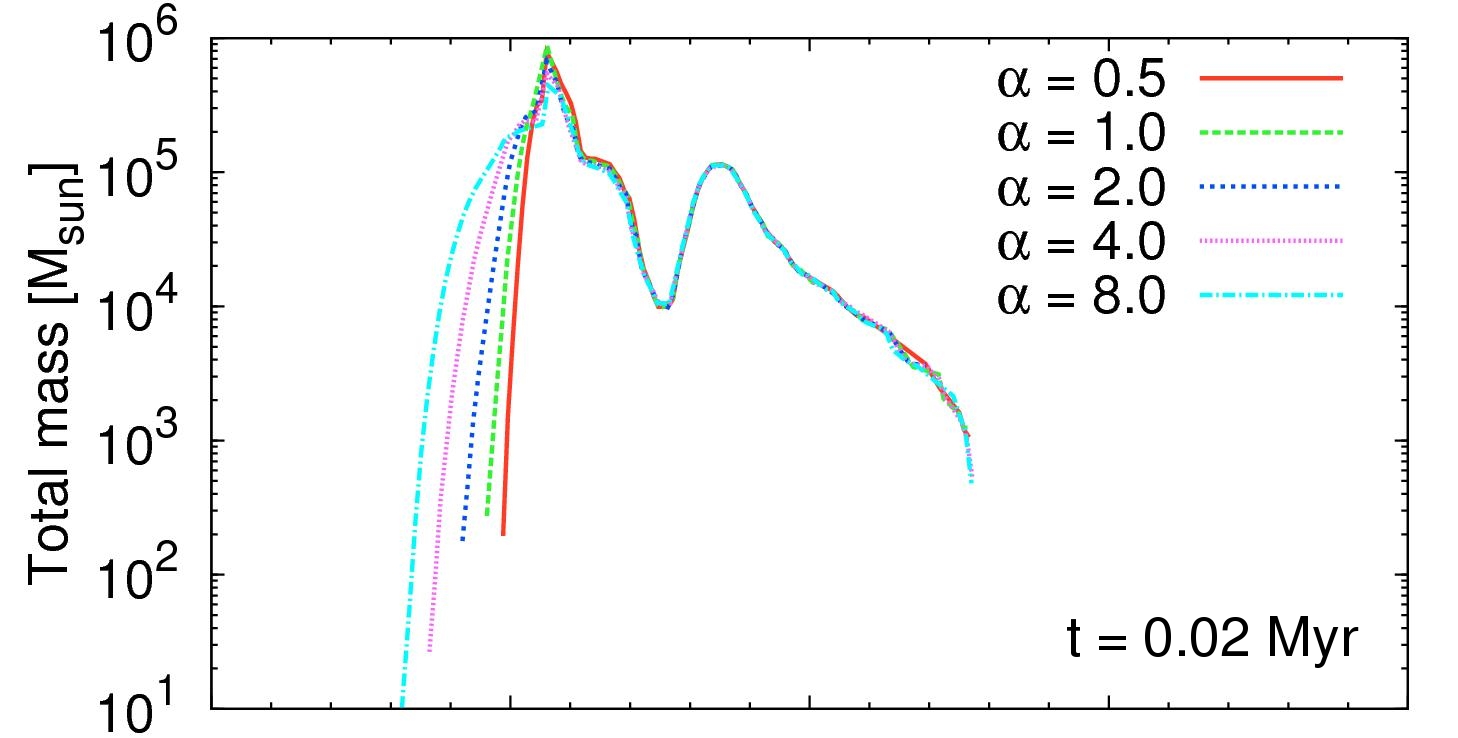}
}
\centerline{
\includegraphics[height=0.237\linewidth]{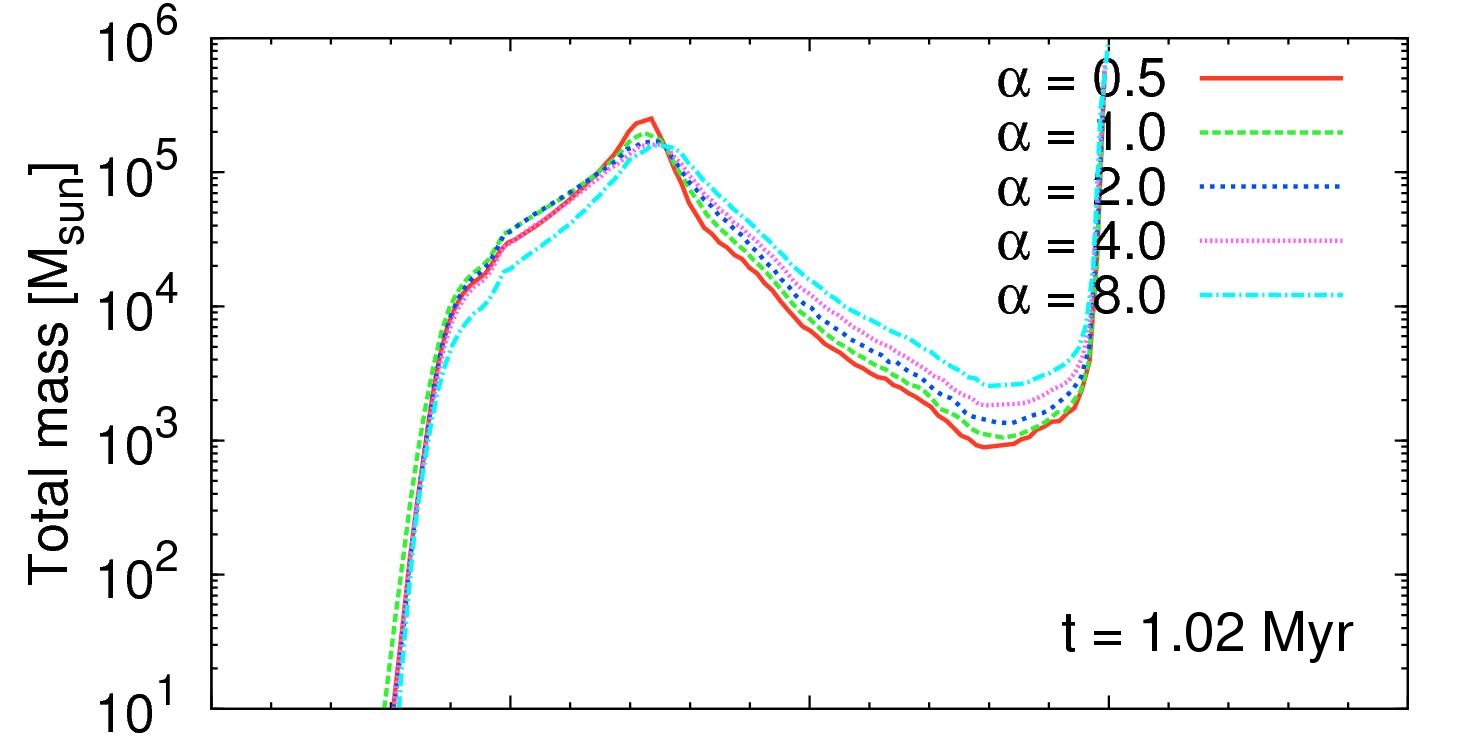}
\includegraphics[height=0.237\linewidth]{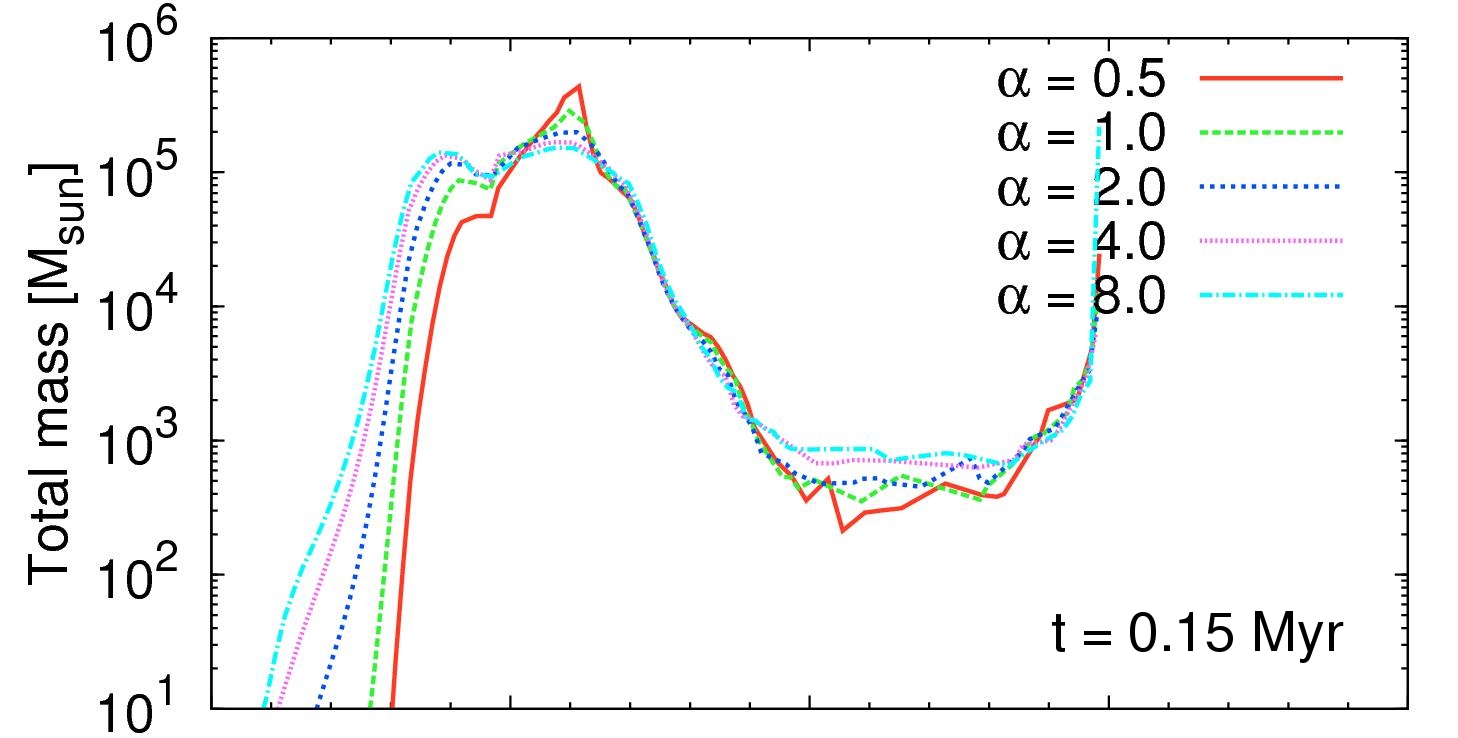}
}
\centerline{
\includegraphics[height=0.237\linewidth]{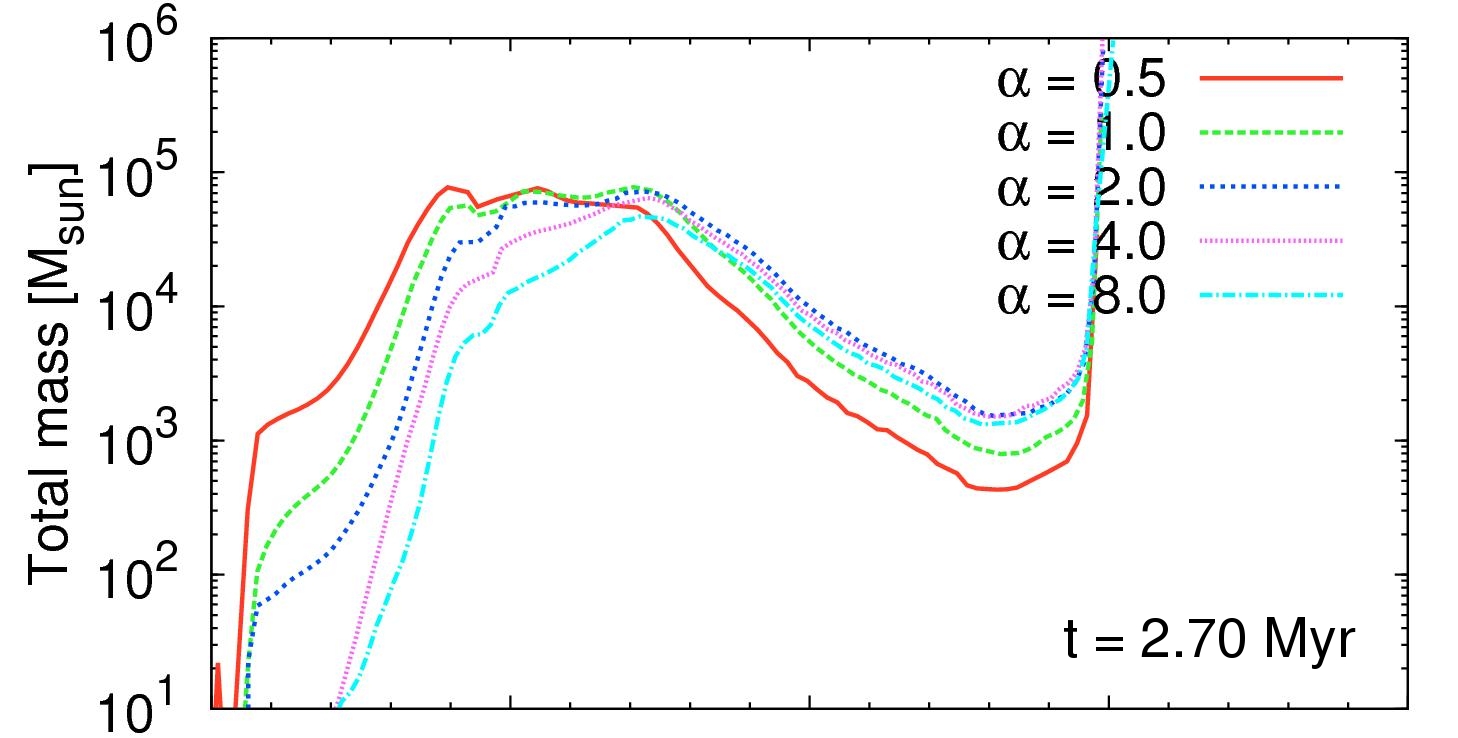}
\includegraphics[height=0.237\linewidth]{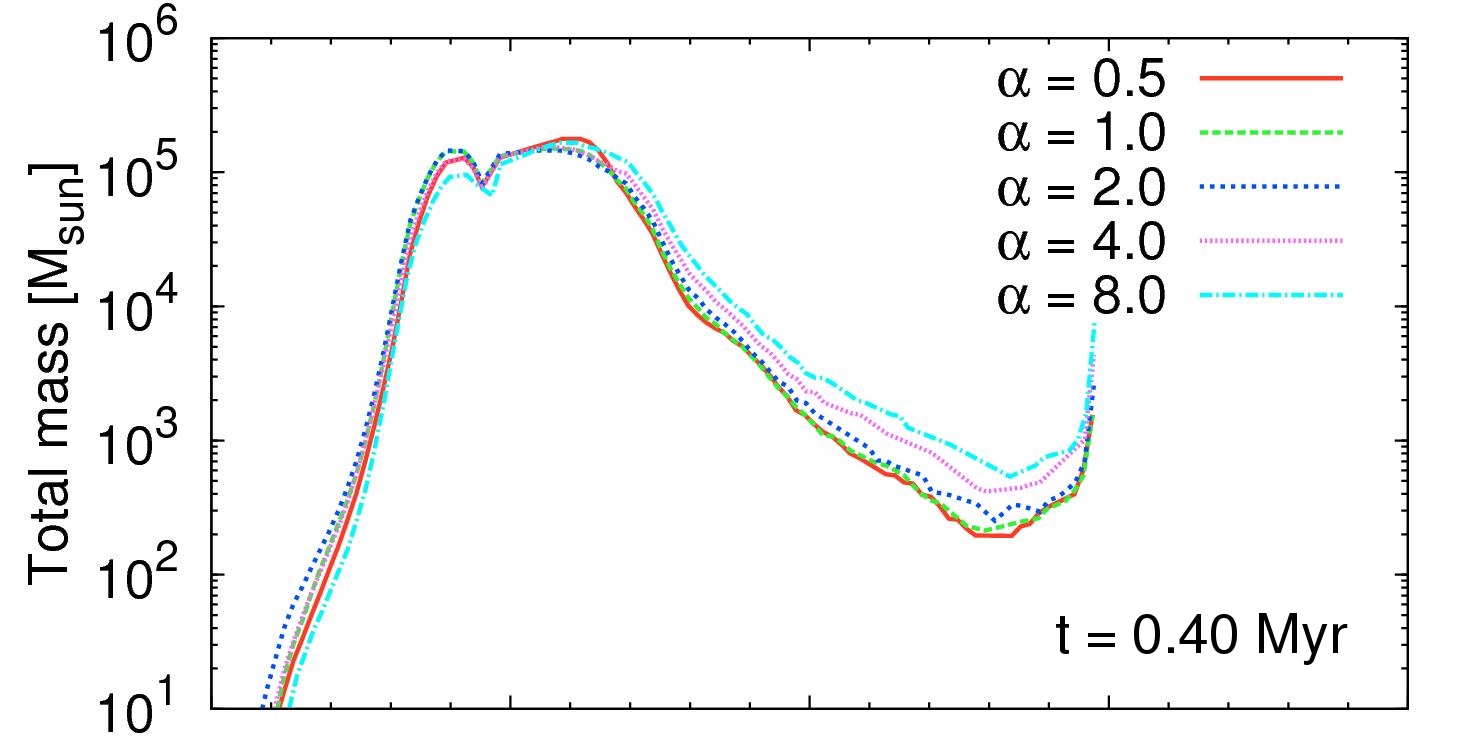}
}
\centerline{
\includegraphics[height=0.237\linewidth]{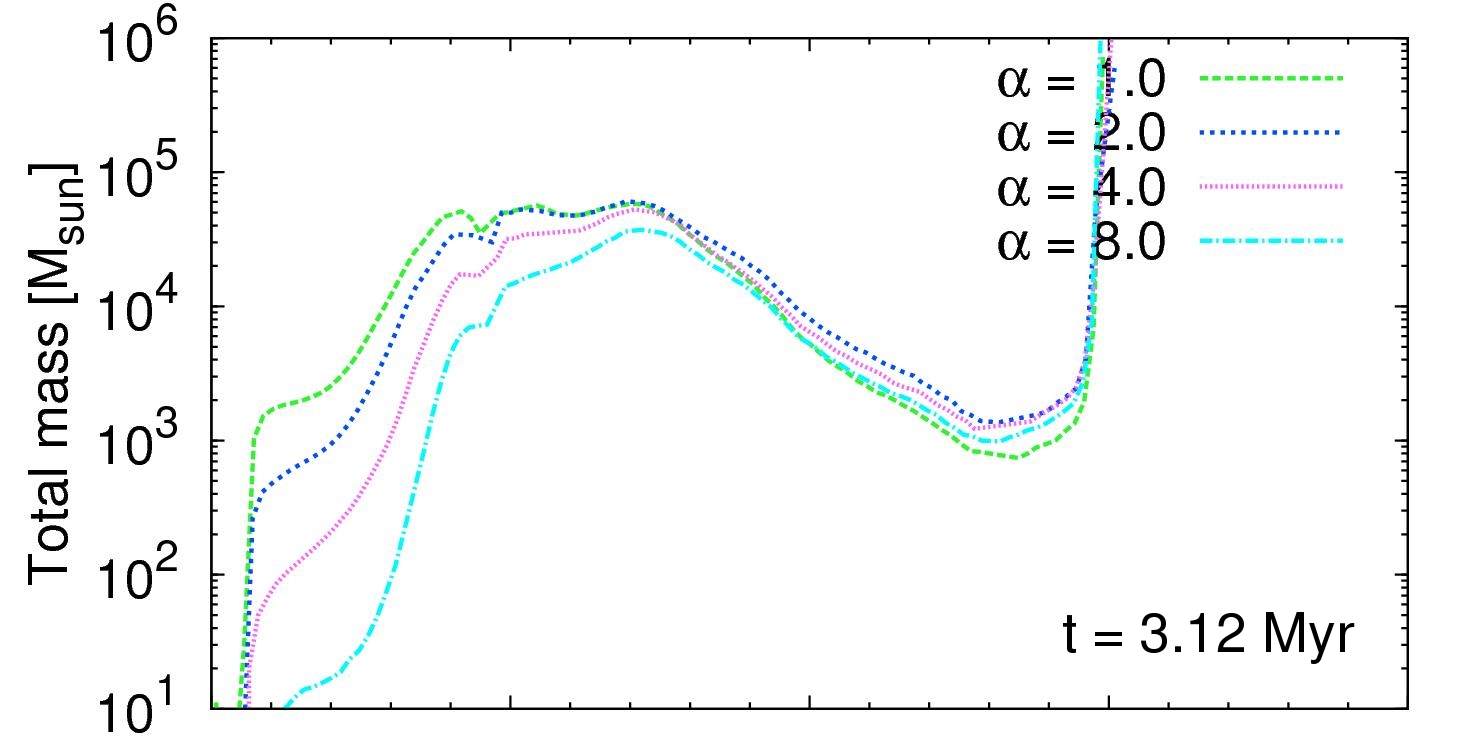}
\includegraphics[height=0.237\linewidth]{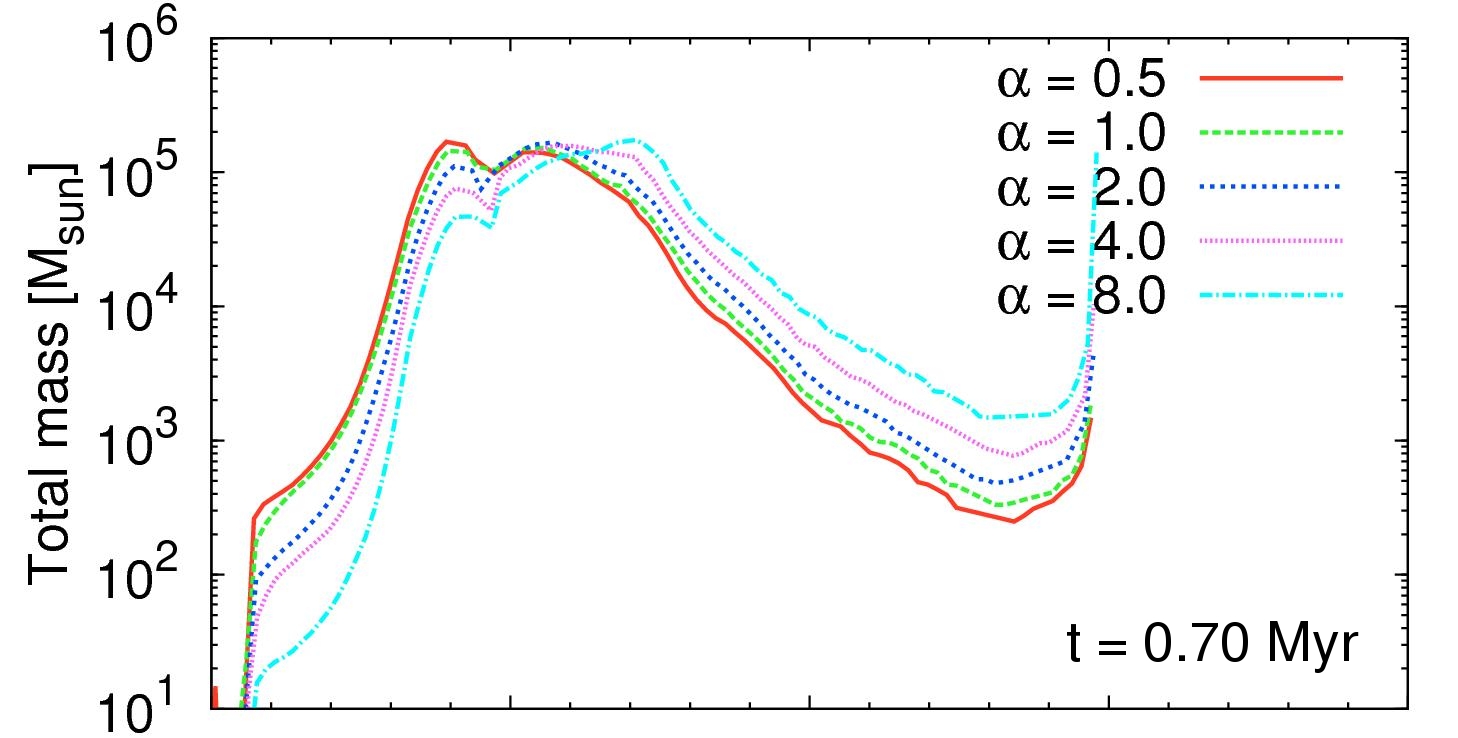}
}
\centerline{
\includegraphics[height=0.284\linewidth]{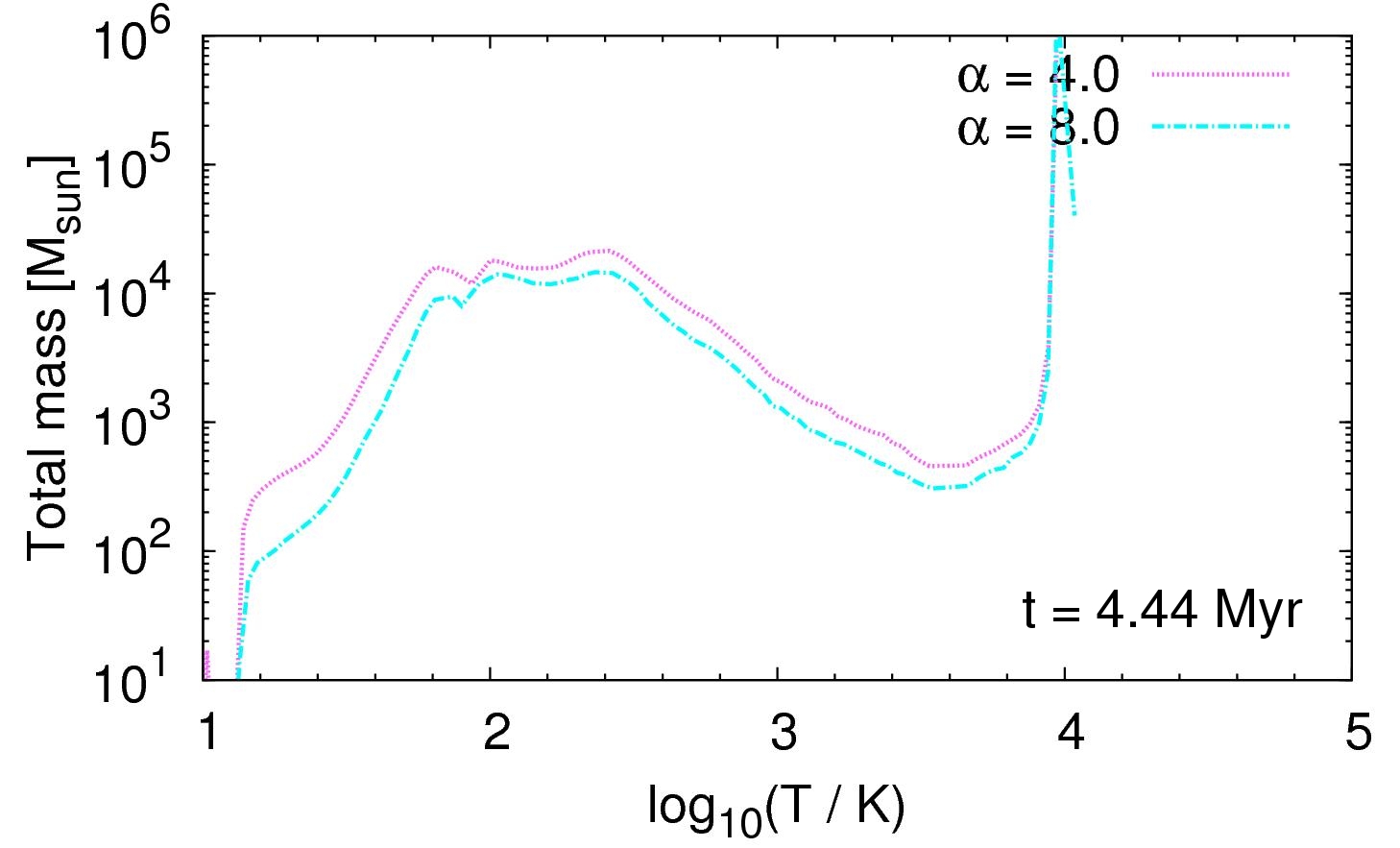}
\includegraphics[height=0.284\linewidth]{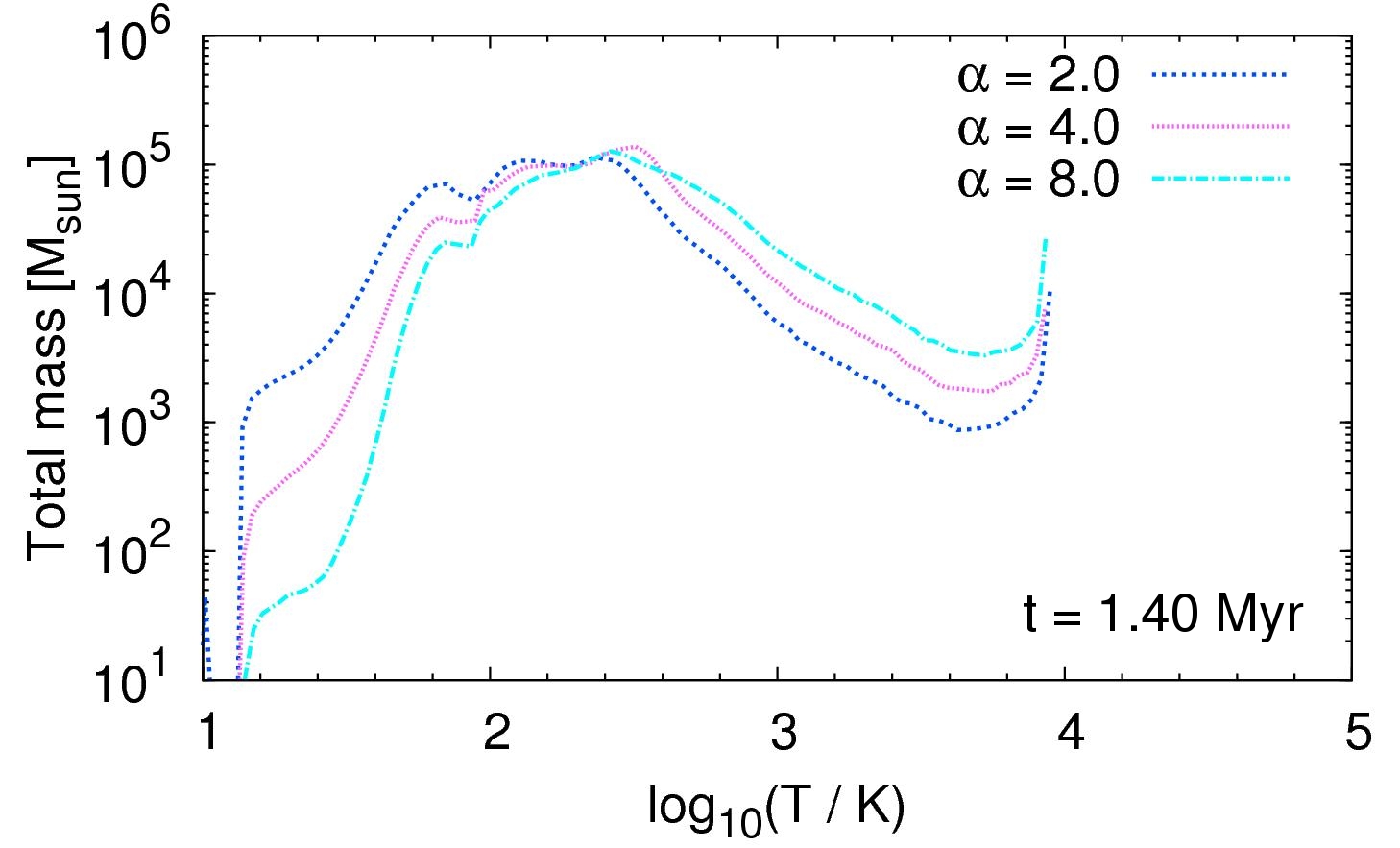}
}
\caption{Mass-weighted temperature PDFs at different times for our five virial $\alpha$ parameters, $\alpha = 0.5, 1.0, 2.0, 4.0$ and $8.0$, for our models with an initial number density of $n_0 = 100\,$cm$^{-3}$ (left column) and $n_0 = 1000\,$cm$^{-3}$ (right column).}
\label{fig:TEMPPDF}
\end{figure*}

\begin{table}
\begin{tabular}{l|c|c|c|c|c|c}
\hline\hline
Model name & $\epsilon_{\Delta \text{t}}$ & $\epsilon_{\text{ff}}$ & $N_{\text{sink}}$ & $t_{*}$ & $t_{\text{end}}$ & $\Delta \text{t}$ \\
 & [\%] & [\%] & & [Myr] & [Myr] & [Myr] \\
\hline
GC-0.5-100 & 8.2 & 25.7 & 3473 & 1.30 & 2.70 & 1.40 \\
GC-1.0-100 & 8.5 & 28.5 & 4699 & 1.80 & 3.12 & 1.31 \\
GC-2.0-100 & 8.5 & 19.1 & 4166 & 1.84 & 3.80 & 1.96 \\
GC-4.0-100 & 6.2 & 14.8 & 2511 & 2.56 & 4.44 & 1.84 \\
GC-8.0-100 & 2.3 & 6.1 & 644 & 2.75 & 4.44 & 1.65 \\
\hline
GC-0.5-1000 & 4.5 & 10.9 & 2966 & 0.42 & 1.00 & 0.58 \\
GC-1.0-1000 & 4.6 & 8.0 & 2531 & 0.28 & 1.10 & 0.82 \\
GC-2.0-1000 & 6.8 & 8.0 & 3218 & 0.22 & 1.40 & 1.18 \\
GC-4.0-1000 & 2.6 & 2.9 & 1035 & 0.16 & 1.40 & 1.24 \\
GC-8.0-1000 & 0.6 & 0.7 & 180 & 0.12 & 1.40 & 1.28 \\
\hline
GC-16.0-10000 & 0.7 & 0.7 & 1406 & 0.02 & 0.40 & 0.38 \\
\hline
SOL-0.5-1000 & 5.1 & 13.2 & 16248 & 0.21 & 0.75 & 0.54 \\
SOL-8.0-1000 & 4.3 & 4.6 & 8617 & 0.08 & 1.40 & 1.32 \\
\hline
\end{tabular}
\caption{The star formation efficiencies $\epsilon_{\Delta \text{t}}$ give the amount of gas being converted to sink particles within a time $\Delta t = t_{\text{end}} - t_{*} < t_{\text{ff}}$, where $t_{\text{end}}$ denotes the end of our simulation and $t_{*}$ the time when the first star forms. $N_{\text{sink}}$ gives the total number of sink particles formed during the time $\Delta t$. To calculate the SFEs per free-fall time, $\epsilon_{\text{ff}}$, we extrapolate based on $\epsilon_{\Delta \text{t}}$, assuming that the SFR between $t_{\text{ff}}$ and $t_{\text{end}}$ is the same as between $t_{*}$ and $t_{\text{end}}$. Some simulations with lower $\alpha$ values are not evolved until one free-fall time, which is due to the high computational costs of the individual runs. Nevertheless, we let all those simulations evolve until a $\epsilon_{\Delta \text{t}}$ of at least $\sim4\%$ is reached.}
\label{tab:SFE}
\end{table}

Fig. \ref{fig:images} shows logarithmic column density maps of the different model clouds presented in Table \ref{tab:setup}. As expected, the shape of all the clouds strongly depends on the specific value of $\alpha$ in each simulation. Table \ref{tab:SFE} gives an overview of the star formation efficiencies and the number of sink particles being formed in each model cloud with varying $\alpha$, also illustrated in Fig. \ref{fig:SFEvsAlpha}. In the Figure, we also show the star formation efficiencies that we derive from simulations performed using the same values of $n_{0}$ and $\alpha$, but with a different random seed for the turbulent velocity field, as explained in more detail in Appendix \ref{sec:seed}. Some of the simulations with less turbulent kinetic energy (i.e.\ those with lower $\alpha$ values) were not evolved until the end of one free-fall time. This is because gravitational collapse and star formation become very efficient in these simulations, driving up the computational cost due to the need to refine many high density regions. In these rapidly star-forming runs, we stop our simulations when they either reach one free-fall time or when the computational cost of continuing becomes excessive. In most cases, this occurs once $\sim 8$\% of the gas has formed stars, but in runs GC-0.5-1000 and GC-1.0-1000 the computational cost climbs so steeply as the gas collapses that we are forced to stop when only $\sim 4$\% of the gas has formed stars. We note, however, that we do not include the effects of feedback from young stars (see Section \ref{subsec:limitations}), and so the star formation rate is likely to be overestimated at late times in all of our runs.

We also have to keep in mind that we start with an idealized, spherical and uniform gas distribution at the beginning, rather than with an MC that is already in an evolved physical state. Measuring the star formation efficiencies from the beginning of the simulation is therefore a questionable procedure, since our results would be strongly affected by the initial geometry of the MC. Thus, we give two different star formation efficiencies in Table \ref{tab:SFE}. The first of these, $\epsilon_{\Delta \text{t}}$, denotes the fraction of gas that is converted to stars within a time interval $\Delta t = t_{\text{end}} - t_{*}$, where $t_{\text{end}}$ is the time when we stop the simulation and $t_{*}$ the time when the first star forms. The other, $\epsilon_{\text{ff}}$, is the SFE per free-fall time, which is computed by extrapolating $\epsilon_{\Delta \text{t}}$ to one free-fall time. That means we evaluate
\begin{equation}
\label{eq:eff}
\epsilon_{\text{ff}} = \epsilon_{\Delta \text{t}} \frac{t_{\text{ff}}}{\Delta \text{t}}.
\end{equation}
We also note that the number of sink particles $N_{\text{sink}}$ that form in each simulation might depend on the specific choice of the sink particle formation threshold $n_{\text{thresh}}$. This number therefore has to be treated with caution, since it is probably not converged, given our resolution, while the SFR is converged (see also \citeauthor{GloverAndClark2012a}~2012a). However, we are not aiming at resolving the IMF in this study, but instead want to obtain the total mass that goes into gravitational collapse, which is correctly described given our numerical setup.

Furthermore, we note that the largest source of error in estimating $\epsilon_{\text{ff}}$ is caused by the extrapolation of $\epsilon_{\Delta \text{t}}$ to $\epsilon_{\text{ff}}$. In general, we assume that the mass accretion rates stay constant through the remaining time of extrapolation. Strictly speaking, this is true only for our $\alpha = 8.0$ and $\alpha = 4.0$ models, as we will see later in Section \ref{subsec:sinks}. Further smaller errors in estimating converged values of $\epsilon_{\text{ff}}$ are caused by the concrete realization of the turbulent velocity field (see also Appendix \ref{sec:seed}), as well as by the specific choice of the sink particle threshold and the accretion radius. Altogether, we think that an error of $\sim20-30$\% is a conservative estimate. This is acceptable, because we are primarily interested in analyzing how different levels of turbulence, density and the ISRF/CRF affect the formation of stars instead of measuring exact and converged values of the individual star formation efficiencies.

In general, we find active star formation for all three initial densities and for all virial $\alpha$ parameters, even in clouds which are unbound due to a high value of $\alpha$. The highest efficiencies are found in the models with small virial parameters, which have the lowest turbulent velocity dispersion of all our models (see Table \ref{tab:setup}). The lowest efficiencies are obtained in the GC models with large virial parameters, which are runs with the highest turbulent velocity dispersions. Regarding our extreme GC-16.0-10000 model, we find a SFE of $\epsilon_{\text{ff}} \approx 0.7\%$ even in this case, although the internal velocity dispersion is very high. Moreover, we generally observe a decreasing star formation rate per free-fall time $\epsilon_{\text{ff}}$ with increasing $\alpha$ (see Fig. \ref{fig:SFEvsAlpha}), which we would expect as well due to the increasing amount of turbulent kinetic energy in the simulation domain.

We quantify this and fit an exponential law to the data points shown in Fig. \ref{fig:SFEvsAlpha} using a $\chi^2$-fit. Our fitting function is defined via
\begin{equation}
\label{eq:fitting}
\epsilon_{\text{ff}} \propto \exp(- c \alpha),
\end{equation}
where $\alpha$ is the virial parameter and $c$ is a constant, depending on the model. For our model with $n_0 = 100\,$cm$^{-3}$, we find $c \approx 0.20 \pm 0.02$, while in the $n = 1000\,$cm$^{-3}$ model we obtain a steeper slope $c \approx 0.36 \pm 0.07$ (see Fig. \ref{fig:SFEvsAlpha}). Thus, we find strong evidence that $\epsilon_{\text{ff}}$ depends not only on the virial state of the cloud, but also on its density. Comparing the values for the SFE for our two density models, we find that $\epsilon_{\text{ff}} \sim n_0^{-0.5}$, which means that the SFE per free-fall time is smaller in high density clouds than in clouds with lower density. Interestingly, this is the same scaling with density as the free-fall time itself, which suggests that the change in $\epsilon_{\text{ff}}$ is driven largely by the change in the free-fall time, rather than by a systematic change in the star formation rate.

In addition, we also run two simulations, SOL-0.5-1000 and SOL-8.0-1000, adopting the lower solar neighbourhood values for the ISRF and the CRF, in order to compare the SFEs we measure in this quiescent environment to those measured for the much harsher GC environment. For the virialized cloud (SOL-0.5-1000), we find an SFE per free-fall time that is around 20\% larger than in our corresponding GC model (GC-0.5-1000), demonstrating that for gravitationally bound clouds, the much stronger heating present in the GC has little effect on the star formation rate. It is notable, however, that we form far fewer sink particles in our GC run than in the corresponding solar neighbourhood run, suggesting that the sinks that do form must be systematically larger. Whether this also leads to a systematic change in the initial mass function of the stars forming in this environment remains to be seen; unfortunately, our resolution is too low to allow us to properly address this question. In our runs with $\alpha = 8.0$, we find a much larger difference between the solar neighbourhood and GC runs. In the solar neighbourhood run, increasing $\alpha$ from 0.5 to 8.0 decreases $\epsilon_{\rm ff}$ by less than a factor of three, whereas in the corresponding GC runs, the change in $\epsilon_{\rm ff}$ is closer to a factor of $\sim16$. Therefore, the combination of high turbulent velocities and strong heating is much more effective at suppressing star formation than either effect individually.

\subsection{Analysis of the volume, temperature and column density Probability Distribution Function (PDF)}
\label{subsec:PDF}

Fig. \ref{fig:PDF100} and \ref{fig:PDF1000} show the mass-weighted volume density and the column density PDF for both density models with $n_0 = 100\,$cm$^{-3}$ and $n_0 = 1000\,$cm$^{-3}$ at different times in the cloud evolution for all virial $\alpha$ parameters. In order to compute the column density PDF, our simulations were projected onto a regular $1024^2$ map. We have chosen various time snapshots in the evolution, so that all PDFs properly reflect the different physical states of the cloud. At later times, however, we have removed some of the low $\alpha$ runs from the plots, since these runs are not evolved until the one free-fall time limit. Nevertheless, we can still compare the more evolved PDFs to the PDFs of the low $\alpha$ runs from the snapshot shown above at an earlier time. All plots only include the remaining total gas without the mass already converted to sink particles.

The mass-weighted volume density PDF shows two pronounced peaks in each of the plots, suggestive of a two-phase medium. The peaks are found at densities of $n \approx 1\,$cm$^{-3}$ and $n \approx 10^3\,$cm$^{-3}$ in the $n_0 = 100\,$cm$^{-3}$ model and at $n \approx 10\,$cm$^{-3}$ and $n \approx 10^4\,$cm$^{-3}$ in the $n_0 = 1000\,$cm$^{-3}$ model, respectively. These peaks in the bimodal PDF can be referred to as those of the diffuse ISM and of the denser regions of the MC, which make up most of the mass. We also note that both peaks are well below the density threshold for sink particle formation. Moreover, both the density variance and the mean density generally increase as the simulations evolve with time in our models. In addition, Fig. \ref{fig:TEMPPDF} shows the corresponding mass-weighted temperature PDF for both density models. We find that most of the gas has temperatures of several $100\,$K for all virial parameters, owing to the strong heating by the ISRF and the CRF. This stands in contrast to the situation in local MCs, where a large fraction of the gas mass has $T\sim10-20\,$K.

Analysing the different temporal evolutions of the PDF for the various $\alpha$ values, we find that the high density regions of runs with larger $\alpha$ values are denser than runs with lower $\alpha$ values during the first $\sim25$\% of the free-fall time. This is due to the higher internal velocity dispersion, which can compress the gas more effectively up to higher densities at the beginning of our runs. At later times, the high density regions of simulations with lower $\alpha$ values become denser due to gravitational contraction of the medium. Comparing the PDF of the two density models in Fig. \ref{fig:PDF100} and \ref{fig:PDF1000} in general, we find that both show a similar shape, except a significant shift to larger density values for the $1000\,$cm$^{-3}$ simulations relative to the $100\,$cm$^{-3}$ simulations, which is due to the one order of magnitude difference in the initial number density. We also note that we observe a power-law tail at the higher end of the different column density PDF, once star formation has set in, consistent with previous work in this field, e.g. \citeauthor{Klessen2000a}~(2000a), \citet{KritsukEtAl2011}, \citet{FederrathAndKlessen2013}, \citet{SchneiderEtAl2013} or \citet{RathborneEtAl2014}.

\subsection{Sink particle formation}
\label{subsec:sinks}

\begin{figure}
\centerline{
\includegraphics[height=0.65\linewidth,width=1.0\linewidth]{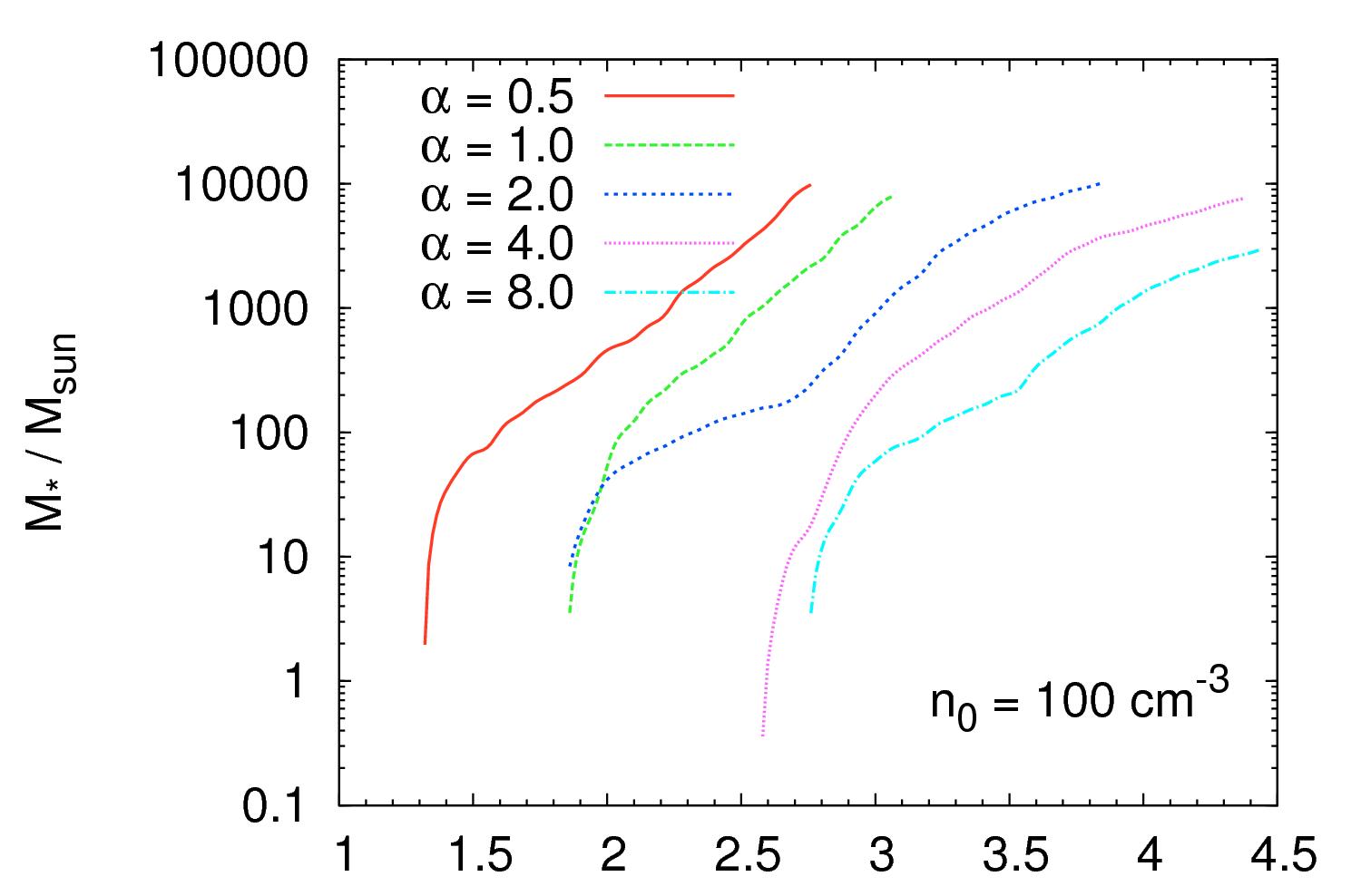} \\
}
\centerline{
\includegraphics[height=0.7\linewidth,width=1.0\linewidth]{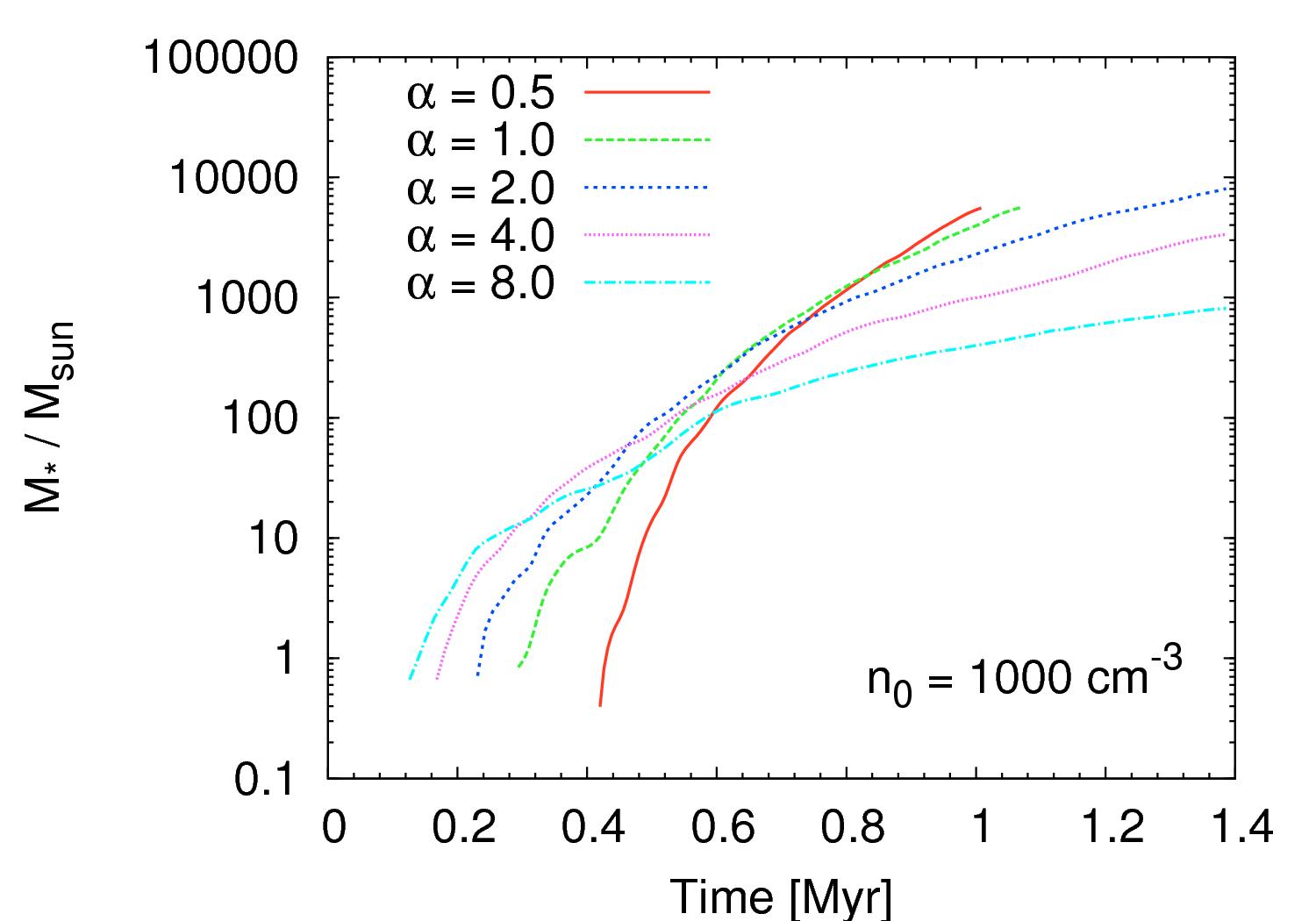} \\
}
\caption{Mass of gas $M_{*}$ converted to stars (sink particles) as a function of time for the different virial $\alpha$ parameters for our models with initial number densities of $n_0 = 100\,$cm$^{-3}$ (top) and $n_0 = 1000\,$cm$^{-3}$ (bottom).}
\label{fig:sinkmass}
\end{figure}

\begin{figure}
\centerline{
\includegraphics[height=0.65\linewidth,width=1.0\linewidth]{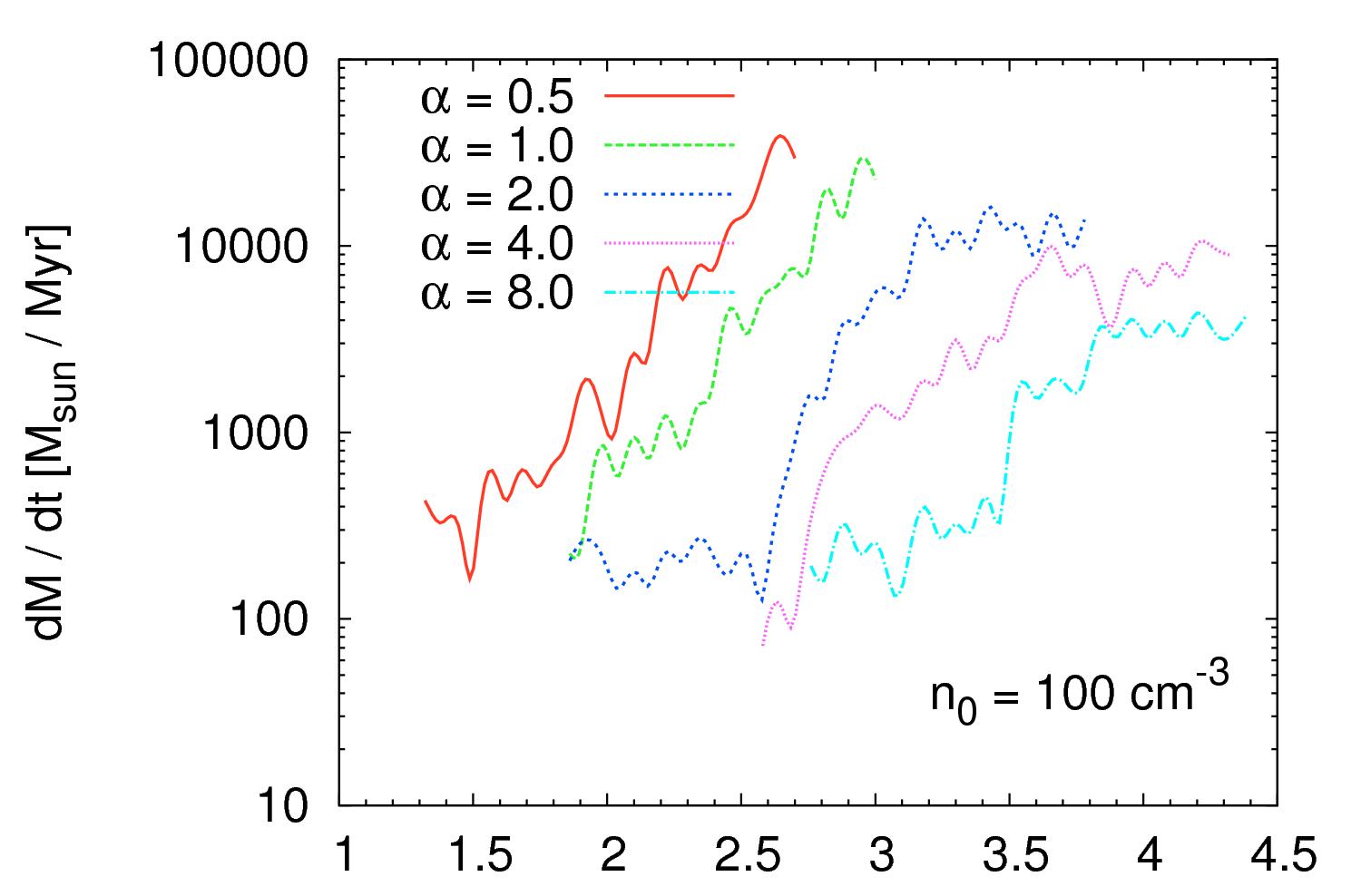} \\
}
\centerline{
\includegraphics[height=0.7\linewidth,width=1.0\linewidth]{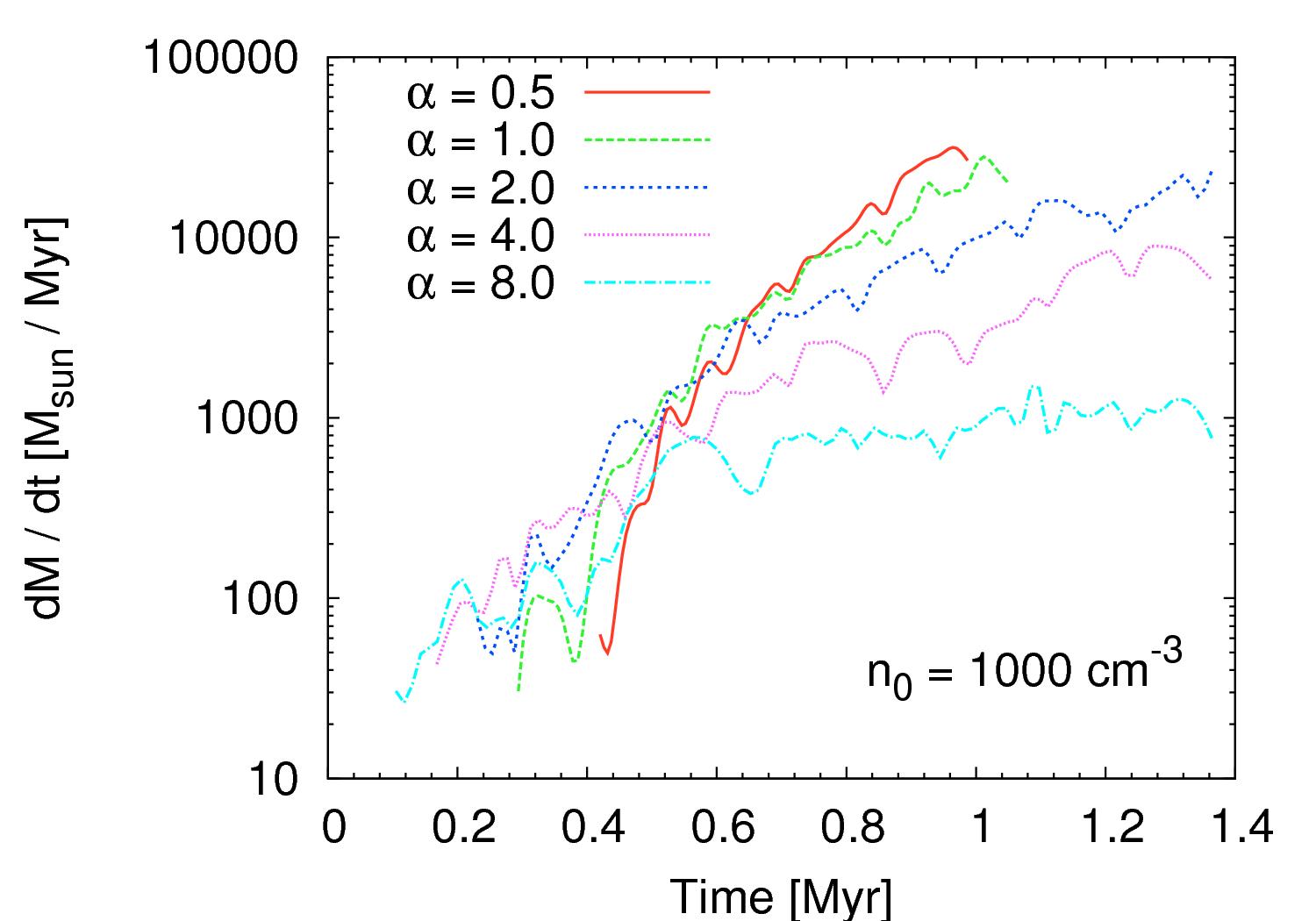} \\
}
\caption{Mass accretion rates $\text{d}M/\text{d}t$ for the corresponding density models and different virial $\alpha$ parameters given in Fig. \ref{fig:sinkmass}.}
\label{fig:rates}
\end{figure}

Fig. \ref{fig:sinkmass} shows the mass of gas that is converted to stars (sink particles) as a function of time for the different $\alpha$ parameters and densities. Depending on the amount of kinetic turbulent energy in the box, we find differences in the temporal evolution of star formation. In our models with a lower initial number density of $n_0 = 100\,$cm$^{-3}$, we find star formation to be triggered by the global collapse of the MC due to gravitational compression. This is more effective for clouds with lower virial parameter. Hence, in this case, turbulence delays and suppresses star formation. For our other models with higher initial number densities of $n_0 = 1000\,$cm$^{-3}$, we find star formation to be triggered by local compression of the gas due to highly turbulent motions, leading to the rapid formation of stars in models with high $\alpha$ values. Overall, star formation is suppressed at high $\alpha$, but not delayed as observed in our low-density model. Instead we note that turbulence can actually trigger and accelerate star formation in localized patches of the cloud \citep[see also the discussion by][]{MacLowAndKlessen2004}.

Sink particles can form once all formation criteria presented in Section \ref{subsec:sinkmethod} are fulfilled. However, highly turbulent motions can locally compress the gas above our sink particle formation threshold more quickly at higher values of the density and the virial parameter. Hence, under our assumptions of sink particle formation, we find that star formation can be triggered by different physical processes, i.e. either by shock compression of the gas leading to overdense regions which fulfill all sink criteria, or by global gravitational collapse. In the latter case, the global collapse dominates the internal velocity dispersions of the cloud, which is e.g. the case in quiescent MCs in the Galaxy. However, we assume that in a CMZ-like environment with highly turbulent motions, star formation is mainly triggered by turbulent shock compression rather than by global gravitational collapse of the cloud as a whole.

Fig. \ref{fig:rates} shows mass accretion rates for the corresponding density models and different virial parameters of our simulations. Depending on the amount of turbulent kinetic energy, we find a strong dependence of the accretion rates on the $\alpha$ parameter. As expected, the gas can be accreted more effectively in case of low $\alpha$ values, i.e. when the kinetic energy is low and the cloud tends to be more gravitationally bound, which holds for all our density models. Furthermore, for those models with a low value of the virial parameter, we find that star formation continues to accelerate during the whole run. If $\alpha$ increases, star formation gets less efficient, the mass accretion rates drop as well and become approximately constant at later times.

\section{Discussion}
\label{sec:discussion}

\subsection{Comparison to star formation in the GC}
\label{subsec:SFinGC}

In this study, we have adopted environmental conditions similar to those experienced by a typical GC cloud in order to see whether high levels of turbulence as well as a much stronger ISRF and CRF will lead to enhanced or reduced star formation rates. As shown by various observations, star formation in the GC is generally thought to be inefficient, i.e.\ having a factor of $\gtrsim10$ smaller star formation rate per free-fall time than what is inferred for local star-forming regions (\citealt{LisEtAl1994}, \citealt{Murray2011}, \citealt{KauffmannEtAl2013}, \citeauthor{LongmoreEtAl2013a}~2013a). If indeed the high level of turbulence and strong heating by the ISRF and by cosmic rays renders GC clouds very inefficient at forming stars, then this would provide a simple explanation for the low efficiency of star formation in the region as a whole.

Our simulations demonstrate that the rate and efficiency at which star formation occurs depends on the $\alpha$ parameter of the cloud. It is therefore useful to look at the typical $\alpha$ values inferred from observations in the GC. It turns out that most clouds and clumps in the GC tend to be rather unbound, i.e.\ having values of $\alpha \gtrsim 1$. For example, we find a mean value $\alpha = 3.0 \pm 1.6$ by computing $\alpha$ for different clumps in the GC cloud GCM-0.02-0.07 based on data given by \citet{TsuboiAndMiyazaki2012}. \citet{KauffmannEtAl2013} evaluated $\alpha$ for the entire GC cloud G0.253+0.016, finding a value of $\alpha \approx 3.8 \pm 1.0$. However, \citet{LongmoreEtAl2012} derived a different value. They estimate this cloud to be roughly in virial equilibrium with $\alpha \approx 1$. In this case, the difference between their estimate and the \citet{KauffmannEtAl2013} value comes from their decision to exclude an additional velocity component in the calculation of $\alpha$. More recently, \citet{RathborneEtAl2015} have computed the virial parameter using data from various observed molecular transitions, finding that the outer regions of G0.253+0.016 may be unbound, while its central region may be bound and collapsing \citep[see Fig. 14 in][]{RathborneEtAl2015}. Furthermore, if we evaluate $\alpha$ for numerous cores from data given in Table 2 in \citet{JohnstonEtAl2014} for the Brick, we also find that most cores tend to be unbound with $\alpha \gtrsim 2.0$. Overall, therefore, it seems plausible that many of the clouds in the GC that are not currently forming stars have $\alpha$ of a few.

Regarding measurements of the star formation efficiencies per free-fall time, \citet{Murray2011} for example observed a Galaxy-wide average value of $\epsilon_{\text{ff}} \approx 0.6\%$, which is comparable to the lowest $\epsilon_{\text{ff}}$ in our simulations for highly unbound clouds. Generally, a typical estimate for the star formation efficiency per free-fall time in the Galaxy is $\sim1\%$, i.e. star formation is quite slow in GMCs \citep{Murray2011,KrumholzEtAl2012}. We recover larger values than this in almost all of our clouds, rather than finding values strongly suppressed compared to the Galactic average, as would be required to explain the low efficiency of star formation at the galactic center. This holds for runs with both low ($\alpha = 0.5$) and high ($\alpha = 8.0$) values of $\alpha$, where $\epsilon_{\text{ff}} \gtrsim 1\%$ for the two models with different initial number density. However, we note that these values reflect observed average efficiencies of star formation in the Galaxy as a whole. Thus, a direct comparison with the SFEs derived from numerical simulations of clouds in isolation presented in this study is complicated, since we are only focussing on one specific physical cloud realization, instead of a larger number of individual MCs.

\subsection{Comparison to previous studies of unbound clouds}
\label{subsec:comparison}

Several previous studies tried to reveal the dependence of the SFE on the virial $\alpha$ parameter. For example, \citet{ClarkEtAl2008} altered the initial level of turbulent support and find that a wide range of SFEs are possible. The SFEs in their study range from up to 60\% to as low as around 0.3\% after two free-fall times. Furthermore, \citet{ClarkEtAl2008} also observed a decreasing SFE with increasing $\alpha$, as confirmed in our study. A similar result was found by \citet{BonnellEtAl2011}, who investigated the formation of young stars in a single MC with a total mass that is a factor of 10 smaller than our clouds. \citet{BonnellEtAl2011} concluded that even small changes in the binding energy of the cloud can cause large variations in terms of the SFE, similar to what we find in this paper. They gave an overall SFE of 15\% when their calculation was terminated. Moreover, \citet{ClarkAndBonnell2005} also showed that unbound clouds result in inefficient star formation, in agreement with \citeauthor{KlessenEtAl2000b}~(2000b), \citet{HeitschEtAl2001}, \citet{ClarkAndBonnell2004}, \citet{ClarkEtAl2008}, \citet{BonnellEtAl2011} and the results in this paper.

Furthermore, \citet{PadoanEtAl2012} analyzed the SFR in supersonic MHD turbulence, also finding that $\epsilon_{\text{ff}}$ decreases exponentially with increasing $\alpha$. In particular, they found that the SFE is insensitive to changes in the sonic Mach number, but sensitive to the Alfv\'enic Mach number. In their study, a decrease in the Alfv\'enic Mach number (equivalent to an increase of the magnetic field) additionally reduced the SFE, but only by a factor less than 2. Hence, although we do not account for MHD turbulence, we conclude that an additional external magnetic field might also slightly reduce our values of $\epsilon_{\text{ff}}$. These findings are also in agreement with previous studies by \citeauthor{KlessenEtAl2000b}~(2000b) and \citet{HeitschEtAl2001}, who analyzed the gravitational collapse in turbulent molecular clouds. They find that star formation cannot be prevented by MHD turbulence, but the magnetic fields delay the local collapse due to the magnetic pressure. Furthermore, they also find that strong turbulence can provide some support on global scales, but may trigger collapse and star formation locally at the stagnation points of convergent shocks, in analogy to the findings in this study.

All these studies analyzed the impact of the turbulent kinetic energy on the SFE. However, regarding the generally more extreme physical conditions in the GC, we have to use a significantly higher initial number density and a stronger ISRF/CRF than what was used in these previous studies in order to better match the environmental parameters found in a typical CMZ-like cloud. Surprisingly, our results are similar to those found in previous studies, although we have made use of a more extreme physical setup regarding the internal velocity, the density and the radiation field.

\subsection{Suppressing star formation}
\label{subsec:supressingSF}

The cloud G0.253+0.016 is supposed to be a typical MC in the GC. However, as shown by \citet{KauffmannEtAl2013} and \citet{JohnstonEtAl2014}, G0.253+0.016 has almost no evidence of current star formation. In contrast to that, our numerical models show active star formation independent of the initial number density and the virial $\alpha$ parameter of the cloud. Indeed, stars form more rapidly in our model clouds than appears to be the case in the Galaxy as a whole. While our neglect of stellar feedback probably explains some of this discrepancy, stellar feedback can only be effective once star formation is ongoing, but cannot explain an almost complete lack of star formation in G0.253+0.016. We are therefore led to conclude that the harsh GC environment and the high level of turbulence present in the GC clouds cannot by themselves produce low enough star formation efficiencies to explain the globally low efficiency of star formation in the GC. They also do not seem to be able to explain why G0.253+0.016 and its neighbouring clouds are not currently forming stars. What then does suppress star formation in the GC?

One possible explanation could be that the turbulent velocity field in many of the GC clouds is composed of a different mixture of solenoidal and compressive modes than in the model clouds in our study (see, e.g. \citeauthor{FederrathEtAl2010a}~2010a, \citeauthor{Federrath2013}~2013, \citeauthor{FederrathAndKlessen2013}~2013). Fundamentally, the reason that high levels of turbulence do not completely suppress star formation in our model clouds is that the same turbulent motions that support the cloud as a whole against collapse also compress some of its gas up to high densities. This high density gas is formed at stagnation points of the turbulent flow and is not supported against gravitational collapse. It is therefore able to form stars efficiently. However, if compressive modes are absent and the velocity field is dominated by purely solenoidal turbulence (generated e.g.\ by the strong shear experienced by the clouds as they orbit the center of the Galaxy), then less gas will be compressed to high densities, and it is plausible that star formation could be more strongly suppressed. In addition, the strong magnetic field present in the GC might reduce the level of star formation by slowing down the collapse of dense cores \citep{PillaiEtAl2015}. However, as mentioned before, this typically leads to a reduction in the star formation rate and efficiency of only about a factor of $\sim2$, see e.g. \citet{PetersEtAl2011}, \citet{HennebelleEtAl2011}, \citet{CommerconEtAl2011} or \citet{SeifriedEtAl2013}.

Alternatively, it might be that the idea that we can explain the low SFE of the GC region as a whole by requiring the individual clouds to all have low SFEs is simply incorrect. Even in our highly turbulent model clouds, star formation does not begin at $t=0$ -- there is a brief period in the evolution of the cloud during which no stars are yet forming. If, as \citeauthor{LongmoreEtAl2013b}~(2013b) suggest, G0.253+0.016 and its neighbouring clouds have only formed very recently, then it may simply be that we are catching them too early in their lives to have started forming significant numbers of stars. In this case, we would need to look elsewhere for an explanation of the galactic center's low star formation rate. In this context, larger-scale effects such as the orbital dynamics of the gas in the deep potential well of the GC, feedback from massive stars in the form of winds and supernovae, or the inflow/outflow of gas into/out of the GC might play important roles by helping to suppress the formation of dense clouds in the GC region. We will explore the possible impact of these various physical processes in a follow-up study.

\subsection{Limitations of the model}
\label{subsec:limitations}

There are a few limitations inherent to our numerical models that one should keep in mind when interpreting our results. Most notably, we do not model feedback from the stars (e.g.\ jets, stellar winds or radiation) that form during the individual runs. This would help to reduce $\epsilon_{\Delta \text{t}}$ and $\epsilon_{\text{ff}}$, but cannot entirely suppress star formation. Furthermore, we do not account for spatial variations in the ISRF or the CRF.

We also deliberately do not account for other important physical effects that might modify the star formation rate and efficiency, such as large-scale dynamics (e.g. spiral arms or spiral instabilities), magnetic fields, galactic tides and shear, supernova feedback or the inflow/outflow of gas. This is because we only want to look at the effects of turbulence and a high ISRF/CRF on cloud dynamics in isolation. We leave the analysis of simulations with further physical effects for future studies.

Moreover, the high computational cost of {\sc Arepo} simulations of clouds with mean densities $n \gg 10^{3}\,$cm$^{-3}$ means that we have to focus on models with number densities $n_0 = 100\,$cm$^{-3}$ and $n_0 = 1000\,$cm$^{-3}$. However, the density of a typical GC cloud like G0.253+0.016 can be even higher than several $\sim10^3\,$cm$^{-3}$, as shown by various studies in the past \citep{LisEtAl2001,ImmerEtAl2012,LongmoreEtAl2012,KauffmannEtAl2013,ClarkEtAl2013,JohnstonEtAl2014}. Therefore, as shown in Section \ref{subsec:ics}, we also run one more extreme model with $n_0 = 10^4\,$cm$^{-3}$ and $\alpha = 16.0$, finding a SFE of $\epsilon_{\text{ff}} = 0.7\%$. Nevertheless, regarding the trends given in Table \ref{tab:SFE}, we expect other high density runs with a virial parameter lower than 16.0 to form stars at an even higher rate due to a lower amount of turbulent kinetic energy.

Fig. \ref{fig:SFEvsAlpha} provides support for this assumption. It clearly shows that $\epsilon_{\text{ff}}$ depends on the mean density $n_0$ roughly as $\epsilon_{\text{ff}} \sim n_0^{-0.5}$ (see Section \ref{subsec:analysis}). If this trend continues to higher density, then typical GC clouds with $n \sim 10^{4}\,$cm$^{-3}$ should have about $\sim1/3$ lower star formation efficiencies per free-fall time than our $n_0 = 1000\,$cm$^{-3}$ model. In this case, we still obtain values of $\epsilon_{\text{ff}} \gtrsim 1\%$, which are higher than the average SFE in the Milky Way (see Section \ref{subsec:SFinGC}). This holds at least for all runs except for the high $\alpha = 8.0$ run, for which we would obtain $\epsilon_{\text{ff}} \sim 0.2\%$, smaller than the value of $\epsilon_{\text{ff}} \approx 0.7\%$ found for the $\alpha = 16.0$ run. However, it is unclear whether this single example can be taken as a systematic evidence for a breaking of our density trends, given the stochasticity with its large uncertainties seen among the results for the $100\,$cm$^{-3}$ and $1000\,$cm$^{-3}$ simulations.

\section{Summary and Conclusions}
\label{sec:summary}

We have performed numerical simulations of molecular clouds with the moving mesh code A{\sc repo} \citep{Springel2010} using environmental properties comparable to those experienced by typical galactic center (GC) clouds. We adopted values for the interstellar radiation field (ISRF) and the cosmic ray flux (CRF) that are a factor of $\sim1000$ larger than the values measured in the solar neighbourhood \citep{ClarkEtAl2013}. We simulated clouds with initial number densities of $n_0 = 100\,$cm$^{-3}$ and $1000\,$cm$^{-3}$ using different virial $\alpha$ parameters of $\alpha = 0.5, 1.0, 2.0, 4.0$ and $8.0$ for each density. The total mass was set to a constant value of $M_{\text{tot}} = 1.3 \times 10^5\,$M$_{\odot}$. In addition, we also ran one more extreme simulation with an initial number density of $10^4\,$cm$^{-3}$ and a virial parameter of $\alpha = 16.0$. Furthermore, we also did two control runs with $n_0 = 1000\,$cm$^{-3}$ as well as $\alpha = 0.5$ and $\alpha = 8.0$ and with ISRF and CRF parameters geared towards the solar neighbourhood. An overview of our model parameters is provided in Table \ref{tab:setup}. We report the following findings:
\begin{itemize}
\item We find active star formation with $\epsilon_{\text{ff}} \gtrsim 1\%$ in all models regardless of the choice of $n_0$ and $\alpha$.
\item Our values are more comparable to Galaxy-wide SFEs than to the inferred SFE in the GC, which observations suggest is a factor $\gtrsim10$ smaller. Star formation is more efficient at lower $\alpha$ values, i.e.\ when the velocity dispersion in the cloud is small.
\item The efficiency of star formation decreases by a factor of $\sim4-10$ as we increase the virial parameter from $\alpha = 0.5$ to $\alpha = 8.0$.
\item We fit exponential functions $\epsilon_{\text{ff}} \propto \exp(- c \alpha)$ to the data, finding $c \approx 0.20 \pm 0.02$ and $c \approx 0.36 \pm 0.07$ for the $n_0 = 100\,$cm$^{-3}$ and $n_0 = 1000\,$cm$^{-3}$ density models, respectively. Thus, we find strong evidence that $\epsilon_{\text{ff}}$ depends not only on the virial state of the cloud, but also on its density. To illustrate this quantitatively, we derive a relation $\epsilon_{\text{ff}} \sim n_0^{-0.5}$ for the star formation efficiencies as a function of the density.
\item For virialized clouds, we find that even a 1000x higher ISRF strength and CRF has only a small effect on the star formation efficiency, decreasing it by around 20\% compared to the value we obtain for a similar cloud in a solar neighbourhood environment. For highly unbound clouds, the stronger ISRF and higher CRF at the galactic center has a much greater effect, decreasing the star formation efficiency by around a factor of 6.
\item Even in our most extreme models, we find star formation efficiencies per free-fall time that are close to 1\%. None of our models produce values that are consistent with the low SFE per free-fall time that is inferred for the galactic center region as a whole.
\end{itemize}
We therefore conclude that the idea tested in this paper -- that the high levels of turbulence present in the GC region, together with the strong ISRF and high CRF combine to yield a persistently low star formation efficiencies within the dense clouds in this region -- does not appear to work in practice. It is possible that including additional physical ingredients (such as magnetic fields, stellar feedback or realistic orbital parameters around the GC) could reduce the star formation efficiencies within individual dense clouds to levels that are consistent with the mean value inferred for the GC region as a whole. Alternatively, it could be that the idea that we can explain the low SFE of the GC region as a whole by requiring the individual clouds to all have low SFEs is incorrect, and that the bottleneck for star formation in the region is actually the assembly of the dense clouds themselves. This is in agreement with \citet{KruijssenEtAl2014}, who speculate that the rate-limiting factor for star formation is the slow evolution of the gas towards collapse. Distinguishing between these two possibilities awaits further work on this topic.

\section*{Acknowledgements}

We thank J. M. Diederik Kruijssen, Katharine Johnston, Rowan Smith, Christian Baczynski, Mark Krumholz and Javier Ballesteros Paredes for informative discussions about the project. We also thank the referee for a timely and very constructive report, which helped to improve the paper a lot. EB, SCOG and RSK acknowledge support from the Deutsche Forschungsgemeinschaft (DFG) via the SFB 881 (sub-projects B1, B2, B5 and B8) ``The Milky Way System'', and the SPP (priority program) 1573, ``Physics of the ISM''. Furthermore, EB acknowledges financial support from the Konrad-Adenauer-Stiftung (KAS) via their ``Promotionsf\"orderung''. The simulations presented in this paper were performed on the Milkyway supercomputer at the J\"ulich Forschungszentrum, funded via SFB 881. Additional simulations were performed on the \textit{kolob} cluster at the University of Heidelberg, which is funded in part by the DFG via Emmy-Noether grant BA 3706, and via a Frontier grant of Heidelberg University, sponsored by the German Excellence Initiative as well as the Baden-W\"urttemberg Foundation. RSK acknowledges support from the European Research Council under the European Community's Seventh Framework Programme (FP7/2007-2013) via the ERC Advanced Grant "STARLIGHT: Formation of the First Stars" (project number 339177).

\begin{appendix}

\section{Simulations with a different random seed}
\label{sec:seed}

\begin{figure}
\centerline{
\includegraphics[height=0.59\linewidth,width=1.0\linewidth]{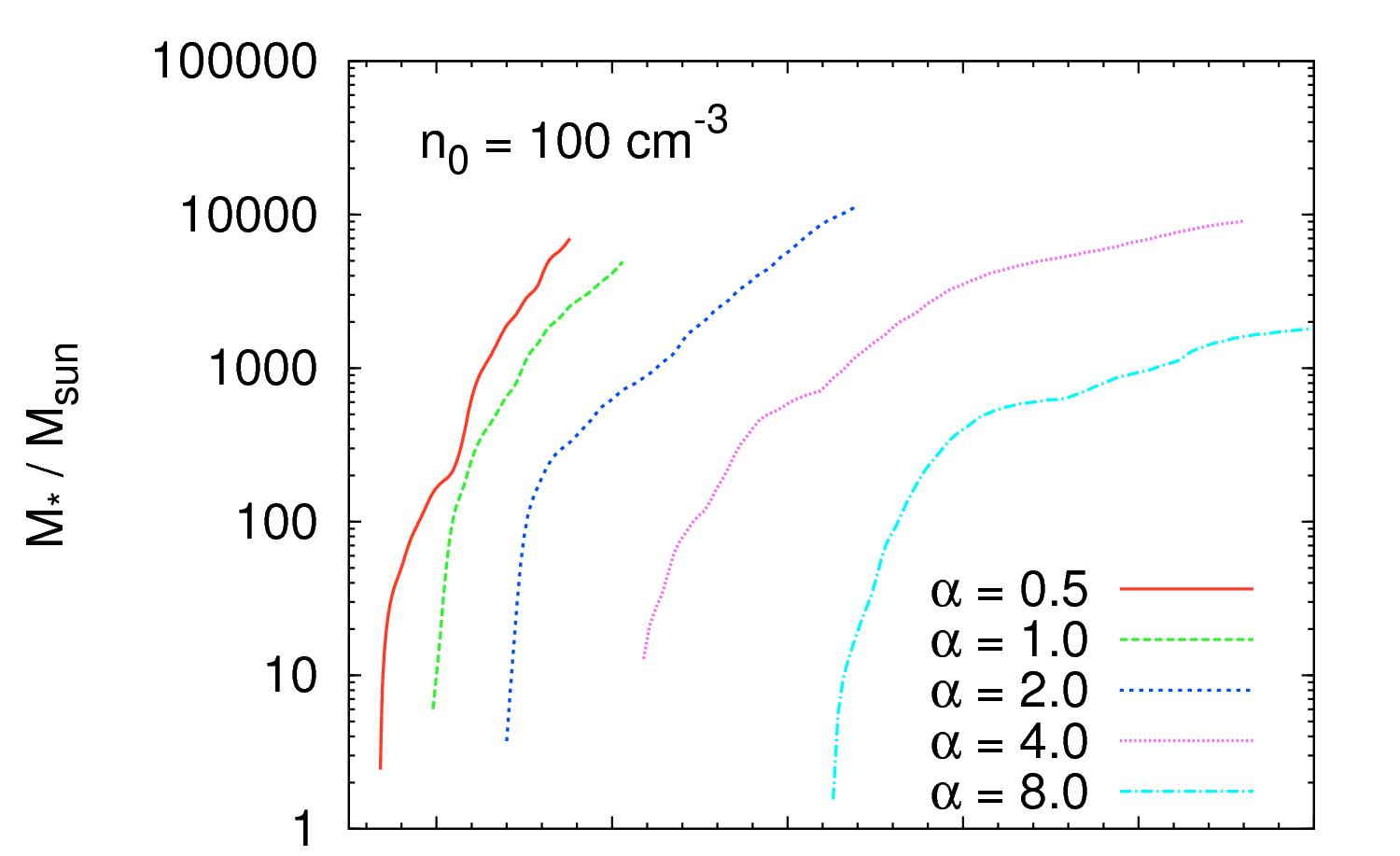} \\
}
\centerline{
\includegraphics[height=0.63\linewidth,width=1.0\linewidth]{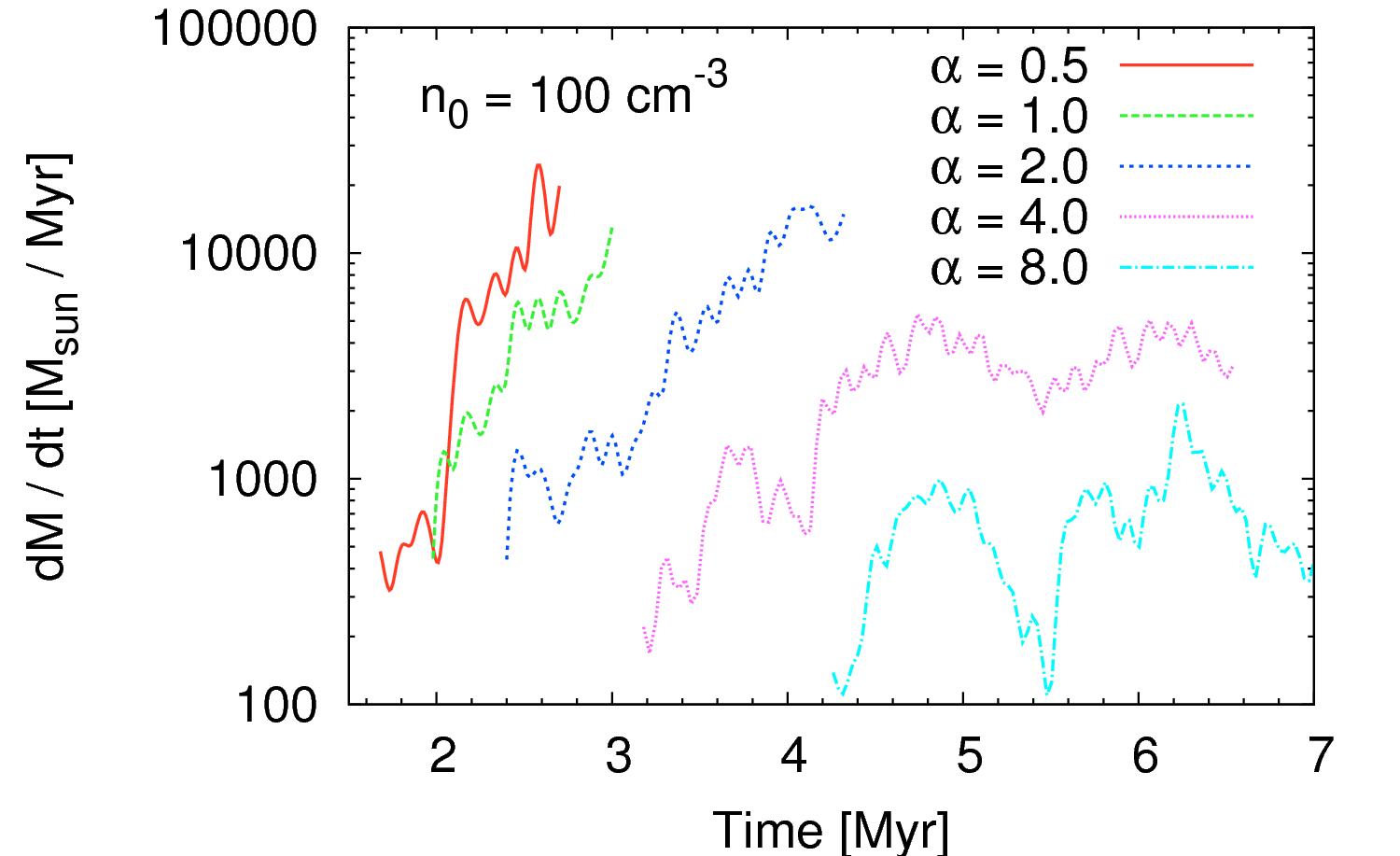} \\
}
\caption{Same as Fig. \ref{fig:sinkmass} and \ref{fig:rates}, but using a different random seed for the turbulent velocity field. The number of sink particles formed by the end of each run, as well as the SFEs, are given in Table \ref{tab:SFE_seed}.}
\label{fig:sinkmass_seed}
\end{figure}

We ran numerical simulations with the same setup described in Section \ref{subsec:ics} for an initial number density of $n_0 = 100\,$cm$^{-3}$ using a different random seed for the turbulent velocity field. Fig. \ref{fig:sinkmass_seed} shows the mass of gas that is converted to stars (sink particles) for the different $\alpha$ parameters as a function of time and their accretion rates. Table \ref{tab:SFE_seed} gives an overview of the star formation efficiencies and the number of sink particles that are formed in each cloud. Since star formation sets in later in this turbulent environment compared to our fiducial models presented in Section \ref{subsec:analysis}, we let the two runs GC-4.0-100-SEED and GC-8.0-100-SEED evolve beyond one free-fall time in order to get a significant number of sink particles. Although star formation starts at later times, caused by the different statistical flows of the turbulent velocity field, we again measure large efficiencies $\epsilon_{\text{ff}}$ for all numerical models even in highly turbulent environments. We also find the same statistical trends as already observed in Table \ref{tab:SFE}, i.e.\ that the SFEs per free-fall time strongly depend on the virial parameter. However, a direct comparison of the $\epsilon_{\text{ff}}$ values in models with various random seeds is complicated due to the statistics of the turbulent velocity fields, leading to a completely different star formation history and thus to variable SFE.

\begin{table}
\begin{tabular}{l|c|c|c|c|c|c}
\hline\hline
Model name & $\epsilon_{\Delta \text{t}}$ & $\epsilon_{\text{ff}}$ & $N_{\text{sink}}$ & $t_{*}$ & $t_{\text{end}}$ & $\Delta \text{t}$ \\
 & [\%] & [\%] & & [Myr] & [Myr] & [Myr] \\
\hline
GC-0.5-100-SEED & 5.7 & 24.3 & 2483 & 1.70 & 2.73 & 1.03 \\
GC-1.0-100-SEED & 4.1 & 15.0 & 1420 & 1.98 & 3.10 & 1.12 \\
GC-2.0-100-SEED & 9.4 & 20.7 & 2717 & 2.40 & 4.40 & 2.00 \\
GC-4.0-100-SEED & 7.6 & 10.0 & 2067 & 3.20 & 6.56 & 3.36 \\
GC-8.0-100-SEED & 1.1 & 1.8 & 405 & 4.25 & 7.00 & 2.75 \\
\hline
\end{tabular}
\caption{Same as Table \ref{tab:SFE}, but using a different random seed for the turbulent velocity field.}
\label{tab:SFE_seed}
\end{table}

\bibliographystyle{mn2e}
\bibliography{lit/literature}

\begin{thebibliography}{}

\bibitem[\protect\citeauthoryear{{Ballesteros-Paredes}, {Klessen}, {Mac Low} \&
  {Vazquez-Semadeni}}{{Ballesteros-Paredes}
  et~al.}{2007}]{Ballesteros-ParedesEtAl2007}
{Ballesteros-Paredes} J.,  {Klessen} R.~S.,  {Mac Low} M.-M.,
  {Vazquez-Semadeni} E.,  2007, Protostars and Planets V, pp 63--80

\bibitem[\protect\citeauthoryear{{Bate}, {Bonnell} \& {Price}}{{Bate}
  et~al.}{1995}]{BateEtAl1995}
{Bate} M.~R.,  {Bonnell} I.~A.,    {Price} N.~M.,  1995, \mnras, 277, 362

\bibitem[\protect\citeauthoryear{{Bonnell}, {Smith}, {Clark} \&
  {Bate}}{{Bonnell} et~al.}{2011}]{BonnellEtAl2011}
{Bonnell} I.~A.,  {Smith} R.~J.,  {Clark} P.~C.,    {Bate} M.~R.,  2011,
  \mnras, 410, 2339

\bibitem[\protect\citeauthoryear{{Clark} \& {Bonnell}}{{Clark} \&
  {Bonnell}}{2004}]{ClarkAndBonnell2004}
{Clark} P.~C.,  {Bonnell} I.~A.,  2004, \mnras, 347, L36

\bibitem[\protect\citeauthoryear{{Clark}, {Bonnell} \& {Klessen}}{{Clark}
  et~al.}{2008}]{ClarkEtAl2008}
{Clark} P.~C.,  {Bonnell} I.~A.,    {Klessen} R.~S.,  2008, \mnras, 386, 3

\bibitem[\protect\citeauthoryear{{Clark}, {Bonnell}, {Zinnecker} \&
  {Bate}}{{Clark} et~al.}{2005}]{ClarkAndBonnell2005}
{Clark} P.~C.,  {Bonnell} I.~A.,  {Zinnecker} H.,    {Bate} M.~R.,  2005,
  \mnras, 359, 809

\bibitem[\protect\citeauthoryear{{Clark}, {Glover} \& {Klessen}}{{Clark}
  et~al.}{2012}]{ClarkEtAl2012}
{Clark} P.~C.,  {Glover} S.~C.~O.,    {Klessen} R.~S.,  2012, \mnras, 420, 745

\bibitem[\protect\citeauthoryear{{Clark}, {Glover}, {Ragan}, {Shetty} \&
  {Klessen}}{{Clark} et~al.}{2013}]{ClarkEtAl2013}
{Clark} P.~C.,  {Glover} S.~C.~O.,  {Ragan} S.~E.,  {Shetty} R.,    {Klessen}
  R.~S.,  2013, \apjl, 768, L34

\bibitem[\protect\citeauthoryear{{Commer{\c c}on}, {Hennebelle} \&
  {Henning}}{{Commer{\c c}on} et~al.}{2011}]{CommerconEtAl2011}
{Commer{\c c}on} B.,  {Hennebelle} P.,    {Henning} T.,  2011, \apjl, 742, L9

\bibitem[\protect\citeauthoryear{{Draine}}{{Draine}}{1978}]{Draine1978}
{Draine} B.~T.,  1978, \apjs, 36, 595

\bibitem[\protect\citeauthoryear{{Elmegreen} \& {Scalo}}{{Elmegreen} \&
  {Scalo}}{2004}]{ElmegreenAndScalo2004}
{Elmegreen} B.~G.,  {Scalo} J.,  2004, \araa, 42, 211

\bibitem[\protect\citeauthoryear{{Federrath}}{{Federrath}}{2013}]{Federrath2013}
{Federrath} C.,  2013, \mnras, 436, 1245

\bibitem[\protect\citeauthoryear{{Federrath} \& {Klessen}}{{Federrath} \&
  {Klessen}}{2013}]{FederrathAndKlessen2013}
{Federrath} C.,  {Klessen} R.~S.,  2013, \apj, 763, 51

\bibitem[\protect\citeauthoryear{{Federrath}, {Roman-Duval}, {Klessen},
  {Schmidt} \& {Mac Low}}{{Federrath} et~al.}{010a}]{FederrathEtAl2010a}
{Federrath} C.,  {Roman-Duval} J.,  {Klessen} R.~S.,  {Schmidt} W.,    {Mac
  Low} M.-M.,  2010a, \aap, 512, A81

\bibitem[\protect\citeauthoryear{{Glover} \& {Clark}}{{Glover} \&
  {Clark}}{012a}]{GloverAndClark2012a}
{Glover} S.~C.~O.,  {Clark} P.~C.,  2012a, \mnras, 421, 9

\bibitem[\protect\citeauthoryear{{Glover} \& {Clark}}{{Glover} \&
  {Clark}}{012b}]{GloverAndClark2012b}
{Glover} S.~C.~O.,  {Clark} P.~C.,  2012b, \mnras, 421, 116

\bibitem[\protect\citeauthoryear{{Glover}, {Federrath}, {Mac Low} \&
  {Klessen}}{{Glover} et~al.}{2010}]{GloverEtAl2010}
{Glover} S.~C.~O.,  {Federrath} C.,  {Mac Low} M.-M.,    {Klessen} R.~S.,
  2010, \mnras, 404, 2

\bibitem[\protect\citeauthoryear{{Glover} \& {Mac Low}}{{Glover} \& {Mac
  Low}}{2007}]{GloverAndMacLow2007}
{Glover} S.~C.~O.,  {Mac Low} M.-M.,  2007, \apjs, 169, 239

\bibitem[\protect\citeauthoryear{{Greif}, {Springel}, {White}, {Glover},
  {Clark}, {Smith}, {Klessen} \& {Bromm}}{{Greif} et~al.}{2011}]{GreifEtAl2011}
{Greif} T.~H.,  {Springel} V.,  {White} S.~D.~M.,  {Glover} S.~C.~O.,  {Clark}
  P.~C.,  {Smith} R.~J.,  {Klessen} R.~S.,    {Bromm} V.,  2011, \apj, 737, 75

\bibitem[\protect\citeauthoryear{{G\"usten}, {Walmsley} \& {Pauls}}{{G\"usten}
  et~al.}{1981}]{GuestenEtAl1981}
{G\"usten} R.,  {Walmsley} C.~M.,    {Pauls} T.,  1981, \aap, 103, 197

\bibitem[\protect\citeauthoryear{{Habing}}{{Habing}}{1968}]{Habing1968}
{Habing} H.~J.,  1968, \bain, 19, 421

\bibitem[\protect\citeauthoryear{{Heitsch}, {Mac Low} \& {Klessen}}{{Heitsch}
  et~al.}{2001}]{HeitschEtAl2001}
{Heitsch} F.,  {Mac Low} M.-M.,    {Klessen} R.~S.,  2001, \apj, 547, 280

\bibitem[\protect\citeauthoryear{{Hennebelle}, {Commer{\c c}on}, {Joos},
  {Klessen}, {Krumholz}, {Tan} \& {Teyssier}}{{Hennebelle}
  et~al.}{2011}]{HennebelleEtAl2011}
{Hennebelle} P.,  {Commer{\c c}on} B.,  {Joos} M.,  {Klessen} R.~S.,
  {Krumholz} M.,  {Tan} J.~C.,    {Teyssier} R.,  2011, \aap, 528, A72

\bibitem[\protect\citeauthoryear{{Immer}, {Menten}, {Schuller} \&
  {Lis}}{{Immer} et~al.}{2012}]{ImmerEtAl2012}
{Immer} K.,  {Menten} K.~M.,  {Schuller} F.,    {Lis} D.~C.,  2012, \aap, 548,
  A120

\bibitem[\protect\citeauthoryear{{Jappsen}, {Klessen}, {Larson}, {Li} \& {Mac
  Low}}{{Jappsen} et~al.}{2005}]{JappsenEtAl2005}
{Jappsen} A.-K.,  {Klessen} R.~S.,  {Larson} R.~B.,  {Li} Y.,    {Mac Low}
  M.-M.,  2005, \aap, 435, 611

\bibitem[\protect\citeauthoryear{{Johnston}, {Beuther}, {Linz}, {Schmiedeke},
  {Ragan} \& {Henning}}{{Johnston} et~al.}{2014}]{JohnstonEtAl2014}
{Johnston} K.~G.,  {Beuther} H.,  {Linz} H.,  {Schmiedeke} A.,  {Ragan} S.~E.,
    {Henning} T.,  2014, \aap, 568, A56

\bibitem[\protect\citeauthoryear{{Kauffmann}, {Pillai} \& {Zhang}}{{Kauffmann}
  et~al.}{2013}]{KauffmannEtAl2013}
{Kauffmann} J.,  {Pillai} T.,    {Zhang} Q.,  2013, \apjl, 765, L35

\bibitem[\protect\citeauthoryear{{Kennicutt} \& {Evans}}{{Kennicutt} \&
  {Evans}}{2012}]{KennicuttAndEvans2012}
{Kennicutt} R.~C.,  {Evans} N.~J.,  2012, \araa, 50, 531

\bibitem[\protect\citeauthoryear{{Kennicutt}
  Jr.}{{Kennicutt}}{1998}]{Kennicutt1998}
{Kennicutt} Jr. R.~C.,  1998, \apj, 498, 541

\bibitem[\protect\citeauthoryear{{Klessen}}{{Klessen}}{000a}]{Klessen2000a}
{Klessen} R.~S.,  2000a, \apj, 535, 869

\bibitem[\protect\citeauthoryear{{Klessen} \& {Glover}}{{Klessen} \&
  {Glover}}{2014}]{KlessenAndGlover2014}
{Klessen} R.~S.,  {Glover} S.~C.~O.,  2014, ArXiv e-prints, 1412.5182

\bibitem[\protect\citeauthoryear{{Klessen}, {Heitsch} \& {Mac Low}}{{Klessen}
  et~al.}{000b}]{KlessenEtAl2000b}
{Klessen} R.~S.,  {Heitsch} F.,    {Mac Low} M.-M.,  2000b, \apj, 535, 887

\bibitem[\protect\citeauthoryear{{Kritsuk}, {Norman} \& {Wagner}}{{Kritsuk}
  et~al.}{2011}]{KritsukEtAl2011}
{Kritsuk} A.~G.,  {Norman} M.~L.,    {Wagner} R.,  2011, \apjl, 727, L20

\bibitem[\protect\citeauthoryear{{Kruijssen}, {Dale} \& {Longmore}}{{Kruijssen}
  et~al.}{2015}]{KruijssenEtAl2015}
{Kruijssen} J.~M.~D.,  {Dale} J.~E.,    {Longmore} S.~N.,  2015, \mnras, 447,
  1059

\bibitem[\protect\citeauthoryear{{Kruijssen}, {Longmore}, {Elmegreen},
  {Murray}, {Bally}, {Testi} \& {Kennicutt}}{{Kruijssen}
  et~al.}{2014}]{KruijssenEtAl2014}
{Kruijssen} J.~M.~D.,  {Longmore} S.~N.,  {Elmegreen} B.~G.,  {Murray} N.,
  {Bally} J.,  {Testi} L.,    {Kennicutt} R.~C.,  2014, \mnras, 440, 3370

\bibitem[\protect\citeauthoryear{{Krumholz}, {Dekel} \& {McKee}}{{Krumholz}
  et~al.}{2012}]{KrumholzEtAl2012}
{Krumholz} M.~R.,  {Dekel} A.,    {McKee} C.~F.,  2012, \apj, 745, 69

\bibitem[\protect\citeauthoryear{{Krumholz} \& {Kruijssen}}{{Krumholz} \&
  {Kruijssen}}{2015}]{KrumholzEtAl2015}
{Krumholz} M.~R.,  {Kruijssen} J.~M.~D.,  2015, ArXiv e-prints, 1505.07111

\bibitem[\protect\citeauthoryear{{Krumholz} \& {McKee}}{{Krumholz} \&
  {McKee}}{2005}]{KrumholzAndMcKee2005}
{Krumholz} M.~R.,  {McKee} C.~F.,  2005, \apj, 630, 250

\bibitem[\protect\citeauthoryear{{Krumholz} \& {Tan}}{{Krumholz} \&
  {Tan}}{2007}]{KrumholzAndTan2007}
{Krumholz} M.~R.,  {Tan} J.~C.,  2007, \apj, 654, 304

\bibitem[\protect\citeauthoryear{{Lis} \& {Menten}}{{Lis} \&
  {Menten}}{1998}]{LisAndMenten1998}
{Lis} D.~C.,  {Menten} K.~M.,  1998, \apj, 507, 794

\bibitem[\protect\citeauthoryear{{Lis}, {Menten}, {Serabyn} \& {Zylka}}{{Lis}
  et~al.}{1994}]{LisEtAl1994}
{Lis} D.~C.,  {Menten} K.~M.,  {Serabyn} E.,    {Zylka} R.,  1994, \apjl, 423,
  L39

\bibitem[\protect\citeauthoryear{{Lis}, {Serabyn}, {Zylka} \& {Li}}{{Lis}
  et~al.}{2001}]{LisEtAl2001}
{Lis} D.~C.,  {Serabyn} E.,  {Zylka} R.,    {Li} Y.,  2001, \apj, 550, 761

\bibitem[\protect\citeauthoryear{{Longmore}, {Bally}, {Testi}, {Purcell},
  {Walsh}, {Bressert}, {Pestalozzi}, {Molinari}, {Ott}, {Cortese}, {Battersby},
  {Murray}, {Lee}, {Kruijssen}, {Schisano} \& {Elia}}{{Longmore}
  et~al.}{013a}]{LongmoreEtAl2013a}
{Longmore} S.~N.,  {Bally} J.,  {Testi} L.,  {Purcell} C.~R.,  {Walsh} A.~J.,
  {Bressert} E.,  {Pestalozzi} M.,  {Molinari} S.,  {Ott} J.,  {Cortese} L.,
  {Battersby} C.,  {Murray} N.,  {Lee} E.,  {Kruijssen} J.~M.~D.,  {Schisano}
  E.,    {Elia} D.,  2013a, \mnras, 429, 987

\bibitem[\protect\citeauthoryear{{Longmore}, {Kruijssen}, {Bally}, {Ott},
  {Testi}, {Rathborne}, {Bastian}, {Bressert}, {Molinari}, {Battersby} \&
  {Walsh}}{{Longmore} et~al.}{013b}]{LongmoreEtAl2013b}
{Longmore} S.~N.,  {Kruijssen} J.~M.~D.,  {Bally} J.,  {Ott} J.,  {Testi} L.,
  {Rathborne} J.,  {Bastian} N.,  {Bressert} E.,  {Molinari} S.,  {Battersby}
  C.,    {Walsh} A.~J.,  2013b, \mnras, 433, L15

\bibitem[\protect\citeauthoryear{{Longmore}, {Rathborne}, {Bastian}, {Alves},
  {Ascenso}, {Bally}, {Testi}, {Longmore}, {Battersby}, {Bressert}, {Purcell},
  {Walsh}, {Jackson}, {Foster}, {Molinari}, {Meingast}, {Amorim} \&
  {Lima}}{{Longmore} et~al.}{2012}]{LongmoreEtAl2012}
{Longmore} S.~N.,  {Rathborne} J.,  {Bastian} N.,  {Alves} J.,  {Ascenso} J.,
  {Bally} J.,  {Testi} L.,  {Longmore} A.,  {Battersby} C.,  {Bressert} E.,
  {Purcell} C.,  {Walsh} A.,  {Jackson} J.,  {Foster} J.,  {Molinari} S.,
  {Meingast} S.,  {Amorim} A.,    {Lima} J.,  2012, \apj, 746, 117

\bibitem[\protect\citeauthoryear{{Mac Low} \& {Klessen}}{{Mac Low} \&
  {Klessen}}{2004}]{MacLowAndKlessen2004}
{Mac Low} M.-M.,  {Klessen} R.~S.,  2004, Reviews of Modern Physics, 76, 125

\bibitem[\protect\citeauthoryear{{McKee} \& {Ostriker}}{{McKee} \&
  {Ostriker}}{2007}]{McKeeAndOstriker2007}
{McKee} C.~F.,  {Ostriker} E.~C.,  2007, \araa, 45, 565

\bibitem[\protect\citeauthoryear{{Molinari}, {Bally}, {Noriega-Crespo},
  {Compi{\`e}gne}, {Bernard}, {Paradis}, {Martin}, {Testi}, {Barlow}, {Moore},
  {Plume}, {Swinyard}, {Zavagno}, {Calzoletti}, {Di Giorgio} \&
  {Elia}}{{Molinari} et~al.}{2011}]{MolinariEtAl2011}
{Molinari} S.,  {Bally} J.,  {Noriega-Crespo} A.,  {Compi{\`e}gne} M.,
  {Bernard} J.~P.,  {Paradis} D.,  {Martin} P.,  {Testi} L.,  {Barlow} M.,
  {Moore} T.,  {Plume} R.,  {Swinyard} B.,  {Zavagno} A.,  {Calzoletti} L.,
  {Di Giorgio} A.~M.,    {Elia} 2011, \apjl, 735, L33

\bibitem[\protect\citeauthoryear{{Murray}}{{Murray}}{2011}]{Murray2011}
{Murray} N.,  2011, \apj, 729, 133

\bibitem[\protect\citeauthoryear{{Nelson} \& {Langer}}{{Nelson} \&
  {Langer}}{1997}]{NelsonAndLanger1997}
{Nelson} R.~P.,  {Langer} W.~D.,  1997, \apj, 482, 796

\bibitem[\protect\citeauthoryear{{Padoan}, {Haugb{\o}lle} \&
  {Nordlund}}{{Padoan} et~al.}{2012}]{PadoanEtAl2012}
{Padoan} P.,  {Haugb{\o}lle} T.,    {Nordlund} {\AA}.,  2012, \apjl, 759, L27

\bibitem[\protect\citeauthoryear{{Peters}, {Banerjee}, {Klessen} \& {Mac
  Low}}{{Peters} et~al.}{2011}]{PetersEtAl2011}
{Peters} T.,  {Banerjee} R.,  {Klessen} R.~S.,    {Mac Low} M.-M.,  2011, \apj,
  729, 72

\bibitem[\protect\citeauthoryear{{Pillai}, {Kauffmann}, {Tan}, {Goldsmith},
  {Carey} \& {Menten}}{{Pillai} et~al.}{2015}]{PillaiEtAl2015}
{Pillai} T.,  {Kauffmann} J.,  {Tan} J.~C.,  {Goldsmith} P.~F.,  {Carey} S.~J.,
     {Menten} K.~M.,  2015, \apj, 799, 74

\bibitem[\protect\citeauthoryear{{Rathborne}, {Longmore}, {Jackson}, {Foster},
  {Contreras}, {Garay}, {Testi}, {Alves}, {Bally}, {Bastian}, {Kruijssen} \&
  {Bressert}}{{Rathborne} et~al.}{2014}]{RathborneEtAl2014}
{Rathborne} J.~M.,  {Longmore} S.~N.,  {Jackson} J.~M.,  {Foster} J.~B.,
  {Contreras} Y.,  {Garay} G.,  {Testi} L.,  {Alves} J.~F.,  {Bally} J.,
  {Bastian} N.,  {Kruijssen} J.~M.~D.,    {Bressert} E.,  2014, \apj, 786, 140

\bibitem[\protect\citeauthoryear{{Rathborne}, {Longmore}, {Jackson},
  {Kruijssen}, {Alves}, {Bally}, {Bastian}, {Contreras}, {Foster}, {Garay},
  {Testi} \& {Walsh}}{{Rathborne} et~al.}{2015}]{RathborneEtAl2015}
{Rathborne} J.~M.,  {Longmore} S.~N.,  {Jackson} J.~M.,  {Kruijssen} J.~M.~D.,
  {Alves} J.~F.,  {Bally} J.,  {Bastian} N.,  {Contreras} Y.,  {Foster} J.~B.,
  {Garay} G.,  {Testi} L.,    {Walsh} A.~J.,  2015, ArXiv e-prints,
  arXiv:1501.07368

\bibitem[\protect\citeauthoryear{{Scalo} \& {Elmegreen}}{{Scalo} \&
  {Elmegreen}}{2004}]{ScaloAndElmegreen2004}
{Scalo} J.,  {Elmegreen} B.~G.,  2004, \araa, 42, 275

\bibitem[\protect\citeauthoryear{Schmidt}{Schmidt}{1959}]{Schmidt1959}
Schmidt M.,  1959, Astrophys.J., 129, 243

\bibitem[\protect\citeauthoryear{{Schneider}, {Andr{\'e}}, {K{\"o}nyves},
  {Bontemps}, {Motte}, {Federrath}, {Ward-Thompson}, {Arzoumanian},
  {Benedettini}, {Bressert}, {Didelon}, {Di Francesco}, {Griffin}, {Hennemann}
  \& {Hill}}{{Schneider} et~al.}{2013}]{SchneiderEtAl2013}
{Schneider} N.,  {Andr{\'e}} P.,  {K{\"o}nyves} V.,  {Bontemps} S.,  {Motte}
  F.,  {Federrath} C.,  {Ward-Thompson} D.,  {Arzoumanian} D.,  {Benedettini}
  M.,  {Bressert} E.,  {Didelon} P.,  {Di Francesco} J.,  {Griffin} M.,
  {Hennemann} M.,    {Hill} T.,  2013, \apjl, 766, L17

\bibitem[\protect\citeauthoryear{{Seifried}, {Banerjee}, {Pudritz} \&
  {Klessen}}{{Seifried} et~al.}{2013}]{SeifriedEtAl2013}
{Seifried} D.,  {Banerjee} R.,  {Pudritz} R.~E.,    {Klessen} R.~S.,  2013,
  \mnras, 432, 3320

\bibitem[\protect\citeauthoryear{{Sembach}, {Howk}, {Ryans} \&
  {Keenan}}{{Sembach} et~al.}{2000}]{SembachEtAl2000}
{Sembach} K.~R.,  {Howk} J.~C.,  {Ryans} R.~S.~I.,    {Keenan} F.~P.,  2000,
  \apj, 528, 310

\bibitem[\protect\citeauthoryear{{Shetty}, {Kelly} \& {Bigiel}}{{Shetty}
  et~al.}{2013}]{ShettyEtAl2013}
{Shetty} R.,  {Kelly} B.~C.,    {Bigiel} F.,  2013, \mnras, 430, 288

\bibitem[\protect\citeauthoryear{{Shetty}, {Kelly}, {Rahman}, {Bigiel},
  {Bolatto}, {Clark}, {Klessen} \& {Konstandin}}{{Shetty}
  et~al.}{2014}]{ShettyEtAl2014a}
{Shetty} R.,  {Kelly} B.~C.,  {Rahman} N.,  {Bigiel} F.,  {Bolatto} A.~D.,
  {Clark} P.~C.,  {Klessen} R.~S.,    {Konstandin} L.~K.,  2014, \mnras, 437,
  L61

\bibitem[\protect\citeauthoryear{{Smith}, {Glover}, {Clark}, {Klessen} \&
  {Springel}}{{Smith} et~al.}{014a}]{SmithEtAl2014a}
{Smith} R.~J.,  {Glover} S.~C.~O.,  {Clark} P.~C.,  {Klessen} R.~S.,
  {Springel} V.,  2014a, \mnras, 441, 1628

\bibitem[\protect\citeauthoryear{{Smith}, {Glover} \& {Klessen}}{{Smith}
  et~al.}{014b}]{SmithEtAl2014b}
{Smith} R.~J.,  {Glover} S.~C.~O.,    {Klessen} R.~S.,  2014b, \mnras, 445,
  2900

\bibitem[\protect\citeauthoryear{{Springel}}{{Springel}}{2010}]{Springel2010}
{Springel} V.,  2010, \mnras, 401, 791

\bibitem[\protect\citeauthoryear{{Truelove}, {Klein}, {McKee}, {Holliman} II,
  {Howell}, {Greenough} \& {Woods}}{{Truelove} et~al.}{1998}]{TrueloveEtAl1998}
{Truelove} J.~K.,  {Klein} R.~I.,  {McKee} C.~F.,  {Holliman} II J.~H.,
  {Howell} L.~H.,  {Greenough} J.~A.,    {Woods} D.~T.,  1998, \apj, 495, 821

\bibitem[\protect\citeauthoryear{{Tsuboi} \& {Miyazaki}}{{Tsuboi} \&
  {Miyazaki}}{2012}]{TsuboiAndMiyazaki2012}
{Tsuboi} M.,  {Miyazaki} A.,  2012, \pasj, 64, 111

\bibitem[\protect\citeauthoryear{{Yusef-Zadeh}, {Wardle} \&
  {Roy}}{{Yusef-Zadeh} et~al.}{2007}]{Yusef-ZadehEtAl2007}
{Yusef-Zadeh} F.,  {Wardle} M.,    {Roy} S.,  2007, \apjl, 665, L123

\end{thebibliography}

\end{appendix}
\end{document}